\documentclass{article}
\usepackage{graphicx} 
\usepackage{geometry}
\usepackage{amsfonts}
\usepackage{amsmath}
\usepackage{amsthm}
\usepackage{amssymb}
\usepackage{braket}
\usepackage{biblatex}
\usepackage{xcolor}
\usepackage{tikz}
\usepackage{hyperref}
\usetikzlibrary{positioning}
\usepackage{ragged2e}

\newcommand{\norm}[1]{\left\lVert #1 \right\rVert}
\newcommand{\ketbra}[2]{\left| #1 \right\rangle \left\langle #2 \right|}

\newtheorem{thm}{Theorem}

\newtheorem{prop}{Proposition}
\newtheorem{defn}{Definition}

\author{\begingroup
\hypersetup{urlcolor=navyblue}
\href{https://orcid.org/0009-0006-5387-5291}{Emily Beatty~\footnote{\href{mailto:emily.beatty@ens-lyon.fr}{emily.beatty@ens-lyon.fr}}
\endgroup} \\ {\em Univ Lyon, ENS Lyon, UCBL, CNRS, Inria, LIP, F-69342, Lyon Cedex 07, France}
}
\date{}

\title{Wasserstein Distances on Quantum Structures: an Overview}

\begin{document}

\maketitle

\begin{abstract}
    
The theory of optimal transport of probability measures has wide-ranging applications across a number of different fields, including concentration of measure, machine learning, Markov chains, and economics. The generalisation of optimal transport tools from probability measures to quantum states has shown great promise over the last few years, particularly in the development of the theory of Wasserstein-style distances and divergences between quantum states. Such distances have already led to a broad range of developments in the quantum setting such as functional inequalities, convergence of solutions in many-body physics, improvements to quantum generative adversarial networks, and more. However, the literature in this field is quite scattered, with very few links between different works and no real consensus on a `true' quantum Wasserstein distance. The aim of this review is to bring these works together under one roof and give a full overview of the state of the art in the development of quantum Wasserstein distances. We also present a variety of open problems and unexplored avenues in the field, and examine the future directions of this promising line of research. This review is written for those interested in quantum optimal transport in coming from both the fields of classical optimal transport and of quantum information theory, and as a resource for those working in one area of quantum optimal transport interested in how existing work may relate to their own.

\end{abstract}

\section{Introduction} 
\label{section:intro}
The theory of optimal transport was introduced by Monge in his memoir ``{\em Sur la théorie des déblais et des remblais}'' \cite{monge-1781} in 1781, covering the practically motivated issue of efficiently transporting piles of resources from one arrangement to another. The most straightforward problem of optimal transport begins with a space, a starting distribution, a target distribution, and a cost function on the space defining how costly it is to transport from one place in the space to another. The basic aim is to work out the minimal overall cost to transport the starting distribution to the target distribution, and in some cases to work out the transport function which gives this minimal cost.

This quickly transforms into a problem in measure theory. The space is naturally a measurable space, the starting and target distributions naturally probability distributions, the cost function naturally a well-defined map from the product of the space and itself to the non-negative reals, the transport function naturally a function from the space to itself which pushes the starting distribution forward to the target distribution, and the overall cost of such a function naturally an integral over the starting distribution of the cost between a point in the space and its target.
The central object of this area is the family of classical Wasserstein distances \cite{wasserstein-encyc-maths}. These are metrics on the set of probability distributions which encode the optimal transport cost with respect to some underlying metric on the state space.

The theory of optimal transport has numerous applications in economics~\cite{villani2008optimal,santambrogio-2009,sinkhorn-1964,buttazzo-2009,15_min_city}, functional inequalities~\cite{otto-villani-2000,lott-villani-2005,gentil-hwi-2020,milman-1983,erbar-maas-2012}, Markov chains \cite{path_coupling}, heat flows \cite{otto-jordan-1988,schrodinger-connections-2013,schrodinger-connections-meanfield-2020} and more \cite{pmlr-v70-arjovsky17a,classical_applications_signals_machinelearning}, and recent years have shown a flurry of interest in generalising this field to the non-commutative setting to try and replicate these applications in the theory of quantum states. This follows a general pattern of research in quantum information theory: we have many tools for research in probability theory and functional analysis, and few tools for research in quantum information. However, as quantum states can be considered as a non-commutative analogue of probability distributions, the translation of concepts from probability to quantum information gives us new tools to enhance our understanding of quantum states and unlocks new applications in all areas related to quantum physics: computing, chemistry, theormodynamics, machine learning, and more.

As will be discussed in Section \ref{subsection:quantumbg_obstacles}, despite much progress and research over the last decade it appears that there is yet no `true' quantum generalisation of the classical Wasserstein distances. Differences between the classical and quantum settings, grounded in non-commutativity and entanglement, mean that verbatim translations of definitions are not straightforward. The quantum marginal problem \cite{Haapasalo_2021} is a barrier to a quantum version of the gluing lemma which is a key step in the proof of the triangle inequality in the classical setting \cite[page 94]{villani2008optimal}, and defining a cost matrix for a given underlying geometric structure of a quantum system proves to be a difficult task (see Section \ref{section:coupling}). Each of the current definitions has chosen something to sacrifice - such as the triangle inequality \cite{SC-19-GAN}, flexibility in the underlying geometry of the system \cite{GdP-20-W1}, or faithfulness \cite{GPDT-19-QC} - in return for a definition which satisfies the rest of the properties. This means that each of the proposals can be applied in different scenarios, depending on which properties are relevant to the task at hand. This has led to exciting results in the quantum setting whose classical counterparts rely on only a few properties of the classical Wasserstein distances, but that applications relying on many different properties of the classical Wasserstein distances, such as rapid mixing of Markov chains, are as yet untouched in the classical setting.

In particular, the difficulty of naïvely generalising the classical definition has led to promising proposals to generalise equivalent definitions for special cases, such as a dual formulation for first-order Wasserstein distances \cite{GdP-20-W1,GdP-23-ST} or a dynamical formulation of second-order Wasserstein distances \cite{CM-16-GF}. These allow previously intractable properties like the triangle inequality to be exploited in certain frameworks for applications such as concentration inequalities, lower bounds on computational complexity, simulation of open quantum systems, and more.
This does mean, however, that different definitions of quantum Wasserstein distances bear little to no resemblance to one other, and it is difficult to draw links from one framework to another. The aim of this work is to bring together existing definitions of quantum Wasserstein distances, to give an introduction to the topic, to encourage investigations of links between different definitions to develop unifying frameworks of quantum Wasserstein distances, and to explore existing open problems and the landscape of future directions of this field. The varied results stemming from the existing constructions show the breadth of potential in quantum generalisations of Wasserstein distances, though the existing theory is nowhere near complete.

This review is organised as follows.
Section \ref{section:notation_refs} introduces notation and gives some standard references in this area. The following introductory parts, Sections \ref{section:classical} and \ref{section:quantumbg}, are written with two different readers in mind. Section \ref{section:classical} covers classical optimal transport and is intended for those familiar with quantum mechanics but new to optimal transport theory. Section \ref{section:quantumbg} covers the quantum information theoretical prerequisites and thus is intended for those familiar with classical optimal transport theory but with no background in quantum mechanics. The end of this section also discusses desirable properties of a quantum generalisation of Wasserstein distances and the obstacles to a good generalisation arising from non-commutativity and entanglement.

In Sections \ref{section:coupling}, \ref{section:dynamics}, and \ref{section:lipschitz}, we discuss existing generalisations of quantum Wasserstein distances, split by the theory defining their approach. Section \ref{section:coupling} covers those definitions which follow a {\em coupling} approach, Section \ref{section:dynamics} those which follow a {\em dynamical approach}, and Section \ref{section:lipschitz} a {\em Lipschitz} approach. For more details of this split, see Section \ref{subsection:quantumbg_obstacles}. In each of these sections we discuss the applications brought out by each of the different approaches, focusing on those which have not been replicated by any other method. In Section \ref{section:globalapps} we discuss applications of the quantum Wasserstein distances common to all three broad approaches.
We conclude in Section \ref{section:perspectives} with a review of the open directions in the field, and in Section \ref{section:conclusion} with an overview of the state of the art.

There are also broader topics in the theory of non-commutative optimal transport which may be of interest to the reader, but which we will not cover in this overview. Notably, we will not discuss generalisations of optimal transport from the theory of free probability \cite{BV-01-FP}, nor entropy regularised optimal transport \cite{caputo2024quantumoptimaltransportconvex, gerolin2024noncommutativeoptimaltransportsemidefinite, MW-21-DF}, nor distances defined between quantum channels or unitary operations \cite{XQ-23-QW, RD-23-QC}. Nor will we discuss the distance defined in \cite{Zyczkowski_1998} via the pullback of the classical Wasserstein distance through the Husimi Q-representation, though this may also be of interest to the reader. We focus here on fundamentally quantum generalisations of classical Wasserstein metrics to the space of quantum states.

\subsection{A note for readers and authors}

This article is intended to be a resource for the community surrounding quantum optimal transport. Its aim is threefold. Firstly, it serves to introduce those familiar with classical optimal transport to the quantum generalisation of this problem, assuming no knowledge of the mathematical formalisms of quantum mechanics. Secondly, it serves to introduce those familiar with quantum information theory to the problem of generalising classical optimal transport to the quantum setting, assuming no knowledge of the classical problem beyond basic probability. Thirdly, it serves to bring together a fairly fractured field in which papers approach the central problem in vastly different ways, with different viewpoints, motivations, and aims. It acts to bring these frameworks together under one roof, highlighting fundamental similarities and differences between them.

Though many attempts have been made to include all relevant definitions of quantum Wasserstein distances, it is inevitable that some of the existing definitions will have slipped through the cracks of this search. If you believe any definitions or articles have been overlooked, you are strongly encouraged to contact the author at the email address provided.

\section{Notation and standard references}
\label{section:notation_refs}
\subsection{The Monge-Kantorovich problem}

The Monge problem~\cite{monge-1781}
discusses the cost of transport between probability distributions. For a discussion of optimal transport in the classical setting, the book and its revised version \cite{villani2008optimal} by Villani are very detailed, though for a more gentle introduction one might consider the set of lecture notes \cite{Ollivier_Pajot_Villani_2014} from a summer school in Grenoble in 2009. Those familiar with the classical problem of optimal transport are encouraged to skip to Section \ref{subsection:notation_density_operators}

\begin{defn}
    For measurable spaces $\mathcal{X}$ and $\mathcal{Y}$ equipped with probability measures $\mu, \nu$ respectively, a {\em transport map} is a map $T : \mathcal{X} \to \mathcal{Y}$ for which the pushforward measure $T_{\#} \mu $, given by $T_{\#}\mu (B) = \mu(T_{-1}(B))$, satisfies $T_{\#}\mu  = \nu$. Given then a {\em cost function} $c : \mathcal{X} \times \mathcal{Y} \to \mathbb{R}_{\geq 0}$, in which $c(x,y)$ represents the cost of moving one unit of mass from point $x \in \mathcal{X}$ to $y \in \mathcal{Y}$, the {\em Monge optimal transport cost} is then

\begin{equation}
    \mathcal{T}^c_M(\mu,\nu) = \inf_{T: \mathcal{X} \to \mathcal{Y}, T_{\#} \mu = \nu} \int_{\mathcal{X}} c(x,T(x)) \text{d} \mu(x).
\end{equation}
\end{defn}

It is possible, for some $\mu, \nu$, that no map $T$ exists. For example, in discrete spaces $\mathcal{X} = \mathcal{Y} = \{0,1\}$, with $\mu = \delta_0$ as a point mass and $\nu = \frac{1}{2}(\delta_0 + \delta_1)$ uniform, there is no map $T$ with $T_{\#} \mu = \nu$. To avoid this we instead study the relaxation of the problem due to Kantorovich~\cite{kantorovich-1942}
where we replace transport maps with {\em couplings}.

\begin{defn}
A coupling of measures $\mu$ and $\nu$ is a measure $\omega$ on the product space $\mathcal{X} \times \mathcal{Y}$ whose marginals on $\mathcal{X}$ and $\mathcal{Y}$ are $\mu$ and $\nu$ respectively. Formally, for the subspace projections $\pi_{\mathcal{X}} : \mathcal{X} \times \mathcal{Y} \to \mathcal{X}$ and $\pi_{\mathcal{Y}} : \mathcal{X} \times \mathcal{Y} \to \mathcal{Y}$, $\omega$ has $(\pi_{\mathcal{X}})_{\#} \omega = \mu$ and $(\pi_{\mathcal{Y}})_{\#} \omega = \nu$.
\end{defn}
For sets $\mathcal{A} \subseteq \mathcal{X}, \mathcal{B} \subseteq \mathcal{Y}$, we interpret $\omega(\mathcal{A},\mathcal{B})$ as the amount of measure transported from $\mathcal{A}$ to $\mathcal{B}$, and the marginal conditions ensure that we do indeed transport $\mu$ onto $\nu$. Write $\mathcal{C}(\mu,\nu)$ for the set of couplings of $\mu$ with $\nu$.

\begin{defn} \label{defn:kantorovich_relaxation_classical_cost}
    
For couplings, the {\em optimal transport cost} is then
\begin{equation} \label{eq:defn_tc_couplings}
    \mathcal{T}^c(\mu,\nu) = \inf_{\omega \in \mathcal{C}(\mu,\nu)} \int_\mathcal{X} c(x,y) \text{d} \omega(x,y).
\end{equation}
\end{defn}

This is a relaxation in the sense that any transport map $T$ leads to a coupling $(\text{id},T)_{\#} \mu$ of $\mu$ and $\nu$. In many cases, the Kantorovich problem reduces to the Monge problem. The exact conditions for this reduction vary wildly depending on $c$, and are discussed in detail in \cite[Chapter 10, e.g. Theorems 10.28, 10.38, 10.42]{villani2008optimal}. A typical example \cite[Chapter 1]{villani2008optimal} would be in the case where $\mathcal{X}$ is a Euclidean space, $c$ is strictly convex, and $\mu$ and $\nu$ are absolutely continuous with respect to the Lebesgue measure. However, when studying quantum states we often work with quantum analogues of discrete probability spaces, for which there is often no transport map $T$, and so from here on when referring to classical optimal transport we refer exclusively to the Kantorovich relaxation of the problem. We also assume from here on that all our measurable spaces are Polish and all costs $c$ are lower semicontinuous, to avoid being distracted by pathological examples with zero relevance to the quantum setting. In this case, \cite{figalli-book-2023}
guarantees the existence of an optimal coupling.

\subsection{The Wasserstein distances}

Much of the theory of classical optimal transport focuses specifically on the case where $\mathcal{X} = \mathcal{Y}$, and $\mathcal{X}$ is equipped with a metric $d$. For any $p \geq 1$, we can take our cost function $c$ to be defined by $c(x,y) = d(x,y)^p$. This leads to the key definition of the classical {\em Wasserstein distances}.
\begin{defn}
    Let $d$ be a metric on $\mathcal{X}$. The \textbf{classical Wasserstein distance}\cite{villani2008optimal} of order $p$ with respect to $d$ between two probability measures $\mu$, $\nu$ is
\begin{equation}
    \mathcal{W}_p^{d}(\mu,\nu) = \left(\mathcal{T}^{d^p}(\mu,\nu) \right)^{1/p}.
\end{equation}
\end{defn}
Similarly, the classical Wasserstein distance for $p = \infty$ is defined as
\begin{equation}
    \mathcal{W}_\infty^d(\mu,\nu) = \inf_{\omega \in \mathcal{C}(\mu,\nu)} \sup_{(x,y) \in \text{supp}(\omega)} c(x,y).
\end{equation}
In the case where $\mathcal{X}$ is equipped with the discrete metric $d(x,y) = 1-\delta_{xy}$, we recover the total variation distance $\mathcal{W}_1^d(\mu, \nu) = |\mu-\nu|_{\text{TV}}$ for $p = 1$ \cite[page 10]{villani2008optimal}. Much of the literature surrounding quantum optimal transport discusses quantum generalisations of these quantities, as (see Section \ref{section:classical}) most of the applications of classical optimal transport come directly from the classical Wasserstein distances and so it seems likely that applications to quantum states will follow suit.
For Wasserstein distances between classical probability distributions, we always write calligraphic $\mathcal{W}_p^d$ for order $p$ with respect to distance $d$.

\subsection{Density operator formalism} \label{subsection:notation_density_operators}

The theory in this section comes from the book Quantum Computation and Quantum Information by Nielsen and Chuang \cite{Nielsen_Chuang}, which is also the standard reference for any background in quantum information theory in the so-called density operator formalism, specifically those topics with a computational flavour. Those familiar with quantum information theory are encouraged to skip to Section \ref{section:classical}.

The model used most commonly in quantum information theory is that which models quantum states as operators on a Hilbert space. Given a Hilbert space $\mathcal{H}$, a quantum state is a postitive semi-definite linear operator $\rho$ of trace $1$. Such operators are called {\em density operators} and the set of such operators is notated $\mathcal{D}(\mathcal{H})$. Where $\mathcal{H}$ is separable, via the spectral theorem any such state can be written in the form
\begin{equation}
    \rho = \sum_{i} \lambda_i \ket{v_i} \bra{v_i}
\end{equation}
for some orthonormal basis $\{\ket{v_i}\}_i$ of $\mathcal{H}$, writing $\bra{v_i}$ for the adjoint. The properties $\sum_i \lambda_i = 1$, $\lambda_i \geq 0$ mean these states behave similarly to probability distributions. One of the broad aims of quantum information theory is therefore, among other things, to exploit this link to translate well-established tools from classical probability theory into the realm of quantum states, in order to replicate and apply their applications to improve our understanding of quantum physics.

The choice of Hilbert space informs our perspective on the physical system involved. Two common choices are $\mathcal{H} = L^2(\mathbb{R}^d)^{\otimes n}$ and $\mathcal{H} = \left(\mathbb{C}^d \right)^{\otimes n}$. The first, encompassing continous-variable quantum information theory, allows us to model quantum states in a way that highlights their physical properties of position and momentum. The second, referred to as a quantum spin system, treats each copy of $\mathbb{C}^d$ as a qudit (quantum $d$-dimensional bit) and so allows us to model $n$-qudit computational systems.

States with rank one are called {\em pure} states, written as $\ket{\psi}\bra{\psi}$ for some vector $\ket{\psi}$. We also refer to such a $\ket{\psi}$ as a pure state, though the distinction is always clear from context. These $\ket{\psi}$ refer interchangeably to vectors themselves and vectors representing elements of $\mathbb{P}{\mathcal{H}}$, with the convention that the inner product $\braket{\psi | \psi} = 1$. The notation for a vector $\ket{\psi}$ is known as a \textit{ket}, and its adjoint $\bra{\psi}$ as a \textit{bra}, with the whole notational system referred to as \textit{bra-ket notation}.  A state $\rho$ which is not pure is called {\em mixed}.

Important to the theory of optimal transport is the definition of the marginal, given in quantum information theory by the partial trace. For the amalgamation of two systems $\mathcal{H}_1$ and $\mathcal{H}_2$, we take quantum states $\rho_{12}$ on the tensor product $\mathcal{H}_1 \otimes \mathcal{H}_2$. The equivalent of the marginal on the first space is then the partial trace $\text{Tr}_2$ over the second, where $\text{Tr}_2 : \mathcal{B}(\mathcal{H}_1 \otimes \mathcal{H}_2) \to \mathcal{B}(\mathcal{H}_1)$ is the linear map uniquely defined such that for all product operators $X \otimes Y \in \mathcal{B}(\mathcal{H}_1 \otimes \mathcal{H}_2)$, we have $\text{Tr}_2[X \otimes Y] = X \text{Tr}[Y]$.

We present in Table \ref{tab:cq-dictionary} a `classical-quantum dictionary' for the convenience of those readers less familiar with the quantum setting.

\begin{table}[htbp]
    \centering
    \begin{tabular}{|c|c||c|c|} \hline
       Classical object  & Notation & Quantum object & Notation   \\ \hline
       Probability space  & $\Omega$ & Hilbert space & $\mathcal{H}$ \\
       Probability measure & $\mu, \nu \in \mathcal{P}(\Omega)$ & Quantum state & $\rho, \sigma \in \mathcal{D}(\mathcal{H})$ \\
       Point mass & $\delta_x$ & Pure quantum state & $\ketbra{\psi}{\psi}$ \\
       Functions & $X, Y$ & Hermitian operator & $O$ \\
       Integral & $\int$ & Trace & $\text{Tr}$ \\
       Expectation value & $\mathbb{E}_\mu [X] = \int_\Omega X(\omega) \text{d}\mu(\omega)$ & Expectation value & $\langle O \rangle_{\rho} = \text{Tr}[O\rho]$ \\
       Product of spaces & $\Omega_1 \times \Omega_2$ & Product of spaces & $\mathcal{H}_A \otimes \mathcal{H}_B$ \\
       Marginal on $\Omega_1$ & $\mu_1 = (\pi_1)_{\#} \mu_{12}$ & Partial trace & $\rho_1 = \text{Tr}_2[\rho_{12}]$\\
       Information entropy & $H(\mu) = -\sum_{x \in \Omega} \mu(x) \log \mu(x) $ & von Neumann entropy & $S(\rho) = -\text{Tr}[\rho \log \rho]$ \\
       Relative entropy & $D(\mu \| \nu) = \sum_{x \in \Omega} \mu(x) \log \frac{\mu(x)}{\nu(x)}$ & Relative entropy & $D(\rho \| \sigma) = \text{Tr}[\rho(\log \rho - \log \sigma)]$ \\
       \hline

    \end{tabular}
    \caption{A classical-quantum dictionary: common objects in classical probability and their quantum counterparts.}
    \label{tab:cq-dictionary}
\end{table}

Generally speaking, this correspondence allows us to translate tools from probability theory into the language of quantum mechanics, in order to better explain the latter. However, this is not free. The non-commutativity of $\mathcal{D}(\mathcal{H})$ means these tools cannot always be naïvely generalised whilst maintaining their important properties, and aspects of entanglement in quantum mechanics means they may not always behave in the way we expect. On the other hand, this can unveil new properties or relations which help us better understand the nature of quantum objects.

\subsection{Von Neumann algebra formalism}

A more powerful but also more involved model used in quantum information theory is that in which quantum states are modelled as linear functionals on a von Neumann algebra. This is not strictly necessary for anything we discuss in this review. However, we find that some of the quantities defined, particularly those from Section \ref{subsection:coupling_modular_plans}, rely on the von Neumann algebra model for their intuition. A von Neumann algebra $M$ is a $*$-subalgebra of the algebra $\mathcal{B}(\mathcal{H})$ of bounded operators on a Hilbert space $\mathcal{H}$ that contains the unit $1$ and is closed in the weak operator topology. We say that a linear functional $\omega : \mathcal{B}(\mathcal{H}) \to \mathbb{C}$ is {\em positive} when, for all $A \in \mathcal{B}(\mathcal{H})$, we have $\omega(A^*A) \geq 0$, and \textit{normalised} when $\omega(1) = 1$. We then model quantum states as positive normalised linear functionals on $M$. Such a state is said to be \textit{faithful} or \textit{nondegenerate} if $\omega(A) = 0 \implies A = 0$. For a more detailed introduction to von Neumann algebras, see \cite{vnas}.

For any density operator $\rho \in \mathcal{D}(\mathcal{H})$, a state on $M$ can be defined by $\omega_\rho (A) = \text{Tr}[\rho A]$. Such states are called {\em normal} states on the von Neumann algebra. It is not the case that all states on $\mathcal{B}(\mathcal{H})$ are normal (unless $\mathcal{H}$ is finite-dimensional). However, if we impose a continuity assumption on states (namely that they are $\sigma$-weakly continuous), then all states on $\mathcal{B}(\mathcal{H})$, and consequently on $M$, are indeed normal. In many cases, we use this correspondence freely.

When extending the von Neumann algebra formalism to the infinite-dimensional case, we move to the language of $C^*$-algebras. For ease of presentation, everything presented in this review will use either the density matrix formalism or the von Neumann algebra formalism. Note, however, that many of the results presented in the von Neumann algebra framework do also extend to the framework of $C^*$-algebras, another mathematical framework describing quantum mechanics which we will not discuss here for brevity. For more details, the lecture notes by Landsman \cite{landsman1998lecturenotescalgebrashilbert} give a good introduction to $C^*$-algebras and their use in quantum mechanics.

An important area~\cite{opalgs} from the language of von Neumann algebras used in quantum mechanics is modular theory. The key object used is the {\em modular operator} and its associated group of {\em modular automorphisms}.
\begin{defn} \label{defn:modular_operator_group}
    Let $\sigma$ be a nondegenerate density matrix on a von Neumann algebra $M$. The \textbf{modular operator} associated to $\sigma$ is 
    \begin{equation}
        \Delta_\sigma A = \sigma A \sigma^{-1}.
    \end{equation}
    This gives the \textbf{modular automorphism group} associated to $\sigma$, which is $(\alpha_t)_{t \in \mathbb{C}}$ for $\alpha_t = \Delta_\sigma^{-it}$.
\end{defn}

\section{Classical optimal transport and its applications} 
\label{section:classical}
Again, those familiar with classical optimal transport and its various reformulations are encouraged to skip to Section \ref{section:quantumbg}. Again, for more detail on the topics discussed in this section we refer to Villani's book \cite{villani2008optimal}, and the lecture notes from a summer school in Grenoble in 2009 \cite{Ollivier_Pajot_Villani_2014}.

\subsection{Kantorovich duality} \label{subsection:classical_kantorovich}

An important reformulation of the optimal transport cost is the dual form due to Kantorovich. Indeed for cost $c:\mathcal{X}\times\mathcal{Y}\to\mathbb{R}_{\geq 0}$, and sets $C_b(\mathcal{X})$ and $C_b(\mathcal{Y})$ the continuous bounded functions on $\mathcal{X}$ and $\mathcal{Y}$ respectively, the following dual form holds:

\begin{equation} \label{eq:kant_duality_fg}
    \mathcal{T}^c(\mu,\nu) = \sup_{f \in C_b(\mathcal{X}), g \in C_b(\mathcal{Y}), f + g \leq c } \int_\mathcal{X} f(x) \text{d} \mu(x) + \int_\mathcal{Y} g(y) \text{d} \nu(y).
\end{equation}

In the particular case of the $\mathcal{W}_1^d$ distance, when $\mathcal{X} = \mathcal{Y}$ and is equipped with metric $d$, the optimisation over $f + g \leq d$ reduces to the optimisation over $f$ 1-Lipschitz with respect to $d$, and $g = -f$. In this case,

\begin{equation} \label{eq:classical-lipschitz}
    \mathcal{W}_1^d(\mu,\nu) = \sup_{\norm{f}_L \leq 1} \int_{\mathcal{X}} f(x)\text{d}(\mu-\nu)(x).
\end{equation}

For the supremum \eqref{eq:kant_duality_fg}, it is also possible to show under minimal assumptions that this is in fact a maximum. It follows that, for an optimal coupling $\hat{\omega}$ and optimal $\hat{f}, \hat{g}$ that $c(x,y) = \hat{f}(x) + \hat{g}(y)$ $\omega$-almost everywhere. Assuming differentiability, this gives $\nabla_x c(x,y) = \nabla f(x)$ for $\omega$-a.e. $(x,y)$. In the quadratic case ($\mathcal{X} = \mathcal{Y} = \mathbb{R}^d$), this leads to the well-known Brenier theorem.

\begin{thm}
    Let $\mathcal{X} = \mathcal{Y} = \mathbb{R}^d$ and let $c(x,y) = \frac{1}{2}\norm{x-y}^2$. Let $\mu, \nu$ be probability measures on $\mathbb{R}^d$ such that $\norm{\cdot}$ has finite second moment under $\mu, \nu$ and such that $\mu \ll \text{d}x$. Then there exists a unique optimal transport plan $\hat{\omega}$ from $\mu$ to $\nu$ and $\hat{\omega} = (\text{id} \times \nabla \phi )_{\#} \mu$ for some convex function $\phi$.
\end{thm}

\subsection{Benamou-Brenier} \label{subsection:classical_benamou_brenier}

When $\mathcal{X} = \mathcal{R}^d$ or a convex subset $\Omega \subseteq \mathbb{R}^d$, we can further study the $\mathcal{W}_2$ distance via the dynamics of the flow.
For a smooth bounded vector field $v : [0,T] \times \Omega \to \mathbb{R}^d$ tangent to $\partial \Omega$ we can define the flow $X(t,x)$ of $v$ via
\begin{equation}
    \begin{cases}
        & \dot{X}(t,x) = v(t,X(t,x)) \\
        & X(0,x) = x.
    \end{cases}
\end{equation}

For a probability density $\mu$ on $\Omega$ with finite second moment, we can then define its flow under $v$ via $\mu_t = (X(t,\cdot))_{\#} \mu$. The tangency condition means that the flow stays inside $\Omega$ so that $\mu_t$ is supported on $\Omega$.

As a result of this definition, the flow $\mu_t$ solves the {\em continuity equation}
\begin{equation} \label{eq:continuity_equation}
    \partial_t \mu_t + \text{div}(v_t \mu_t) = 0.
\end{equation}

This idea of flow of probability measures turns out to be crucial to the Wasserstein distances, as shown in the following formula by Benamou and Brenier~\cite{benamou-brenier-2000}.

\begin{thm}
Given convex $\Omega \subseteq \mathbb{R}^d$ and probability measures $\mu, \nu$ with finite second moment, we have
\begin{equation}
    \mathcal{W}_2(\mu, \nu)^2 = \inf \left\{ \int_0^1 \int_{\Omega} |v_t(x)|^2 \mu_t(x) \text{{\em d}} x \text{{\em d}} t : \mu_0 = \mu, \mu_1 = \nu, \partial_t \mu_t + \text{{\em div}} (v_t \mu_t ) = 0, \mu_t v_t \text{ tangent to } \partial \Omega \right\}.
\end{equation}
\end{thm}
The optimal $\mu_t, v$ in this case is given by $X(t,x) = x + t(T(x) - x)$ where $T = \nabla \phi$ is an optimal transport plan from $\mu$ to $\nu$, and by choosing $v(t,x)$ such that $\dot{X}(t) = v(t,X(t))$. With this choice we can set $\mu_t = (X(t, \cdot))_{\#} \mu$.

From this infimum, we can also define the {\em Wasserstein norm} of the path derivative of $\mu_t$ by
\begin{equation}
  \norm{\partial_t \mu_t}_{\mu_t}^2 = \inf_{v_t} \left\{ \int_{\Omega} |v_t|^2 \mu_t \text{d} x : \partial_t \mu_t + \text{div}(v_t \mu_t) = 0,  \mu_t v_t \text{ tangent to } \partial \Omega  \right\}.
\end{equation}

\subsection{Applications of classical optimal transport}

\subsubsection{Economics and urban planning}

The most obvious application of the theory of optimal transport is the one it was originally conceived for: the physical transport of objects. The original formulation of the problem is associated with construction: you have some piles of material, and some holes to fill, and you wish to fill the holes with the material in the most efficient way. Nowadays, this appears in every aspect of supply chain logistics ~\cite{villani2008optimal}.
Efficient calculation of optimal maps via the Sinkhorn algorithm
\cite{sinkhorn-1964}
allows this to be implemented incredibly efficiently: a computational complexity of $O(n^2 \log(n) \epsilon^{-3})$ \cite{COTFNT} compared to $O(n^3 \log (n))$ for standard linear programming algorithms such as network simplex or interior point methods \cite{lucys_blog}.

In slightly more abstract economics, Wasserstein distances and the theory of optimal transport play an important role in urban planning. Two important factors in the planning of an urban area are the distribution of residents, and the distribution of services. For example, the concept of the 15-minute city
\cite{15_min_city}
sets $\mathcal{W}^t_\infty (\mu,\nu) \leq 15$ where the metric $t(x,y)$ is the time in minutes to walk from $x$ to $y$. Works
\cite{buttazzo-2009}
and
\cite{santambrogio-2009}
discuss this in great detail, adding the additional parameters of density penalisation (as crowdedness is undesireable) and concentration of services (as transport between urban areas is more efficient when services are concentrated).

\subsubsection{Heat flows and Schrödinger evolutions}
\label{subsubsection:classicalapplicationheatflows}
In a seminal work, \cite{otto-jordan-1988}
established that the solution to the Fokker-Planck equation follows the gradient of the associated free energy functional, where gradient is taken with respect to the Wasserstein distance. In the case of the heat equation, this is the gradient of the relative entropy. Concretely, a solution to a Fokker-Planck equation represents the probability density of the position (or velocity) of a particle under evolution described by a stochastic differential equation. This link to allows the solution to previously intractable physical systems to be approached from a new angle.

We are concerned specifically by equations of the form
\begin{equation}
    \frac{\partial \mu}{\partial t} = \text{div} (\nabla V(x)\mu) + \beta^{-1} \Delta \mu
\end{equation}
with initial conditions $\mu(x,0) = \mu_0(x)$. Here $V : \mathbb{R}^n \to [0,\infty)$ is a smooth potential, $\beta > 0$, and $\mu_0$ is a probability density. This guarantees that $\mu_t = \mu(\cdot, t)$ is a probability density for almost all $t$. It was shown in 
\cite{otto-jordan-1988}
that at each time instance $t$, the evolution of $\mu_t$ follows the direction of steepest descent of the free energy functional
\begin{equation}
    F(\mu) = E(\mu) + H(\mu)
\end{equation}
where $E$ is the energy and $H$ the entropy:
\begin{equation}
    E(\mu) = \int_{\mathbb{R}^n} \nabla V  \text{d}\mu(x) \qquad \text{and} \qquad H(\mu) = \int_{\mathbb{R}^n} \mu \log  \text{d}\mu(x).
\end{equation}
Here, ``steepest descent'' is defined with respect to the second-order Wasserstein distance $\mathcal{W}_2$ on probability distributions on $\mathbb{R}^n$ whose metric on $\mathbb{R}^n$ is the Euclidean metric.

Similarly, the Benamou-Brenier formulation of optimal transport has been linked closely to Schrödinger problems. The links to the standard Schrödinger problem are discussed in detail in \cite{schrodinger-connections-2013}, and those to the mean-field Schrödinger problem are discussed in~\cite{schrodinger-connections-meanfield-2020}.

The Schrödinger problem discusses the most likely paths of a single particle under Brownian motion on space given prescribed initial and final probability distributions $\mu_0$ and $\mu_1$ of its position, and is given by $\inf_{\pi \in \mathcal{C}(\mu_0,\mu_1)} H(\pi \| R_{01})$ for $R_{01}$ the joint law of initial and final positions of Brownian motion. This relative entropy can be rewritten in the form of a Benamou-Brenier optimal transport problem \cite{schrodinger-connections-2013}.
    
For the mean field Schrödinger problem (MSFP), we try to find the most likely path of a set of $N$ interacting particles under Brownian motion given an initial position $\mu_{\text{in}}$ and an `unlikely' final position $\mu_{\text{fin}}$. This reduces to minimising a large deviations principle rate function over all paths whose positions are $\mu_{\text{in}}$ and $\mu_{\text{fin}}$ at times $0$ and $T$ respectively. This, again, can be formulated as an optimal transport problem in the style of Benamou-Brenier \cite[Theorem 1.3]{schrodinger-connections-meanfield-2020}.

\subsubsection{Rapid mixing of Markov chains}

For an irreducible aperiodic finite Markov chain with a unique stationary distribution $\omega$, we define the {\em mixing time to $\epsilon$}, notated $t_{\text{mix}}(\epsilon)$ \cite{mcmt}, to be the number of iterations after which the total variation distance between the Markov chain and its stationary distribution is at most $\epsilon$. The {\em path coupling} method \cite{path_coupling} gives a straightforward way of obtaining upper bounds on these mixing times using optimal transport.

Suppose a Markov chain with transition function $P$ is defined on a graph on vertices $\mathcal{X}$ equipped with path metric $d$. Let $\mathcal{W}_1^d$ be the first-order Wasserstein distance associated to this metric. Suppose for each edge $(v,u)$ there exists a coupling $\tau$ of $P_{\#}\delta_u$ and $P_{\#} \delta_v$ such that
\begin{equation}
    \mathcal{T}_1^d(\tau) \leq e^{-\alpha} d(v,u).
\end{equation}
Then the Markov chain has a unique stationary distribution, with a mixing time to $\epsilon$ satisfying
\begin{equation}
    t_{\text{mix}}(\epsilon) \leq \frac{\log(1/\epsilon) + \log \text{diam}_d(\mathcal{X})}{\alpha}.
\end{equation}
In other words, to logarithmically bound the mixing time of the Markov chain, it suffices to find couplings between the distributions of neighbours in $\mathcal{X}$ after one iteration of the transition matrix.

 The study of mixing times is a broad area of probability theory and obtaining bounds on mixing times has many practical applications. Perhaps the most important is bounds on the error in Markov chain Monte Carlo (MCMC) methods to sample from otherwise intractable high-dimensional distributions in mathematical physics~\cite{mcmc_mathphys}, computational biology~\cite{mcmc_sysbio}, and numerical integration~\cite{mcmc_numerint}.

\subsubsection{Concentration of measure} \label{subsubsection:classicalapplicationCIs}
A large class of concentration inequalities, known as {\em transportation-cost inequalities}, come directly from or have strong connections to the theory of optimal transport. We give a list here:

\begin{itemize}
    \item \label{item:classical_CI_TCp}
        A measure $\mu$ verifies a {\em transportation-entropy} or {\em transport-cost} inequality with constant $C$, notated $TC_p(C)$, if for all measures $\nu$ we have \cite{marton-1996}
        \begin{equation} \tag{$TC_p(C)$}
            \mathcal{W}_p(\mu,\nu) \leq \sqrt{2 C D(\nu||\mu)}
        \end{equation}
        where $D(\nu \| \mu) = \int \mu(x) \frac{\log \mu (x)}{\log \nu(x)} \text{d}x$ is the Kullback-Leibler divergence between $\mu$ and $\nu$.
    \item \label{item:classical_CI_LSI_Poincare}
        Given a norm $\norm{\cdot}$ and a derivative $\nabla$, a measure $\mu$ verifies a {\em log-Sobolev inequality} with constant $C$ if for all functionals $f$ on its state space \cite{otto-villani-2000},
        \begin{equation} \tag{$LSI(C)$}
            H_\mu(f^2) \leq C\mathbb{E}_{\mu}\norm{\nabla f}^2
        \end{equation}
        where $H$ is the entropy functional with respect to $\mu$, defined by $H_\mu(g) = \int g \log g \text{ d} \mu$,
        and a {\em Poincaré inequality} with constant $\lambda > 0$ if for all functionals $f$ on its state space \cite{jog-2024},
        \begin{equation}\tag{$PI(\lambda)$}
            \lambda \text{Var}_{\mu}f \leq \mathbb{E}_{\mu}\norm{\nabla f}^2.
        \end{equation}
    \item \label{item:classical_CI_gaussian_exponential}
        Given a norm $\norm{\cdot}_{\text{Lip}}$ representing the Lipschitz constant of a function, we say that a measure $\mu$ verifies a {\em Gaussian concentration inequality} if there exist constants $c_1, c_2 > 0$ such that for all functionals $f$ on its state space with median $m(f)$,
        \begin{equation} \tag{$Gauss(c_2)$}
            \mathbb{P}_{\mu}[|f - m(f)| > r] \leq c_1\exp \left( -c_2 r^2 \norm{f}_{\text{Lip}}^2\right),
        \end{equation}
        and an {\em exponential concentration inequality} if
        \begin{equation} \tag{$Exp(c_2)$}
            \mathbb{P}_{\mu}[|f - m(f)| > r] \leq c_1\exp \left( -c_2 r \norm{f}_{\text{Lip}}^2\right).
        \end{equation}        
\end{itemize}
 We can allow, in weaker versions of these last two inequalities, the constants $c_1, c_2$ to depend on $f$.

From \cite{CRND-17-CI}, we reproduce the following diagram showing implications of concentration inequalities and references to their proofs.

\centering
\begin{tikzpicture}
\node (LSI)    {LSI($C$)};
\node (TC2) [right=3cm of LSI] {TC$_2(C_2)$};
\node (PI) [right=3cm of TC2] {PI($\lambda$)};
\node (TC1) [below=of TC2] {TC$_1(C_1)$};
\node (Gauss) [right=3cm of TC1] {Gauss($c_2$)};
\node (Exp) [right=3cm of PI] {Exp($c_2$)};
\draw[->] (LSI.east) -- (TC2.west) node[midway, above] {\cite{otto-villani-2000, erbar-maas-2012}, $C_2 = C$};
\draw[->] (TC2.east) -- (PI.west) node[midway, above] {\cite{otto-villani-2000, erbar-maas-2012}};
\draw[->] (PI.east) -- (Exp.west) node[midway, above] {\cite{milman-1983}};
\draw[->] (TC1.east) -- (Gauss.west) node[midway, above] {\cite{otto-villani-2000, tc_gotze}};
\draw[->] (TC2.south) -- (TC1.north) node[midway, left] {\cite{erbar-maas-2012}, $C_1 \geq C_2$};
\end{tikzpicture}

\justifying
More specifically for $TC_2(C_2) \implies PI(\lambda)$, for classical Markov chains we take $\lambda$ the spectral gap of the generator $L$ and norm $\norm{\nabla f_\omega}^2 = -f(\omega) L(f)(\omega)$.
The work
\cite{otto-villani-2000}
also produced the celebrated HWI inequality on $\mathbb{R}^n$ relating the relative entropy $H(\cdot || \mu)$ to some reference measure $\mu$, second-order Wasserstein distance with respect to the Euclidean distance $W_2^2(\cdot, \mu)$, and Fisher information $I(\cdot || \mu)$. Taking the reference measure $\mu = e^{-V}\text{vol}^n$ where $\text{vol}^n$ is the Lebesgue measure and $V$ is some potential whose Hessian has smallest eigenvalue at least some curvature parameter $\kappa$, we get that for all $\nu$,
\begin{equation}
    H(\nu || \mu) \leq \mathcal{W}_2(\nu,\mu)\sqrt{I(\nu||\mu)} - \frac{\kappa}{2}\mathcal{W}_2^2(\nu,\mu).
\end{equation}
Similar statements hold for the case where $\mathbb{R}^n$ is replaced by a Riemannian manifold whose Ricci tensor $\text{Ric}_{V,N}$ satisfies a suitable lower bound: see \cite{gentil-hwi-2020}
for details. The notion of optimal transport has also been used \cite{lott-villani-2005}
to extend the notion of Ricci curvature from Riemannian manifolds to more general metric spaces, and then to show that Ricci curvature lower bounds lead to similar HWI inequalities.
Some other distributions have also been shown to verify $TC$ inequalities: for example, the $\mathcal{W}_1$ distance on the Hamming hypercube $\{0,1,\dots, d\}^n$ verifies $TC_1\left(\frac{n}{2}\right)$.

\subsubsection{Image manipulation and machine learning}
Some of the earliest applications of optimal transport were in image manipulation \cite{classical_applications_signals_machinelearning}. An advantage of Wasserstein distances in this context is that they can give distances between image signatures of different formats (for example, between histograms with different binning). Matching images via Wasserstein distance gives higher precision in image retrieval than any other common metric. Work \cite{classical_applications_signals_machinelearning} also discusses the success of Wasserstein distance methods in colour transference (for digital image enhancements), texture synthesis (for, among other things, age-progression images of missing people), resolution improvements, and image restoration from noisy or incomplete data.

In a similar vein, Wasserstein distances have also been used with great success as distinguishability metrics or loss functions in machine learning. Work \cite{pmlr-v70-arjovsky17a} introduced the Wasserstein Generative Adversarial Network (WGAN), a variant on the traditional GAN, which takes advantage of Kantorovich duality to use Wasserstein distances and Lipschitz functions in the generation process. This allows for greater precision in the training and generation process, but also improves GAN algorithms by eliminating mode collapse and enhancing stability.

\section{Quantum information framework}
\label{section:quantumbg}
As with the previous section, those coming from the quantum information theory side are encouraged to skip to Section \ref{subsection:quantumbg_desirable_properties} for a discussion of the desirable properties of a proposed quantum Wasserstein distance. For a more detailed treatment of the topics discussed in this section, see the book \cite{Nielsen_Chuang} by Nielsen and Chuang for the density operator perspective and the collection of notes \cite{funcanaquantinfo} for the operator algebras perspective.

\subsection{Notable operators, objects, and transforms} \label{subsubsection:quantumbg_notable_operators}

The first key set of objects in quantum information theory is the family of Schatten $p$-norms on traceless operators.

\begin{defn}
    For a traceless operator $X$ on Hilbert space $\mathcal{H}$, define
    \begin{equation}
        \norm{X}_p = \left( \text{{\em Tr}} [|X|^p] \right)^{1/p}
    \end{equation}
    where $|X| = \sqrt{X^*X}$.
\end{defn}
These are commonly used to describe differences between states, for example the \textit{trace distance} between states $\rho$ and $\sigma$ is given by $\frac{1}{2}\norm{\rho-\sigma}_1$.

When in the realm of quantum computation, using Hilbert space $\mathcal{H} = \left(\mathbb{C}^d \right)^{\otimes n}$, we fix an orthonormal basis $\ket{0}, \ket{1}, \dots, \ket{d-1}$ of $\mathbb{C}^d$. This gives the following definition.
\begin{defn}
    The \textbf{computational basis} of $(\mathbb{C}^d)^{\otimes n}$ is the set $\left\{ \ket{k_1} \otimes \dots \otimes \ket{k_n} : k_i \in \{0, 1, \dots, d-1\} \right\}$.
\end{defn}

We introduce here two sets of operators commonly used in the realm of quantum information theory.

\begin{defn}
    The matrices
\begin{equation} \label{eq:defn_paulis}
\sigma_x = \begin{pmatrix}
0 & 1\\
1 & 0
\end{pmatrix} \qquad
\sigma_y = \begin{pmatrix}
    0 & -i \\ i & 0
\end{pmatrix} \qquad
\sigma_z = \begin{pmatrix}
    1 & 0 \\ 0 & -1
\end{pmatrix}
\end{equation} are known as the \textbf{Pauli} matrices, commonly referred to as simply $X$, $Y$ and $Z$. 
\end{defn}
Together with the identity $I$ these form a real basis of the set of $2 \times 2$ Hermitian matrices, they anticommute, they each have eigenvalues $1$ and $-1$ with equiangular eigenvectors. In $d$ dimensions, these generalise to the Heisenberg-Weyl matrices, where we have the set $\{X^iZ^j\}_{i,j = 0}^{d-1}$ for $X$ the ``phase shift" operator sending $\ket{k} \mapsto e^{\omega k} \ket{k}$ where $\omega = 2\pi i /d$, and $Z$ the ``clock shift" sending $\ket{k} \mapsto \ket{k+1}$ (modulo $d$).

\begin{defn}
   The \textbf{SWAP operator} on $\mathcal{H} \otimes \mathcal{H}$ is defined by its action on a product basis:
\begin{equation} \label{eq:defn_swap}
    \text{SWAP}\ket{i}\ket{j} = \ket{j}\ket{i}.
\end{equation}
This operator has eigenvalues $1$ and $-1$, and we call its eigenspaces the symmetric subspace and the antisymmetric subspace respectively. The operators of importance here are the \textbf{symmetric projector} $P_{\text{sym}}$, which is the orthogonal projector onto the symmetric subspace, and the \textbf{antisymmetric projector }$P_{\text{asym}}$ likewise for the antisymmetric subspace. 
\end{defn}

Additionally, for a self-adjoint operator $O$ and a set $I \subseteq \mathbb{R}$, we write $\mathbf{1}_{I}(O)$ for the projection onto the span of eigenvectors of $O$ whose eigenvalues lie in $I$.

For transforms, we introduce three main transforms used in continuous-variable quantum information theory. For more detail on these objects and objects specific to continous-variable quantum information, see the book \cite{serafini} by Serafini.
\begin{defn}
    Let $\rho$ be a state on Hilbert space $\mathcal{H}$.
    \begin{itemize}
        \item For $\mathcal{H} = L^2(\mathbb{C}^d)$,
        \begin{itemize}
            \item A \textbf{Töplitz operator at scale $\epsilon$} with respect to distribution $\mu$ on $\mathbb{C}^d$ is 
            \begin{equation}
                \rho_\mu = \frac{1}{(2\pi \epsilon)^d} \int_{\mathbb{C}^d} \ketbra{z,\epsilon}{z,\epsilon} \text{{\em d}}\mu(z)
            \end{equation}
            where as a functional on $\mathbb{R}^d$,
            \begin{equation}
                \left|q + ip, \epsilon \right\rangle (x) = (\pi \epsilon)^{-d/4} e^{-(x-q)^2/2\epsilon} e^{ip\dot x/\epsilon}.
            \end{equation}
        \end{itemize}
        \item For $\mathcal{H} = L^2(\mathbb{R}^n)$,
        \begin{itemize}
            \item the \textbf{Wigner transform at scale $\epsilon$} of $\rho$ is $f_\rho : \mathbb{R}^n \otimes \mathbb{R}^n \to \mathbb{R}$ given by
        \begin{equation}
            f_\rho(q,p) = \left(\frac{1}{2\pi \epsilon} \right)^n \int_{\mathbb{R}^n} e^{-\frac{i}{\epsilon}p \cdot y} \left\langle q + \frac{1}{2}y \right| \rho \left| q - \frac{1}{2} y \right\rangle \text{{\em d}}y,
        \end{equation}
        \item the \textbf{Weyl transform} is the inverse of the Wigner transform,
        \item the \textbf{Husimi transform at scale $\epsilon$} is the $\star$-product of a Gaussian $G_{\epsilon/2}^{2n}$ centered at $0$ on $\mathbb{R}^{2n}$ with the Wigner transform, defined by
        \begin{equation}
            \tilde{f}_\rho = G_{\epsilon/2}^{n/2} \star_{q,p} f_\rho.
        \end{equation}
        \end{itemize}
    \end{itemize}
\end{defn}

\subsection{Channels and semigroups}

As mentioned in Section \ref{section:classical}, transformations of probability distributions are a key part of classical optimal transport. The Monge optimal transport cost $\mathcal{T}_M^c$ is defined with respect to a transport map, and in many cases (see the discussion following Definition \ref{defn:kantorovich_relaxation_classical_cost}) the solution to Kantorovich relaxation of optimal transport admits a transport map.

The quantum equivalent of a classical stochastic map is a {\em quantum channel}. A quantum channel is a map $\mathcal{N}: \mathcal{B}(\mathcal{H}_1) \to \mathcal{B}(\mathcal{H}_2)$ which is both trace preserving (for all $\rho$, $\text{Tr} [\mathcal{N}(\rho)] = \text{Tr} [\rho]$) and completely positive (for all auxiliary systems $\mathcal{H}'$, $\mathcal{N} \otimes \text{id}_{\mathcal{H}'}$ sends positive semi-definite operators to positive semi-definite operators). The upshot of this is that it sends states to states. The definition within the von Neumann algebra framework is equivalent: an endomorphism of the linear functionals on $M$ which is trace preserving and completely positive.

In terms of continuous time evolution of quantum systems, the quantum equivalent of a classical Markov chain is the {\em quantum Markov semigroup} (QMS). For this, we usually use the von Neumann algebra framework. A collection $(\mathcal{P}_t)_{t \geq 0}$ of bounded operators on $M$ is a QMS if,
\begin{itemize}
    \item For all $A \in M$, $\mathcal{P}_0(A) = A$,
    \item For all $t, s \geq 0$, $\mathcal{P}_{t+s} = \mathcal{P}_t(\mathcal{P}_s)$,
    \item For all $t \geq 0$, $\mathcal{P}_t(\mathbf{1}) = \mathbf{1}$,
    \item For all $t$, $\mathcal{P}_t$ is completely positive,
    \item For all $t$, $\mathcal{P}_t$ is a $\sigma$-weakly continous operator in $M$, and for all $A$, $t \mapsto \mathcal{P}_t(A)$ is continous with respect to the $\sigma$-weak topology on $M$.
\end{itemize}
The dual $\mathcal{P}_t^\dagger$ is then a quantum channel, where complete continuity is preserved under the dual, and trace preservation comes from the dual of the condition $\mathcal{P}_t(\mathbf{1}) = \mathbf{1}$.

In finite dimension, QMSs can be written as $\mathcal{P}_t = e^{t \mathcal{L}}$ for an infinitesimal generator $\mathcal{L}$ called the {\em Lindbladian}. Under uniform continuity of the map $t \mapsto \mathcal{P}_t$, the Lindbladian has the form
\begin{equation}
    \mathcal{L}(A) = i[H,A] + \sum_j \left(V_j^* A V_j - \frac{1}{2} \left\{V_j^{*}V_j, A \right\} \right)
\end{equation}
where $H$ is the Hamiltonian associated to $(\mathcal{P}_t)_{t \geq 0}$, and $[\cdot , \cdot]$ and $\{ \cdot , \cdot \}$ are the commutator and anticommutator respectively. For many discussions, we focus specifically on the case where $\mathcal{P}_t$ has a unique invariant state, which is full-rank (nondegenerate), and unless otherwise stated we assume this is the case.

It is important to note that this definition is given in what we call the \textit{Heisenberg picture}: that is, the viewpoint of physical evolution where we let measurements evolve and states remain static, rather than the \textit{Schrödinger picture} in which measurements remain static and states evolve. From a mathematical viewpoint the distinction makes little difference, though some papers (such as \cite{CM-12-2W, CM-16-GF}) make use of the Heisenberg picture while others (such as \cite{araiza2025resourcedependentcomplexityquantumchannels, Ding_2025}) make use of the Schrödinger picture. The two are dual to one another, and so the Schrödinger evolution of a quantum state under a physical system represented by quantum Markov semigroup with Lindbladian $\mathcal{L}$ is given by
\begin{equation}
\frac{\partial \rho}{\partial t} = \mathcal{L}^\dagger \rho = -i[H,\rho] + \sum_j \left(V_j^* \rho V_j - \frac{1}{2} \left\{V_j^{*}V_j,\rho \right\} \right).
\end{equation}

An important notion characterising semigroups is the GNS inner product, or more generally the $\varphi$-inner product. For a nondegenerate density matrix $\sigma$, elements $a, b \in M$, and $s \in \mathbb{R}$, we have
\begin{equation} \label{eq:defn_s_IP}
    \langle A, B \rangle_{s, \sigma} = \text{Tr}[A^* \Delta_{\sigma}^{1-s} B].
\end{equation} The special case $s = 1$ gives the {\em GNS inner product} associated to $\sigma$, and $s = 1/2$ the corresponding {\em KMS inner product}. The \textit{BKM inner product} is $\langle \cdot , \cdot \rangle_{BKM,\sigma} = \int_0^1 \langle \cdot , \cdot \rangle_{s,\sigma} \textrm{d}s$. More generally for a function $\varphi: (0, \infty) \to (0, \infty)$, we have
\begin{equation}\label{eq:defn_varphi_IP}
    \langle A, B \rangle_{\varphi, \sigma} = \text{Tr}[A^* R_{\sigma} \circ \varphi( \Delta_{\sigma}) B].
\end{equation}
This then gives the definition of the {\em 2-Dirichlet form} of a QMS:
\begin{equation}
    \mathcal{E}_{\varphi, 2}(A,B) = -\langle A, \mathcal{L}(B) \rangle_{\varphi, \sigma}.
\end{equation}

This gives a key definition in QMSs:
\begin{defn}
    We say that a QMS $\mathcal{P}_t$ satisfies the $\sigma$\textbf{-detailed balance condition} (\textbf{$\sigma$-DBC}, or more specifically \textbf{$\sigma$-GNS-DBC}) if its generator $\mathcal{L}$ is self-adjoint with respect to the GNS inner product $\langle \cdot, \cdot \rangle_{1,\sigma}$. \textbf{KMS detailed balance} and \textbf{BKM detailed balance} are defined equivalently.
\end{defn}
Note that this means $\mathcal{P}_t^\dagger \sigma = \sigma$, and so under the assumption of a unique invariant state we may simply refer to $\mathcal{P}_t$ satisfying the DBC. Furthermore, in the presence of detailed balance, $\mathcal{P}_t$ commutes with the modular automorphism group. Any QMS satisfying the DBC then has a standard form, due to Alicki's theorem \cite{ALICKI1976249}.
\begin{prop}
   Let $\mathcal{P}_t$ satisfy the $\sigma$-DBC for some nondegenerate $\sigma$. Then its generator $\mathcal{L}$ has form
   \begin{equation} \label{eq:qms-dbc-standard-form-alicki}
       \mathcal{L}(A) = \sum_j e^{-\omega_j / 2}\left(V_j^*[A,V_j]  + + [V_j^*, A]V_j \right)
   \end{equation}
   where $\omega_j$ are real and the set $\{V_j\}_j$ is a set of orthogonal traceless eigenvectors of $\Delta_\sigma$ each with eigenvalue $\omega_j$, which is closed under $*$. Note the operators $V_j$ do not need belong to $M$ itself.
\end{prop}

\subsection{Concentration inequalities} \label{subsection:quantumbg_conc_ineqs}

It is natural, at this point, to introduce the notation for a broad class of applications of the upcoming quantum Wasserstein distances. As discussed in Section \ref{subsubsection:classicalapplicationCIs}, classical concentration inequalities are fundamental in fields such as statistics~\cite{ci_states}, convex geometry \cite{ci_geo}, information theory and communications~\cite{raginsky2015concentrationmeasureinequalitiesinformation}, and many more \cite{ci_concmeasure_book}, and generalising these to the quantum setting would be desirable. Furthermore, existing quantum concentration inequalities have shown to be useful in distinguishability \cite{Benoist_2022} and parameter estimation \cite{CRND-17-CI, Rouz__2024, caro-learning}.

The same concepts of concentration of measure exist in the realm of quantum states, in the sense of concentration of eigenvalues and eigenspaces rather than mass. Hence we can define many of the same concentration inequalities seen in the classical setting for quantum states. As we will see in Sections \ref{subsection:dynamical_concentration} and \ref{subsection:lipschitz_applications}, their satisfaction is one of the key applications of the quantum Wasserstein distances defined thus far.

As we will see later, a quantum Wasserstein distance in its simplest terms is a generalisation of the distances $\mathcal{W}_p$ to give a distance between quantum states. For now, write a generic quantum Wasserstein distance as $W_p$: superscripts will later be used to distinguish different definitions.

We recall the standard entropic quantities,
\begin{itemize}
    \item the {\em quantum relative entropy}, also known as {\em Umegaki's relative entropy}:
    \begin{equation}
        D(\rho \| \sigma) = \text{Tr}[\rho (\log \rho - \log \sigma)]
    \end{equation}
    \item for a QMS $\mathcal{P}_t$ with generator $\mathcal{L}$, the {\em quantum Fisher information} or {\em entropy production}:
    \begin{equation}
        \mathcal{I}_{\sigma}(\rho) = -\text{Tr} [(\log \rho - \log \sigma)\mathcal{L}^\dagger \rho].
    \end{equation}
\end{itemize}

We can then define the following transportation-cost inequalities.
\begin{defn}
    We say that a quantum Wasserstein distance $W_p$ satisfies:
    \begin{enumerate}
        \item
            a {\em $p^{th}$-order transportation cost inequality} with constant $C > 0$ if for all states $\rho$, $\sigma$:
            \begin{equation} \label{eq:defn_tcp} \tag{$TC_p(C)$}
                W_p(\rho,\sigma) \leq \sqrt{C D(\rho \| \sigma)},
            \end{equation}
        \item
            a {\em Gaussian concentration inequality} if, for some definition $\text{{\em Lip}}(O)$ of the Lipschitz constant of an operator $O$ linked to $W_1$ via Kantorovich duality, there exist constants $c_1, c_2$ such that for all states $\rho$ we have
            \begin{equation} \label{eq:defn_q_gaussconc} \tag{$Gauss(c_2)$}
                \text{{\em Tr}}[\rho\mathbf{1}_{[r,\infty)}(O -\text{{\em Tr}}[O\rho])] \leq c_1 \exp \left( -c_2r^2 \text{{\em Lip}}(O) \right)
            \end{equation}
        \item
            an {\em exponential concentration inequality} if, under the same conditions as for Gaussian concentration,
            \begin{equation} \label{eq:defn_gaussconc} \tag{$Exp(c_2)$}
                \text{{\em Tr}}[\rho\mathbf{1}_{[r,\infty)}(O -\text{{\em Tr}}[O\rho])] \leq c_1 \exp \left( -c_2r \text{{\em Lip}}(O) \right).
            \end{equation}
    \end{enumerate}
    Some definitions of the Gaussian and exponential inequalities may vary, but the crux of each is that the log of the measure of the tail of the distribution decays quadratically, or linearly, respectively with $r$. Some of the definitions, in particular in Section \ref{subsection:lipschitz_applications}, may also use the moment-generating function in place of the eigenvalue indicator function, for equations such as 
    \begin{equation}
        \text{Tr}\exp [t(O - \text{Tr}[O]\mathbb{I}/d)] \leq c_1 \exp c_2 f(t) g(\text{{\em Lip}}(O))
    \end{equation}
    where $f$ is some function which is linear for exponential concentration and quadratic for Gaussian concentration, and $g$ some monotone function.

\end{defn}

\begin{defn} Let $\mathcal{P}_t$ be a QMS with generator $\mathcal{L}$ and unique fixed point $\sigma$. We say that it satisfies:
\begin{enumerate}
    \item[4.]
        a {\em modified logarithmic Sobolev inequality} with constant $\lambda > 0$ if for all states $\rho$, we have 
        \begin{equation} \label{eq:defn_mlsi} \tag{$MLSI(\lambda)$}
            D(\rho \| \sigma) \leq \frac{1}{2\lambda} \mathcal{I}_{\sigma} (\rho).
        \end{equation}
\end{enumerate}
If $W_p$ is then a quantum Wasserstein distance on the state space of this QMS, we say it satisfies~\cite{CRND-17-CI}
\begin{enumerate}
    \item[5.] 
        a {\em HWI inequality} with constant $\kappa \in \mathbb{R}$ if for all states $\rho$, we have
        \begin{equation} \label{eq:defn_hwi} \tag{$HWI(\kappa)$}
            D(\rho \| \sigma) \leq W_2(\rho,\sigma) \sqrt{\mathcal{I}_\sigma (\rho)} - \frac{\kappa}{2} W_2(\rho,\sigma)^2.
        \end{equation}
    \item[6.]
        a {\em $TC_{p}$ inequality} with constant $C > 0$ if for all states $\rho$, equation \eqref{eq:defn_tcp} holds.
    \item[7.]
        a {\em Poincaré inequality} with constant $\lambda \in \mathbb{R}$ with respect to $\varphi:(0,\infty) \to (0, \infty)$ if for all $a \in M$,
        \begin{equation} \label{eq:defn_poincare} \tag{$PI(\lambda)$}
            \lambda \norm{a}_{\varphi, \sigma}^2 \leq \mathcal{E}_{\varphi, 2}(a,a).
        \end{equation}
\end{enumerate}
\end{defn}

\subsection{Desirable properties of a quantum Wasserstein distance}
\label{subsection:quantumbg_desirable_properties}
Works centering around the generalisation of the classical Wasserstein distances to the quantum setting do so with varying goals in mind. Some are centred around directly generalising some of the applications arising from the classical distances, and others are centred around finding a ``true" or ``correct" generalisation which replicates as many of the properties of the classical distances as possible, with the view that such a definition should naturally give rise to generalisations of its applications.

For prior discussions of the field of quantum optimal transport, \cite{dario-invitation} gives a shorter introduction than the present work to the current state of the art, and the set of lecture notes \cite{budapest-school} from a summer school in Budapest in 2022 gives a more detailed overview of select works in this field. The key references for quantum optimal transport distances will be introduced in Sections \ref{section:coupling}, \ref{section:dynamics}, and \ref{section:lipschitz}.

Many of the approaches have been curated to allow the distances defined to satisfy a number of desirable properties of such a $W_p$. While this list is by no-means exahustive, these properties include (each in a very general sense):
\begin{itemize}
    \item $W_p$ being a true distance; that is
    \begin{itemize}
        \item Non-negativity and non-degeneracy: $W_p(\rho,\sigma) \geq 0$ with equality $\iff \rho = \sigma$,
        \item Symmetry: $W_p(\rho,\sigma) = W_p(\sigma,\rho)$,
        \item The triangle inequality: $W_p(\rho, \sigma) + W_p(\sigma, \tau) \geq W_p(\rho,\sigma)$.
    \end{itemize}
    \item Classical links: links to the classical definition in the following ways:
        \begin{itemize}
            \item Standard definition via couplings,
            \item Dual definition via Kantorovich duality,
            \item Dynamical formulation via Benamou-Brenier.
        \end{itemize}
    \item Continuity: $W_p$ being continuous for $p \neq \infty$.
    \item Agreement with trace distance: in the specific case where $p = 1$ and the distance $W_1$ has been constructed as a generalisation of the classical Wasserstein distance arising from the discrete metric, we have $W_1(\rho,\sigma) = \frac{1}{2}\norm{\rho-\sigma}_1$.
    \item Data processing: for quantum channels $\Phi$ which ``respect'' the cost function (for some sensible definition of ``respect''), $W_p(\rho,\sigma) \geq W_p(\Psi(\rho),\Psi(\sigma))$.
    \item Ease of calculation: for effective use of these distances, there should be an algorithm to calculate or approximate $W_p(\rho,\sigma)$ for a given $\rho, \sigma$. This algorithm should be efficient, in the sense that for $\rho$, $\sigma$ admitting a $\text{poly}(\log \dim \mathcal{H})$ description this algorithm should ideally run in $\text{poly}(\log \dim \mathcal{H})$ time, though an algorithm running in $\text{poly}(\dim \mathcal{H})$ would also be useful for studying small cases.
    \item Flexibility in definition: that a $W_1$, $W_2$, and potentially a $W_p$ for general $p$ are defined from the same framework and thus can be related or linked.
    \item Concentration inequalities: recovery of some or all of the implications and satisfaction of some or all of the concentration inequalities discussed in Section \ref{subsection:quantumbg_conc_ineqs}.
    \item Applicability: that the definition of $W_p$ lead to results about quantum states,
    \item Variety of application: that $W_p$ be defined in, and therefore applicable to, as many underlying state spaces and physical settings as possible.
\end{itemize}

\subsection{Obstacles to generalisation} \label{subsection:quantumbg_obstacles}
The most natural place to start when attempting to define a quantum version of optimal transport is from quantum couplings. Rewriting \eqref{eq:defn_tc_couplings} in the language of quantum information, we could for some `cost operator' $C$ try to define
\begin{equation} \label{eq:naive_generalisation}
    T^C(\rho,\sigma) = \inf_{\pi \in \mathcal{C}(\rho,\sigma)} \text{Tr}[C\pi]
\end{equation}
where the set of couplings is defined as $\mathcal{C}(\rho,\sigma) = \{\pi \in \mathcal{D}(\mathcal{H} \otimes \mathcal{H}) : \text{Tr}_1[\pi] = \sigma, \text{Tr}_2[\pi] = \rho \}$. For the Wasserstein distances in particular, we can take either a matrix $D$ which represents the distance between pure states, and then use $C = D^p$, or we can choose $C$ as a $p^{\text{th}}$-order cost matrix directly. This gives

\begin{equation}
    W_p(\rho,\sigma) = \left(\inf_{\pi \in \mathcal{C}(\rho,\sigma)} \text{Tr}[D^p\pi]\right)^{1/p} \qquad \text{or} \qquad    W_p(\rho,\sigma) = \left(\inf_{\pi \in \mathcal{C}(\rho,\sigma)} \text{Tr}[C\pi]\right)^{1/p}.
\end{equation}

However, this direct translation of the original definition to the quantum setting does not lend itself easily to distances which satisfy desired properties. Indeed, it was proven in \cite{NY-18-QE}
that no cost matrix $C$ has $W_1^C(\rho,\sigma) = \frac{1}{2}\norm{\rho-\sigma}_1$. More specifically, they considered the case where we have a bijection $f$, Hermitian $H$, and unitarily invariant distance $d$ such that
\begin{equation}
    d(\rho,\sigma) = f\left(\min_{\pi \in \mathcal{C}(\rho,\sigma)} \text{Tr}[H\pi] \right).
\end{equation}
They showed that any such $f$ must be $f(x) = 2\sqrt{x(1-x)}$, and that any such $H$ must be a linear combination of the symmetric and antisymmetric projectors. They then showed that no such linear combination satisfies this relation for $d$ the trace distance.

Another potential issue with coupling methods for defining quantum Wasserstein distances is the inaccessiblity of the triangle inequality. Classically, the triangle inequality is proved as follows \cite[page 94]{villani2008optimal}: for measures $\mu_1, \mu_2, \mu_3$, take optimal couplings $\omega_{12}$ of $\mu_1$ with $\mu_2$ and $\omega_{23}$ of $\mu_2$ with $\mu_3$. The gluing lemma states that there exists measure $\omega_{123}$ on the three-way product space with $12$-marginal $\omega_{12}$ and $23$-marginal $\omega_{23}$. Then taking the $13$-marginal $\omega_{13}$ comparing its transport cost with that of $\omega_{12}$ and $\omega_{23}$ gives the triangle inequality. However, the key step in this argument is not replicable in the quantum setting. There exist states $\pi_{12}$ on Hilbert space $\mathcal{H}_1 \otimes \mathcal{H}_2$ and $\pi_{23}$ on $\mathcal{H}_2 \otimes \mathcal{H}_3$ such that no state $\pi_{123}$ on $\mathcal{H}_1 \otimes \mathcal{H}_2 \otimes \mathcal{H}_3$ has $12$-marginal $\pi_{12}$ and $23$-marginal $\pi_{23}$. For example, taking $\pi_{12} = \pi_{23} = \ket{\phi^+}\bra{\phi^+}$ where $\ket{\phi^+}$ is the standard 2-qubit Bell state leaves no such $\pi_{123}$. The study of which pairs of states $\pi_{12}$ and $\pi_{23}$ admit such a $\pi_{123}$ is known as the quantum marginal problem \cite{Haapasalo_2021}, area of research in quantum systems.
Without this equivalent to the gluing lemma, there is no straightforward way to construct a coupling $\pi_{23}$ of $\omega_1$ and $\omega_3$ whose cost is directly related to those of $\pi_{12}$ and $\pi_{23}$, so no straightforward comparison of the induced optimal transport cost.

This is closely related to the issue of distribution of mass in a quantum state. Indeed, for a classical probability measure $\mu$ on $\mathcal{X}$, there is exactly one way of interpreting $\mu$ as a distribution of mass over $\mathcal{X}$. But for a quantum state $\rho$ on $\mathcal{H}$, there is an infinite family of ways of interpreting $\rho$ as a distribution of mass over $\mathcal{H}$, in the form $\rho = \int \ketbra{\psi}{\psi} \text{d}\hat{\mu}(\psi)$ for some probability measure $\hat{\mu}$ on $\mathcal{H}$. This lack of clarity in the distribution of mass in a quantum state in turn means that the cost of any transport map could be multiply defined.

Another common problem is the self-distance issue. As was shown in~\cite{KZ-21-QOT}, a general cost matrix $C$ gives rise to a semidistance $W_p$ if and only if $C$ is zero on the symmetric subspace and positive definite on the antisymmetric subspace. Many geometrically natural definitions of such a $C$ do not have this property, meaning that they can never be a semidistance.

Some works have embraced these limitations and defined $W_p$ quantities aiming for applications without focusing directly on properties of $W_p$, while others have adapted the coupling method to avoid these issues. Others further have decided instead to adapt the Kantorovich dual formulation of $\mathcal{W}_1$ or the Benamou-Brenier formulation of $\mathcal{W}_2$ to the quantum setting to avoid the obstacles in the coupling definition. These have the advantage of guaranteeing many of the desirable properties we want, such as the triangle inequality and faithfulness, while removing flexibility in order $p$ and, in the Kantorovich dual case, links to transport plans. These different frameworks motivate our categorisation of the different quantum Wasserstein distances: those which have adapted the formulation in equation \eqref{eq:naive_generalisation} are grouped into a \textit{coupling} approach, those which adapt the Kantorovich dual formula in Section \ref{subsection:classical_kantorovich} a \textit{Lipschitz} approach, and those which adapt the Benamou-Brenier formula in Section \ref{subsection:classical_benamou_brenier} a \textit{dynamical} approach.

\section{Coupling formulations} 
\label{section:coupling}
\subsection{Position-momentum cost matrix} \label{subsection:coupling_position_momentum}
\subsubsection{Definitions}
The most natural place to start is discussion of the distance proposed and developed in a series of papers 
\cite{FG-15-QM,FG-15-SE,FG-17-WP,FG-19-CH,FG-21-MK,FG-21-QS}
by Golse et al. This is an example of a definition which has embraced the limitations of the direct coupling method in order to generalise applications of the classical $\mathcal{W}_p$ without worrying too much about specific properties of $W_p$ itself. These use the Hilbert space $\mathcal{H} = L^2(\mathbb{R}^d)^{\otimes n}$, modelling a physical system in the continuous-variable regime in which the properties of interest are position and momentum. We then have the following:
\begin{defn}
    The \textbf{second-order position-momentum quantum Wasserstein distance} \cite{FG-15-QM} is
    \begin{equation}  \label{eq:defn_mk2_golse}
            W_2^{MK,\epsilon}(\rho,\sigma) = \left(\inf_{\pi \in \mathcal{C}(\rho,\sigma)}\text{{\em Tr}}[(P^\dagger P + Q^\dagger Q)\pi]\right)^{1/2} \qquad \text{ for } \qquad \begin{cases}
                 Q(\omega)(x_1,x_2) = (x_1-x_2) \omega(x_1,x_2), \\
                 P(\omega)(x_1,x_2) = -i\epsilon (\nabla_{x_1} - \nabla_{x_2})\omega(x_1,x_2).
            \end{cases}
    \end{equation}
\end{defn}
These naturally correspond to operators measuring the transport cost with regards to position and momentum respectively. 
\begin{defn}
    The \textbf{semiclassical second-order position-momentum quantum Wasserstein distance} \cite{FG-15-SE} is 
\begin{equation} \label{eq:defn_mk2_semiclassical}
    \mathcal{E}_{\epsilon}(\mu,\sigma)^2 = \inf_{\Pi \in \mathcal{R}(\mu,\sigma)} \int_{\mathbb{R}^{2d}} \text{{\em Tr}}_{L^2(\mathbb{R}^d,\textrm{d}x)}\left[(q-x)^2 + (p + i\epsilon\nabla_x)^2 \right]\textrm{d}q\textrm{d}p.
\end{equation}
where $\mu$ is a probability density on $\mathbb{R}^{2d}$, $\sigma$ a state on $L^2(\mathbb{R}^d)$, and $\mathcal{R}(\mu,\sigma)$ be the set of smooth functions $\Pi$ from $\mathbb{R}^{2d}$ to the set of density operators on $L^2(\mathbb{R}^d)$, such that for each $(q,p) \in \mathbb{R}^{2d}$, we have $\text{{\em Tr}}[\Pi(q,p)] = \mu(q,p)$ and $\int_{\mathbb{R}^{2d}} \Pi(q,p) \textrm{d}q\textrm{d}p = \sigma$.
\end{defn}
This does not appear to have any physical significance, though is incredibly useful for proving properties of $W_2^{MK,\epsilon}$ and relations between $W_2^{MK,\epsilon}$ and the classical $\mathcal{W}_2^{\text{Eucl}}$ distance.

Work \cite{GT-22-SS}
proposes a slight modification to this, where the optimisation in \eqref{eq:defn_mk2_golse} is restricted to separable couplings, rather than the whole of $\mathcal{C}(\rho,\sigma)$. They remark that if there is a separation between the separable and non-separable versions of $W_2^{MK,\epsilon}$ then all optimal couplings for the non-separable version must be entangled. They also link the self-distance of the separable version to quantum Fisher information.

\subsubsection{Properties}
\begin{itemize}
    \item $W_2^{MK,\epsilon}$ is not faithful: all $\rho,\sigma$ have $W_2^{MK,\epsilon}(\rho,\sigma) \geq 2d\epsilon$~\cite{FG-15-QM}.
    \item The definition of $W_2^{MK,\epsilon}$ is closely related to that of the classical $\mathcal{W}_2^{\text{Eucl}}$: namely, for Töplitz operators $\rho, \sigma$ at scale $\epsilon$ based on $\mu, \nu$ respectively, we have~\cite{FG-15-QM}
\begin{equation}
    W_2^{MK,\epsilon} (\rho,\sigma)^2 \leq \mathcal{W}_2^{\text{Eucl}} (\mu,\nu)^2 + 2d\epsilon.
\end{equation}
 \item In the other direction, for states $\rho$ and $\sigma$ on $\mathcal{H}$, let $\tilde{f}_\rho$ and $\tilde{f}_\sigma$ be their Husimi transforms at scale $\epsilon$. Then we have~\cite{FG-15-SE}
\begin{equation}
    W_2^{MK,\epsilon} (\rho,\sigma)^2 \geq \mathcal{E}_\epsilon (\tilde{f}_\rho,\sigma) - d\epsilon \geq \mathcal{W}_2^{\text{Eucl}}(\tilde{f}_\rho,\tilde{f}_\sigma) - 2d\epsilon.
\end{equation}
\item $\mathcal{E}_\epsilon$ also satisfies an analogue of Kantorovich duality. This property is used to prove~\cite{FG-21-QS}
an approximate version of the triangle inequality for $W_2^{MK,\epsilon}:$
\begin{equation}
    W_2^{MK,\epsilon}(\rho,\tau) \leq W_2^{MK,\epsilon}(\rho,\sigma) + W_2^{MK,\epsilon}(\sigma,\tau) + d\epsilon.
\end{equation}
\item The self-distance of $W_2^{MK,\epsilon}$ is bounded~\cite{LL-23-WT} by
\begin{equation} \label{eq:mk2_golse_self_dist_ub}
    W_2^{MK,\epsilon}(\rho,\rho)^2 \leq 4d\epsilon + 4\epsilon^2 \norm{\nabla f_\rho}^2_{L^2(\mathbb{R}^{2d})}
\end{equation} where $f_\rho$ is the Wigner transform of $\rho$. The proof uses $\mathcal{E}_\epsilon$.
\item Quantum optimal transport of this form can be cheaper~\cite{FG-19-CH}, in the sense
that the $W_2^{MK,\epsilon}$ distance between Töplitz operators can be strictly smaller than the distance between the classical distributions they come from.
\end{itemize}
Work \cite{LL-23-WT} also gives explicit calculations of the constants in \eqref{eq:mk2_golse_self_dist_ub} in the case of thermal states, and more specific bounds in the special cases of spectral projections and powers of Töplitz operators.

\subsubsection{Applications}

The main application of these distances is to prove convergence of solutions to the Schrödinger equation in the mean-field limit. Specifically, as discussed in \cite{FG-15-QM} we have the following relation between equations in statistical mechanics:
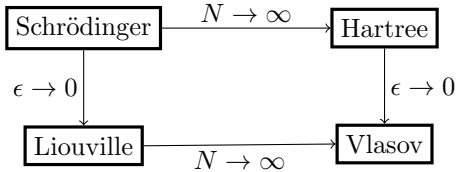
\begin{center}
\begin{tikzpicture}[
squarednode/.style={rectangle, draw=black!100, very thick, minimum size=5mm},
]
\node[squarednode]      (schrodinger)     at (0,1)                         {Schrödinger};
\node[squarednode]        (hartree)       at (4,1) {Hartree};
\node[squarednode]      (liouville)       [below=of schrodinger] {Liouville};
\node[squarednode]        (vlasov)       [below=of hartree] {Vlasov};

\draw[->] (schrodinger.east) -- (hartree.west) node[anchor=south,inner sep=2pt,midway] {$N \to \infty$};
\draw[->] (liouville.east) -- (vlasov.west) node[anchor=north,inner sep=2pt,midway] {$N \to \infty$};
\draw[->] (schrodinger.south) -- (liouville.north) node[anchor=east,inner sep=2pt,midway] {$\epsilon \to 0$};
\draw[->] (hartree.south) -- (vlasov.north) node[anchor=west,inner sep=2pt,midway] {$\epsilon \to 0$};
\end{tikzpicture}
\end{center}
where moving from left to right approaches the mean-field limit (number of particles $N \to \infty$) and from top to bottom approaches the classical setting ($\epsilon \to 0$). An important area of theory in statistical mechanics is the convergence of solutions to these equations following the convergence of the equations themselves. These convergences individually have been proven in other works \cite{nonot_schrodinger_convergence_1,nonot_schrodinger_convergence_2,nonot_schrodinger_convergence_3,nonot_schrodinger_convergence_4}, but the convergence of solutions of the Schrödinger equation to those of the Hartree equation in these papers is not uniform in $\epsilon$, and so convergence in the concurrent limit $N \to \infty$ and $\epsilon \to 0$ had not previously been established. This is the gap that these optimal transport tools fill.

The first result from optimal transport comes from \cite{FG-15-QM}, where they compare solutions of the $N$-body Schrödinger equation and the mean-field Schrödinger equation (Hartree equation), giving a quantitative estimate which is uniform as $\epsilon \to 0$.

\begin{prop}\cite{FG-15-QM}
Letting $t \mapsto \rho_{\epsilon, N}(t)$ be a solution to the $N$-body Schrödinger equation with initial state $\rho_{\epsilon, N}^{\text{in}}$, and $t \mapsto \rho_{\epsilon}(t)$ a solution to the Hartree equation with initial state $\rho_\epsilon^{\text{in}}$, we have for fixed time $t$ that
\begin{equation}
    \frac{1}{N}W_2^{MK,\epsilon} \left( \rho_{\epsilon, N}(t), \rho_{\epsilon}(t)^{\otimes N}\right)^2 \leq \mathcal{O}\left(\frac{1}{N}\right) + \mathcal{O}\left(\frac{1}{N}\right) W_2^{MK,\epsilon} \left( \rho_{\epsilon, N}^{\text{in}}, (\rho_{\epsilon}^{\text{in}})^{\otimes N}\right)^2
\end{equation}
and so for $\rho_{\epsilon}^{\text{in}}$ a Töplitz operator at scale $\epsilon$ and $\rho_{\epsilon, N}^{\text{in}} = (\rho_{\epsilon}^{\text{in}})^{\otimes N}$, this right hand side is $\mathcal{O}\left(\frac{1}{N}\right) + \mathcal{O}\left(2d\epsilon\right)$. Taking $N \to \infty$ and $\epsilon \to 0$ simultaneously gives convergence of solutions.
\end{prop}

The work \cite{FG-15-SE} extends these results, using the semiclassical $\mathcal{E}_{\epsilon}$ to show convergence of solutions of the Schrödinger equation to solutions of the Vlasov equation. 
\begin{prop}\cite{FG-15-SE}
For $\rho_{\epsilon, N}(t)$ as above and $f(t)$ a probability density on $\mathbb{R}^d \otimes \mathbb{R}^d$ which solves the Vlasov equation with initial function $f^{\text{in}}$, we have for fixed $t$ that
\begin{equation}
    \frac{1}{N}\mathcal{E}_\epsilon \left(f(t)^{\otimes N},  \rho_{\epsilon, N}(t)\right)^2 \leq \mathcal{O}\left(\frac{1}{N}\right) + \mathcal{O}\left(\frac{1}{N}\right) \mathcal{E}_\epsilon \left(\left(f^{\text{in}}\right)^{\otimes N}, \rho_{\epsilon, N}^{\text{in}}\right)^2
\end{equation}
and so we get the same convergence results as $N \to \infty$ and $\epsilon \to 0$.
\end{prop}
They also use this $\mathcal{E}_\epsilon$ to show uniformity of convergence in $N$ as $\epsilon \to 0$, leading to convergence of solutions of the Schrödinger equation to those of the Liouville equation.
Finally, they use these quantities to get convergence rate estimates for the already-known convergence of solutions to the Hartree equation to those of the Vlasov equation, which were improved upon in \cite{MI-24-ES}.

The key insight from optimal transport, particularly in uniform convergence of solutions of the Schrödinger equation to those of the Hartree equation, is the use of a metric which encompasses both distance in space and distance in distribution. Previous work using the trace norm \cite{nonot_schrodinger_convergence_1, nonot_schrodinger_convergence_3} does not take into account physical distance between particles, and so does not obtain controls in the large $N$ limit which are uniform as $\epsilon \to 0$. Establishing convergence of solutions to the Liouville equation to those of the Vlasov equation in the classical setting relies on the contruction of an empirical measure, which has no obvious analogue in the quantum setting, and so the construction of this optimal transport metric which behaves well with Töplitz quantisation is key in this final convergence estimate for the problem.

\subsection{Quadratures with partial transpose} \label{subsection:couplings_transpose}
\subsubsection{Definition}

Building on the work of Golse and et. al., \cite{GPDT-19-QC} introduced a similar distance generalising $W_2^{MK,\epsilon}$ to different cost functions, and tweaking slightly to allow couplings to be linked directly to stochastic maps.
\begin{defn}
    Let $\mathcal{C}^*(\rho,\sigma) = \{\pi \in \mathcal{D}(\mathcal{H}\otimes \mathcal{H}^*): \text{{\em Tr}}_{\mathcal{H}}[\pi] = \rho^{T}, \text{{\em Tr}}_{\mathcal{H}^*}[\pi] = \sigma\}$. Then for a set $S = \{R_1, \dots, R_M\}$ of self-adjoint operators on $\mathcal{H}$ we call \textbf{quadratures}, the \textbf{second-order quadratures quantum Wasserstein distance} is
    \begin{equation}
        W_2^S(\rho,\sigma) = \inf_{\pi \in \mathcal{C}^*(\rho,\sigma) }\text{{\em Tr}} \left[\left(  \sum_{i=1}^M \left|R_i \otimes \mathbb{I}_{\mathcal{H}^*} - \mathbb{I}_{\mathcal{H}} \otimes R_i^T \right|^2 \right)\pi \right].
    \end{equation}
\end{defn}
This is very similar to the distance in Section \ref{subsection:coupling_position_momentum}, though they chose the specific case where the $R_i$ were the position and momentum operators, and did not dualise.

This dualisation of the second copy of $\mathcal{H}$ is two-fold. Firstly, it allows for the identification of a canonical purification of a state $\rho$ on $\mathcal{H}$. $\mathcal{H} \otimes \mathcal{H}^*$ is canonically isomorphic to the set $\mathcal{T}_2(\mathcal{H})$ of operators on $\mathcal{H}$ with finite Hilbert-Schmidt norm, via the correspondence $\ket{a} \otimes \bra{b} \mapsto \ketbra{a}{b}$, and so we may express  elements of $\mathcal{D}(\mathcal{H} \otimes \mathcal{H}^*)$ instead as density operators on $\mathcal{T}_2(\mathcal{H})$. In this way, we say the {\em canonical purification of $\rho \in \mathcal{D}(\mathcal{H})$} is $||\sqrt{\rho}\rangle \rangle \langle \langle \sqrt{\rho}||$, where $\| \sqrt{\rho} \rangle$ is the inverse of $\rho$ under this correspondence.

Secondly, suppose $\Phi: \mathcal{D}(\mathcal{H}) \to  \mathcal{D}(\mathcal{H})$ is a quantum channel sending $\rho$ to $\sigma$. Then we may associate coupling
\begin{equation}
    \Pi_{\Phi} = (\Phi \otimes \mathbb{I}_{\mathcal{D}(\mathcal{H}^*)})||\sqrt{\rho}\rangle \rangle \langle \langle \sqrt{\rho} ||
\end{equation}
to channel $\Phi$, in the same way that stochastic maps are associated to classical couplings in the Monge-Kantorovich problem. This map $\Phi \mapsto \Pi_{\Phi}$ is in fact a bijection between couplings of $\rho$ and $\sigma$ and channels sending $\rho$ to $\sigma$.

The authors also proposed a variant of this distance.
\begin{defn}
\begin{equation} \label{eq:couplings_quadratures_tri_ineq_conjecture}
    D_*^S(\rho,\sigma) = \sqrt{W_2^S(\rho,\sigma)^2 - \frac{1}{2}W_2^S(\rho,\rho)^2 - \frac{1}{2}W_2^S(\sigma,\sigma)^2}
\end{equation}
\end{defn}
This was conjectured in~\cite[Remark 5.6]{otqs-palma-trevisan} to be a true distance, motivated by the properties of $W_2^S$ listed below. Up to some non-degeneracy assumptions on $S$ to ensure that $D^*(\rho,\sigma) = 0$ only if $\rho=\sigma$, this is a semidistance.

\subsubsection{Properties and applications}
All properties here come from \cite{GPDT-19-QC} unless otherwise stated.
\begin{itemize}
    \item $W_2^S$ is symmetric, finite as long as $\rho$, $\sigma$ have finite energy, and additive under tensor products.
    \item $W_2^S(\rho,\rho) = \text{Tr}[C_S ||\sqrt{\rho}\rangle \rangle \langle \langle \rho ||]$ were proven. This means that as with~\cite{FG-15-QM}, $W_2^S(\rho,\rho) > 0$ in general.
    \item Similarly to~\cite{FG-21-QS}, there is a variant of the triangle inequality, namely that
\begin{equation}
    W_2^S(\rho,\tau) \leq W_2^S(\rho,\sigma) + W_2^S(\sigma,\sigma) + W_2^S(\sigma, \tau).
\end{equation}
    \item $W_2^S(\rho,\sigma)^2 \geq \frac{1}{2}W_2^S(\rho,\rho)^2 + \frac{1}{2}W_2^S(\sigma,\sigma)^2$, and so $W_2(\rho,\rho) = \min_{\sigma} W_2^S(\rho,\sigma)$. This shows that $D_*^S \geq 0$ with equality when $\rho = \sigma$.
    \item The modified quantity $D_*^S$ satisfies \cite{GB-24-MP} the triangle inequality when either the two outer states, or the middle state, are pure. They also gave numerical evidence for its truth in the general case with $\text{dim}\mathcal{H} = 3,4$.
\end{itemize}

The paper \cite{GPG-22-QWI} showed in the case where $\mathcal{H} = \mathbb{C}^2$ and the quadratures are the three non-identity Pauli matrices, that the isometries of $W_2^S$ are exactly the Wigner symmetries - that is, conjugations by unitary and anti-unitary operators. They also considered the case where only the $X$ and $Z$ Paulis are used, showing that each symmetry can be written as a composition of a quasi-orthogonal map and a quasi-conjugation map.
It is notable here that, due to the non-zero self-distance phenomenon, there exist ``isometries" which are non-surjective, and even those which are non-injective.

As of yet, however, the broader utility of the quantity $W_2^S$ is unknown.
$W_2^S$ was shown to have some applications in the study of quantum Gaussian systems, notably that the optimal coupling between two Gaussian states is itself a Gaussian state. $W_2^S$ is closely related to the classical $\mathcal{W}_2$ on semiclassical states. Specifically, for $\mathcal{H} = L^2(\mathbb{C}^d)$ we have canonical quadratures $S_{\text{can}} = \{R_1, \dots, R_{2d}\}$ 
\begin{equation}
    R_i = Q_i, \qquad R_{d+i} = P_i, \qquad i = 1, \dots, d
\end{equation}
where $Q_i$ is the position operator $(Q_i \psi)(q) = q_i\psi(q)$, and $P_i$ momentum $(P_i \psi)(q) = -i\frac{\partial}{\partial q_i} \psi(q)$ on wavefunction $\psi$.
Indeed, letting the states $\rho_\mu$ and $\rho$ be the P-representations of two classical probability measures $\mu$ and $\nu$ respectively on $\mathbb{C}^d$, we have
\begin{equation}
    W_2^{S_{\text{can}}}(\rho_\mu,\rho_\nu) \leq \mathcal{W}_2^{\text{Eucl}}(\mu,\nu) + d
\end{equation}
and letting $Q_\rho$, $Q_\sigma$ be the Q-representations of two states $\rho$, $\sigma$ respectively on $L^2(\mathbb{C}^d)$, we have
\begin{equation}
    \mathcal{W}_2^{\text{Eucl}}(Q_\rho,Q_\sigma) - d \leq W_2^{S_{\text{can}}}(\rho,\sigma).
\end{equation}

\subsection{Quadratures with modular transport plans} \label{subsection:coupling_modular_plans}

The work of \cite{RD-18-TP, RD-21-WD, RD-22-DB} extends these formulations further, this time in the framework of von Neumann algebras. They give a modification to the cost matrix based on bimodule ideas from Tomita-Takesaki theory which allows us to recover the triangle inequality and faithfulness.

In this setting, a `coupling' between states $\rho$ and $\sigma$ on von Neumann algebras $A$, $B$ respectively, is defined as a state $\omega$ on the algebraic tensor product $A \odot B'$ such that for $a \in A$ and $b' \in B$ we have $\omega(a \otimes 1) = \rho(a)$ and $\omega(1 \otimes b') = \sigma'(b')$. Taking $A = B$, for elements $s = \{r_1, \dots, r_M\}$ of $A$ we define the cost function
\begin{equation} \label{eq:defn_duvenhage_cost}
    c = \sum_{i=1}^M |r_i  \otimes 1 - 1 \otimes S_\sigma r_i^* S_\sigma|^2
\end{equation}
where $S_\sigma$ is the conjugate linear operator with respect to $\sigma$ from Tomita-Takesaki theory. Note that this notation does not strictly make sense in its current form: the way in which \eqref{eq:defn_duvenhage_cost} should be interpreted is discussed rigourously in \cite{RD-18-TP}. The cost of a coupling $\omega$ is then simply $\omega(c)$. It is noted that the dependence of $c$ on $\sigma$ means that this definition is asymmetric. In order to enforce symmetry, they restrict the optimisation to those couplings $\omega$ which satisfy a {\em balance} condition.

For unital completely positive maps $\psi : A \to A$ and $\varphi: B \to B$ and states $\rho$, $\sigma$ on $A$, $B$, respectively, we say that a coupling $\omega$ is in {\em balance} with $(A,\psi,\rho)$ and $(B,\varphi,\sigma)$ (in that order) if for all $a \in A$ and $b' \in B'$ we have
\begin{equation}
    \omega(\psi(a) \otimes b') = \omega(a \otimes \varphi'(b')).
\end{equation}
The set of {\em modular couplings} is then defined as
\begin{equation}
    \mathcal{C}_{\text{mod}}(\rho,\sigma) = \left\{ \omega \in \mathcal{C}(\rho,\sigma) : \omega \text{ in balance with } (A, \alpha_t^\rho, \rho) \text{ and } (B, \alpha_t^\sigma, \sigma) \right\}
\end{equation}
for $\alpha^\rho$ and $\alpha^\sigma$ the modular automorphism groups of $\rho$ and $\sigma$ respectively.
\begin{defn}
The definition of the \textbf{second-order modular quantum Wasserstein distance} is then given by
\begin{equation}
    W_2^s(\rho,\sigma) = \sqrt{ \inf_{\omega \in \mathcal{C}_{\text{mod}}(\rho,\sigma) }\omega(c)}.
\end{equation}
\end{defn}

This restriction to modular transport plans allows the recovery of many interesting properties. Most notably, this construction gives a true metric: assuming that $\{r_1, \dots, r_M\} = \{r_1^*, \dots, r_M^*\}$ and that the set $s$ generates $A$, this quantity is a true metric. Indeed, \cite{RD-18-TP} is able to prove nonnegativity, nondegeneracy, faithfulness, and the triangle inequality. Furthermore, optimal couplings always exist. Work \cite{RD-21-WD} discusses the properties of these distances on reduced systems, and also provides some examples of these distances for specific choices of $s$.

\subsection{Asymmetric projector cost matrix} \label{subsection:coupling_asymmetric_projectors}
\subsubsection{Definition}
In another direction, a series of papers
\cite{SC-19-GAN,KZ-21-QOT,KZ-22-MN,MH-22-MN,SF-21-MK}
have discussed a proposal for a second-order Wasserstein distance where the cost function is the projector onto the asymmetric subspace of $\mathcal{H} \otimes \mathcal{H}$ as defined in Section \ref{subsubsection:quantumbg_notable_operators}. This gives us the following definition~\cite{SC-19-GAN}.
\begin{defn}[Second-order Wasserstein distance from asymmetric projector]
    Let $\mathcal{H}$ be a finite-dimensional Hilbert space and $P_{\text{asym}}$ the projector onto the asymmetric subspace of $\mathcal{H}\otimes \mathcal{H}$. Then
\begin{equation}
    W_2^{\text{asym}}(\rho,\sigma) = \left( \inf_{\pi \in \mathcal{C}(\rho,\sigma)}\text{Tr}[P_{\text{asym}}\pi]  \right)^{1/2}. \label{eq:defn_W2_asym}
\end{equation}
\end{defn}

This quantity is a semidistance, and the infimum here is the primal problem of a semi-definite program (SDP), so calculating it is efficient in small dimension \cite{Gartner2012}.
However, one key property missing is data processing. As the cost function here is an analogue of the discrete metric on a classical space $\mathcal{X}$, we would expect the data processing inequality to hold everywhere. However, \cite{MH-22-MN} showed that in any dimension at least 4, there exist $\rho$, $\sigma$ such that $W_2^{\text{asym}}(\rho,\sigma) > W_2^{\text{asym}}(\rho\otimes \mathbb{I}/2,\sigma \otimes \mathbb{I}/2)$ This means $W_2^{\text{asym}}$ can increase under the partial trace channel.
That same paper
proposed a stabilised modification of this distance to recover the data processing inequality.
\begin{defn}[Stabilised second-order Wasserstein distance from asymmetric projector]
    Let $\mathcal{H}$ be a finite-dimensional Hilbert space. Then
\begin{equation}
    W_2^{\text{{\em asym}},s}(\rho,\sigma) = W_2^{\text{asym}}\left(\rho \otimes \mathbb{I}/{2}, \sigma \otimes \mathbb{I}/{2} \right).
\end{equation}
\end{defn}
Though originally defined as $\inf_\gamma W_2^{\text{asym}}(\rho \otimes \gamma, \sigma \otimes \gamma)$ for all auxiliary systems and states $\gamma$, it was shown 
that this infimum is achieved at $\gamma = \mathbb{I}/2$.

\subsubsection{Properties and applications}

\begin{itemize}
    \item $W_2^{\text{asym}}$ and $W_2^{\text{asym},s}$ are both semidistances \cite{KZ-21-QOT, MH-22-MN}.
    \item $W_2^{\text{asym},s}$ is monotone under CPTP maps \cite{MH-22-MN}.
    \item $W_2^{\text{asym},s}$ is invariant under tensor products, jointly convex, and unitarily invariant \cite{KZ-21-QOT, MH-22-MN}.
    \item The Gibbs manifold for the quantum optimal transport problem for $W_2^{\text{asym}}$ is semialgebraic \cite{DP-22-GM}, which has implications for its calculation via an SDP.
    \item Transport of semiclassical states via $W_2^{\text{asym}}$ can be cheaper than the corresponding classical transport \cite{KZ-21-QOT}.
\end{itemize}
For the triangle inequality, little is known. The paper \cite{SF-21-MK} proved the triangle inequality for dimension $d = 2$, but beyond that it is not clear either way if it holds.

While potientially useful in this context, it unfortunately seems that extending this distance to more interesting geometries of $\mathcal{H}$ is not possible. As discussed in \cite{KZ-21-QOT}, if the cost matrix $C$ is not positive definite on the asymmetric subspace and zero on the symmetric subspace, it will result in a quantity which is not a semidistance. The fairly large asymmetric component of the identity matrix means that any attempt to put the SWAP operator over a subsystem will result in a quantity with non-zero self-distance.

As an application, the work \cite{XS-23-CQ} used the distance $W_2^{\text{asym},2}$ to define a coherence quantifier for quantum states. The coherence measure of a pure state $\ketbra{\psi}{\psi}$ is then
\begin{equation}
    T(\ket{\psi}) = \min_{\delta \in \mathcal{I}_{\mathcal{E}}} W_2^{\text{asym},s} (\ketbra{\psi}{\psi},\delta)
\end{equation}
for an incoherence basis ${\mathcal{E}}$ of $\mathcal{H}$ and the set $\mathcal{I}_{\mathcal{E}}$ of incoherent states. The coherence of a mixed state defined via the convex roof method as
\begin{equation}
    T(\rho) = \min \left\{\sum_j q_j T (\ket{\psi_j}) : \rho = \sum_j q_j \ketbra{\psi_j}{\psi_j}\right\}.
\end{equation}

They show that this measure $T$ satisfies the four postulates~\cite{baumgratz-2014}
of a measure of quantum coherence, and also gave it an operational meaning in terms of the time needed to transform a pure state into an incoherent state under unitary evolution. Coherence is a key ingredient in fields such as quantum algorithms \cite{deutsch-josza}, solid-state physics \cite[Section VIIB]{Hu_2018}, and nanoscale thermodynamics \cite{coherence_use}.

\subsection{Transport plans method} \label{subsection:coupling_transport_plans}
\subsubsection{Definition}

A more recent work by the author of this review gives a generalisation from the coupling point of view which focuses more on flexibility in definition, avoiding the use of cost matrices altogether. Instead, \cite{EB-24-WP}
starts from any distance $d$ on the projective Hilbert space $\mathbb{P}\mathcal{H} = (\mathcal{H} \setminus \{0\}) / \mathbb{C} $, and defines a cost based on the distance itself.
\begin{defn}
A \textbf{quantum transport plan} between density operators $\rho$ and $\sigma$ is defined as any countable set of triples $Q = \{ (q_j, \ket{\psi_j},\ket{\varphi_j})\}_{j \in J}$ with $q_j > 0$ such that $\sum_{j \in J} q_j \ketbra{\psi_j}{\psi_j} = \rho$ and $\sum_{j \in J} q_j \ketbra{\varphi_j}{\varphi_j} = \sigma$. The set of such plans is denoted $\mathcal{Q}(\rho,\sigma)$, and then the \textbf{$p^{\text{th}}$-order quantum Wasserstein distance} can be defined as
\begin{equation}
    W_p^d(\rho,\sigma) = \left( \inf_{Q \in \mathcal{Q}(\rho,\sigma)} \sum q_j d(\ket{\psi_j},\ket{\varphi_j})^p \right)^{1/p}.
\end{equation}
\end{defn}

This gives a definition of a quantum Wasserstein distance for all orders $p \in [1,+\infty]$ (with the infinite-order definition taking $\max_j d(\ket{\psi_j},\ket{\varphi_j})$ instead of the sum), and with respect to any underlying metric $d$ on the set of pure states satisfying some light continuity conditions.

The motivation behind this definition is that in order to define a quantum Wasserstein distance valid for all orders $p$, it is reasonable to expect that, as in the classical case, the distance between pure states is invariant in $p$. This induces a metric on the set of pure states, and so we may instead start from that metric.

\subsubsection{Properties and applications}
Assuming these light continuity conditions on $d$, the following properties are recovered:
\begin{itemize}
    \item Semimetricity: $W_p^d(\rho,\sigma) \geq 0$ with equality if and only if $\rho = \sigma$.
    \item Agreement with $d$: for pure states $\ket{\psi}, \ket{\varphi}$, we have $W_p^d(\ketbra{\psi}{\psi},\ketbra{\varphi}{\varphi}) = d(\ket{\psi},\ket{\varphi})$.
    \item Data processing for mixed unitary channels: for unitaries $U_i$ which are symmetries of $d$, the channel $T(\cdot) = \sum_i p_iU_i \cdot U_i^{\dagger}$ satisfies data processing.
    \item Hierarchy in $p$: For $p_1 < p_2$, we have $W_{p_1}^d \leq W_{p_2}^d$.
    \item Continuity: $W_p^d$ is uniformly continuous.
    \item Infimum achieved at polynomial sizes: For Hilbert space $\mathcal{H}$ of dimension $D$, the infimum in the definition of $W_p^d$ is achieved for a transport plan $Q$ with at most $2D^2$ elements.
\end{itemize}

Additionally, when the $2$-norm is Holder continuous with respect to the metric $d$ with exponent $\alpha = 1$, it is possible to dualise this quantity to get an analogy $L_d(H)$ of the classical Lipschitz constant of an observable $H$. Dualising again gives a norm defined as $\norm{\cdot}_{W_1^d}$, which sits at the bottom of the hierarchy in $p$.

The paper investigates this distance for a few specific cases of the underlying metric $d$, most notably the Riemannian metric on the set of pure states induced by Nielsen's complexity geometry of unitaries \cite{Nielsen_2006}. One open question that remains is if these techniques can be used to extend the analysis of the complexity of quantum circuits (\cite{LL-22-WC}, to be discussed in Section \ref{subsubsection:lipschitz_applications_complexity}) from shallow depth to all circuits.

This viewpoint gives many advantages over other coupling definitions, such as the nontrivial guarantee that $W_p^d$ is a semidistance, and broad flexibility in $p$ and $d$ allowing for a wide range of applications. However, while it is easy to upper bound this distance by finding a good $Q$, it is much more difficult to lower bound it, and reliable approximation is essential for any practical use. It is also unclear, even for $p=1$ and in small dimension, what conditions (if any) on $d$ could allow recovery of the triangle inequality.

The paper \cite{EB-24-WP} goes on to discuss a number of applications of these quantities. Most notably, the flexibility in order $p$ allows comparison of Wasserstein distances of different orders $p_1, p_2$, allowing for a quantum analogue of hypercontractivity. Additionally, this breadth allows for the characterisation of the $p^{\text{th}}$ moment of the distance (with respect to any $d$) of the output of two classical-quantum sources in terms of their average outputs. Finally, they discuss a potential use in the theory of quantum Wasserstein GANs which will be treated in Section \ref{section:globalapps}.

\subsection{Coupling formulations conclusion: desirable properties}

We give in Table \ref{tab:desirable_properties_coupling} the desirable properties detailed in Section \ref{section:quantumbg} and which are satisfied by each of the quantum Wasserstein distances arising from a coupling formulation.

\begin{table}[htbp]
    \centering
    \begin{tabular}{|c|c|c|c|c|c|} \hline \ref{subsection:coupling_asymmetric_projectors}
        Property & \hyperref[subsection:coupling_position_momentum]{$W_2^{MK,\epsilon}$} & \hyperref[subsection:couplings_transpose]{$W_2^{S}$} & \hyperref[subsection:coupling_modular_plans]{$W_2^s$} & \hyperref[subsection:coupling_asymmetric_projectors]{$W_2^{\text{asym}}$} & \hyperref[subsection:coupling_transport_plans]{$W_p^d$} \\ \hline
        Semi-metricity & No & No & Yes & Yes & Yes \\
        Triangle inequality & ? & Modified & Yes & ? & ? \\
        Link to couplings & Yes & Yes & Yes & Yes & Yes \\
        Link to duality & Yes & No & No & No & Yes \\
        Link to dynamics & No & No & Yes & No & No \\
        Continuity & Yes & Yes & Yes & Yes & Yes\\
        Recovers $\norm{\rho-\sigma}_1$ for $p=1$ & N/A & N/A & N/A & N/A & Partially \\
        Data processing & ? & ? & ? & Modified & Yes \\
        Ease of calculation & LP & LP & No & LP & No\\
        Flexibility in definition & No & $p$ & $p$ & No & $p, d$ \\
        Functional inequalities & Yes & No & No & No & Limited \\
        Applicability & Yes & Yes & Limited & Limited & Yes \\
        Variety of application & Limited & No & No & Limited & Yes \\ \hline
        
    \end{tabular}
    \caption{Desirable properties satisfied by those quantum Wasserstein distances given by a coupling formulation}
    \label{tab:desirable_properties_coupling}
\end{table}

\section{Dynamical formulations}
\label{section:dynamics}
\subsection{Construction of the Riemannian metric} \label{subsection:dynamical_construction}

With the limitations of direct generalisation of the classical definition, and the importance of second-order Wasserstein distances for applications such as concentration inequalities as discussed in Section \ref{subsubsection:classicalapplicationCIs}, a series of papers have developed a proposal for a second-order quantum Wasserstein distance by generalising the Benamou-Brenier formulation of the classical quantities.
The definition is motivated by the classical property of the Wasserstein distance discussed in Section \ref{subsubsection:classicalapplicationheatflows}, that the solution of the Fokker-Planck equation associated to the heat equation follows the gradient, with respect to the Wasserstein distance of order 2 on $\mathbb{R}^n$, of the relative entropy. The work of~\cite{CM-16-GF}
gives a construction, with respect to a quantum Markov semigroup satisfying detailed balance, of a Riemannian metric on the set of positive density matrices in dimension $n$ with respect to which the forward Kolmolgorov equation of the semigroup is gradient flow of the quantum relative entropy to the invariant state. This is an extension of the preliminary work in~\cite{CM-12-2W}, treating specifically the Fermi-Ornstein-Uhlenbeck semigroup.

The metric is defined as follows. Let $M$ be a von Neumann subalgebra of $\mathcal{H}$ and suppose that QMS $\mathcal{P}_t = e^{t\mathcal{L}}$ has an extension to a QMS on $\mathcal{H}$ which satisfies detailed balance for its unique invariant state $\sigma$. Regard $\Delta_\sigma$ as an operator on $\mathcal{B}(\mathcal{H})$. By Alicki's theorem, we can write the generator $\mathcal{L}$ in the standard form \eqref{eq:qms-dbc-standard-form-alicki}
\begin{equation} \label{eq:alicki_representation}
    \mathcal{L}A = \sum_{j \in J} e^{-\omega_j / 2} \left( V^*_j [A,V_j] + [V^*_j, A] V_j \right).
\end{equation}

The Riemannian metric is defined through a quantum analogue of the continuity equation, equation \eqref{eq:continuity_equation}.
The paper defines a non-commutative version of `multiplication by $\rho$', $[\rho]_{\omega}$, defined as 
\begin{equation}
    [\rho]_{\omega} = R_\rho \circ f_\omega (\Delta_{\rho})
\end{equation}
where $R$ is right-multiplication and 
\begin{equation}
    f_{\omega}(t) = e^{\omega/2} \frac{t-e^{-\omega}}{\log t + \omega}.
\end{equation}
For the vector $\underline{\omega} = (\omega_1, \dots, \omega_{|J|})$ associated to $\mathcal{L}$, we then define $[\rho]_{\underline{\omega}}$ as an operator on $|J|$ copies of $M$ defined by
\begin{equation}
    [\rho]_{\underline{\omega}} \mathbf{W} = ([\rho]_{\omega_1} W_1, \dots, [\rho]_{\omega_{|J|}} W_{|J|} ).
\end{equation}

The form in equation \eqref{eq:qms-dbc-standard-form-alicki} also induces a natural notion of derivation on the von Neumann algebra. Define the partial derivative operators and the grad operator by commutators
\begin{equation}
    \partial_j A = [V_j, A] \qquad \nabla A = (\partial_1 A, \dots, \partial_{|J|} A)
\end{equation}
which gives 
\begin{equation}
    \text{div}(A_1, \dots, A_{|J|}) = -\sum_{j \in J} \partial_j^{\dagger} A_j = \sum_{j \in J} [A_j, V_j^*]
\end{equation}
where the dagger is with respect to the Hilbert-Schmidt norm.

We are now ready to present the quantum continuity equation.

\begin{defn}
    Let $\mathcal{L}$ have the form in equation \eqref{eq:qms-dbc-standard-form-alicki}, and let $\rho(t)$ be a differentiable path of density operators such that $\rho(0) = \rho_0$. Let $\mathbf{V}$ be a vector field in $M^{|J|}$. We say that $\rho(t)$ and $\mathbf{V}$ satisfy the \textbf{quantum continuity equation for $\mathcal{L}$} if
\begin{equation} \label{eq:quantum_continuity_equation}
    \dot{\rho}(0) + \text{{\em div}}([\rho_0]_{\omega}\mathbf{V}) = 0.
\end{equation}
\end{defn}

We note that for each path $\rho$ there is a unique vector $\mathbf{V}$ of the form $\mathbf{V} = \nabla U$ which satisfies the quantum continuity equation for $\mathcal{L}$.
At each $\rho$, and for generator $\mathcal{L}$, we can then define the associated inner product
\begin{equation}
    \langle \mathbf{V},\mathbf{W} \rangle_{\mathcal{L},\rho} = \sum_{j \in J}  \langle \mathbf{V}_j , [\rho]_{\omega_j} \mathbf{W}_j \rangle_{HS}
\end{equation}
which gives norm $\norm{\mathbf{V}}_{\mathcal{L},\rho}$.
Then we can define the length of tangent vectors in Riemannian metric $g_{\mathcal{L}}$.

\begin{defn}
    The \textbf{dynamical quantum Wasserstein distance of order 2} $W_2^{\mathcal{L}}$ with respect to quantum Markov semigroup satisfying the $\sigma$-DBC with generator $\mathcal{L}$ is the geodesic distance of the Riemannian metric $g_{\mathcal{L}}$ defined by
    \begin{equation} \label{eq:defn_riemannian_metric_L}
        \norm{\dot{\rho}(0)}_{g_{\mathcal{L}},\rho(0)}^2 = \norm{\nabla U}_{\mathcal{L},\rho(0)}^2
    \end{equation}where $\dot{\rho}(0)$ and $\mathbf{V} = \nabla U$ satisfy the quantum continuity equation (equation \eqref{eq:quantum_continuity_equation}) for $\mathcal{L}$.
\end{defn}
The key result of~\cite{CM-16-GF}
is the following theorem.

\begin{thm} \label{thm:qms-dbc-gradient-flow}
    Let $\mathcal{P}_t = e^{t\mathcal{L}}$ be a QMS on $M$ satisfying the $\sigma$-DBC condition for some positive $\sigma$. Then
    \begin{equation} \label{eq:quantum-qms-flow}
        \frac{\partial}{\partial t} \rho = \mathcal{L}^{\dagger} \rho
    \end{equation}
    is the gradient flow for the relative entropy $D(\cdot || \sigma)$ in $g_{\mathcal{L}}$.
\end{thm}

The specific conditions on $\mathcal{L}$ for which $\mathcal{P}_t$ has an associated Riemannian metric satisfying Theorem \ref{thm:qms-dbc-gradient-flow} were explored in~\cite{CM-18-OT}. A more general form of operators giving rise to such metrics $g_{\mathcal{L}}$ was developed, and necessary conditions on $\mathcal{P}_t$ for the existence of $g_{\mathcal{L}}$ were also found. Specifically, for such a $g_{\mathcal{L}}$ to exist it is necessary for $\mathcal{P}_t$ to satisfy $\sigma$-BKM detailed balance.

\subsubsection{Dual dynamical first-order construction}

In another direction, \cite{CM-18-OT} also uses Kantorovich duality to define a $W_1$ based on this dynamical $W_2$. To do this, we define the norm $\norm{\cdot}_{\mathcal{B},2}$ on gradients as follows:
\begin{equation}
    \norm{\nabla A }_{\mathcal{B},2} = \sqrt{\frac{1}{2} \sum_{j \in J} \norm{l_j^{\dagger}(B_j B_j^*)  + r_j(B_j^* B_j) }_{M} }
\end{equation}
for $B_j = \partial_j A$, for $l_j$, $r_j$ left and right multiplication respectively, and for $\norm{\cdot}_M$ the operator norm on $M$. This acts as a measure of the size of the gradient of $A$, and so we consider an operator $A \in M$ to be {\em Lipschitz} in this formulation if $\norm{\nabla A}_{\mathcal{B},2} \leq 1$.

This then gives us, analogously to the classical case, a Kantorovich dual-style $W_1$.

\begin{defn}
    The \textbf{dual dynamical first-order Wasserstein distance} is given by
\begin{equation} \label{eq:defn_dual_dynamical_w1}
    W_1^{\mathcal{L}}(\rho,\sigma) = \sup \left\{ \text{{\em Tr}}[(\rho-\sigma)A] : A \in M, \norm{\nabla A }_{\mathcal{B},2} \leq 1 \right\}.
\end{equation}
\end{defn}

\subsection{Ricci curvature of quantum Markov semigroups}

The latter part of~\cite{CM-16-GF} proved a very important properties of $W_{2}^{\mathcal{L}}$, namely the geodesic convexity of the relative entropy, also known as a quantum Ricci curvature bound. For general quantum Markov semigroups, the concept of Ricci curvature is difficult to generalise, though as mentioned it plays an important role in the theory of classical optimal transport. However, in the classical setting, there is a useful equivalent formulation of a lower Ricci curvature bound~\cite{sturm-2005}
: the Ricci curvature is bounded below by some $\kappa \in \mathbb{R}$ if and only if the entropy $H(\mu)$ is $\kappa$ convex along 2-Wasserstein geodesics.

\begin{defn}
Given a QMS $\mathcal{P}_t$, we say that a function $F$ is \textbf{$\lambda$-geodesically convex} if, for any distance-minimising geodesic $\gamma : [0,1] \to \mathcal{D}(\mathcal{H})$, we have for all $s \in (0,1)$ that
\begin{equation}
    \frac{\text{{\em d}}^2}{\text{{\em d}}^2 s} F(\gamma(s)) \geq \lambda \norm{\dot{\gamma}(s)}_{g_\mathcal{L},\gamma(s)}.
\end{equation}
We say that such a QMS satisfies a \textbf{Ricci curvature bound} with constant $\kappa$, written $\text{\textbf{{\em Ric}}}(\kappa)$, if the relative entropy to $\sigma$ is $\kappa$-geodesically convex, i.e. that
\begin{equation}
    \frac{\text{{\em d}}^2}{\text{{\em d}}^2 s} D(\gamma(s) \| \sigma) \geq \kappa \norm{\dot{\gamma}(s)}_{g_\mathcal{L},\gamma(s)}.
\end{equation}
\end{defn}
Work \cite{CM-16-GF} showed that if $\mathcal{P}_t$ satisfies an {\em intertwining relation}, the details of which we will omit here, then such a Ricci curvature bound does hold.

\subsection{Applications of the dynamical formulation: concentration inequalities} \label{subsection:dynamical_concentration}

Due to its connection with the classical dynamical formulation of optimal transport, this $W_2$ is particularly suited to establishing quantum concentration inequalities.

The papers~\cite{CRND-17-CI, CRND-17-HWI, CM-16-GF,kastoryano-2012}
proved the following implications between concentration inequalities for a quantum Markov semigroup whose associated Hilbert space has dimension $d$ and which satisfy the $\sigma$-DBC:
\begin{itemize}
    \item $MLSI(\lambda) \implies TC_{2,\sigma}(c_2)$ for $c_2 = 1/\lambda$.
    \item $MLSI(\lambda) \implies PI(\lambda)$.
    \item $TC_{2,\sigma}(c_2) \implies PI(\lambda)$ for $\lambda = 1/c_2$.
    \item $PI(\lambda) \implies Exp(c)$ for some constant $c$ depending on both $\lambda$ and the Lipschitz constant.
    \item $TC_{2,\sigma}(c_2) \implies TC_{1,\sigma}(c_1)$ for $c_1 = d c_2$.
    \item $TC_{1,\sigma}(c_1) \implies Gauss(c_1)$ for the following Gaussian concentration inequality for any self-adjoint operator $f$:
    \begin{equation}
        \text{Tr} \left[ \sigma \mathbf{1}_{[r,\infty)}(O - \text{Tr}[\sigma O])\right] \leq \exp \left( - \frac{r^2}{8c_1 \max \left\{ \norm{\left(\Delta_\sigma^{1/2} O\right)_{\text{re}}}_{\text{Lip}}^2 , \norm{\left(\Delta_\sigma^{1/2} O\right)_{\text{im}}}_{\text{Lip}}^2 \right\}} \right).
    \end{equation}
\end{itemize}
The Gaussian concentration inequality has its limits, however: the denominator in the exponential can grow very large unless $O$ and $\sigma$ commute, meaning this inequality can become too loose to be meaningful.
This gives the following chain of inequalities, again from \cite{CRND-17-CI}:

\centering
\begin{tikzpicture}
\node (LSI)    {MLSI($C$)};
\node (TC2) [right=3cm of LSI] {TC$_2(C_2)$};
\node (PI) [right=3cm of TC2] {PI($\lambda$)};
\node (TC1) [below=of TC2] {TC$_1(C_1)$};
\node (Gauss) [right=3cm of TC1] {Gauss($c_2$)};
\node (Exp) [right=3cm of PI] {Exp($c_2$)};
\draw[->] (LSI.east) -- (TC2.west);
\draw[->] (TC2.east) -- (PI.west);
\draw[->] (PI.east) -- (Exp.west);
\draw[->] (TC1.east) -- (Gauss.west);
\draw[->] (TC2.south) -- (TC1.north);
\end{tikzpicture}

\justifying
 As discussed above, for those QMSs for which an intertwining relation holds with constant $\kappa \geq 0$, the condition $\text{Ric}(\kappa)$ also holds. The paper~\cite{CM-16-GF} showed that, as in the classical case~\cite{erbar-2016}, positive Ricci curvature with constant $\kappa > 0$ implies that it satisfies a version of $MLSI(\kappa)$, with some additional constants $c$ and diameter $D$ of $\mathcal{D}(\mathcal{H})$ in the metric $W_{2,\mathcal{L}}$. This allows this chain of concentration inequalities to be applied.
Later, the paper~\cite{AC-20-MLSI} proved that for quantum Markov semigroups generated by embedded Glauber Lindbladians $\mathcal{L}^G$, $MLSI(\lambda)$ holds for some positive constant $\lambda$ and therefore this chain of concentration inequalities can also be applied.

Following on from the implications of positive Ricci curvature, \cite{CRND-17-HWI} also proved that in the case where $\text{Ric}(\kappa)$ holds for $\kappa \in \mathbb{R}$, we can also prove a quantum version of the HWI inequality. Specifically, they proved that if $\text{Ric}(\kappa)$ holds then $HWI(\kappa)$ also holds:
\begin{equation}
    \forall \rho \in \mathcal{D}_+ (\mathcal{H}), \qquad D(\rho \| \sigma) \leq W_{2}^\mathcal{L} (\rho,\sigma) \sqrt{I_\sigma (\rho)} - \frac{\kappa}{2} W_2^{\mathcal{L}} (\rho,\sigma)^2.
\end{equation}
Finally, they also show a direct implication from non-negative Ricci curvature to a Poincaré inequality: $\text{Ric}(\kappa)$ for some $\kappa > 0$ implies $PI(1/(2eD^2))$ where $D = \text{diam}_{\mathcal{L}}(\mathcal{H})$. Work \cite{BRANNAN2022108129} also showed that $HWI(\kappa) \implies MLSI(\kappa)$, completing this chain. The works \cite{MW-20-GE} and \cite[Parts I and II]{BRANNAN2022108129} studied many examples of these Ricci curvature bounds and their implications.

To conclude this section, we see that the development of $W_p^{\mathcal{L}}$ for $p=1$ and $p=2$ has allowed us to recover all of the important implications between concentration inequalities from the classical setting discussed in Section \ref{subsubsection:classicalapplicationCIs} for quantum Markov semigroups satisfying detailed balance. It remains to extend these implications, if possible, to broader classes of semigroups, as well as those with invariant states which are not necessarily faithful. As a precursor to this, work~\cite{DH-19-FDC}
extended these concepts discussed here to the framework of tracial approximate finite $C^*$ algebras, and recreated some of these concentration inequalities in that setting. It also remains to be seen if these methods can extend the application of Ricci curvature and $\mathcal{W}_1$ to rapid mixing of Markov chains discussed in \cite{paulin2015mixingconcentrationriccicurvature} and \cite{path_coupling} to the quantum setting.

\subsection{The infinite-dimensional case}

We will briefly discuss the results of \cite{MW-18-QMS,MW-20-GE,MW-21-DF} which extend the metric here to the infinite-dimensional case. In \cite{MW-18-QMS}, the metric of \cite{CM-16-GF} was extended to the setting of infinite-dimensional quantum systems, or more generally those which are not uniformly bounded. They recovered many of the same key properties as in the original definition: that the semigroup is gradient flow for the entropy with respect to this metric, and geodesic semi-convexity of the entropy leading to Ricci curvature bounds via an analogue of the intertwining method \cite{MW-20-GE}.

\subsection{Dynamical formulations conclusion: desirable properties}

We give in Table \ref{tab:desirable_properties_dynamical} the desirable properties detailed in Section \ref{section:quantumbg} and which are satisfied by the quantum Wasserstein distance arising from a dynamical formulation.

\begin{table}[htbp]
    \centering
    \begin{tabular}{|c|c|} \hline
        Property & \hyperref[subsection:dynamical_construction]{$W_2^{\mathcal{L}}$}  \\ \hline
        Semi-metricity & Yes \\
        Triangle inequality & Yes \\
        Link to couplings & No \\
        Link to duality & Yes \\
        Link to dynamics & Yes \\
        Continuity & Yes \\
        Recovers $\norm{\rho-\sigma}_1$ for $p=1$ & No \\
        Data processing & ? \\
        Ease of calculation & No\\
        Flexibility in definition & $p=1,2$, choice of $\mathcal{L}$ \\
        Functional inequalities & Yes \\
        Applicability & Yes  \\
        Variety of application & Limited  \\ \hline
        
    \end{tabular}
    \caption{Desirable properties satisfied by those quantum Wasserstein distances given by a dynamical formulation}
    \label{tab:desirable_properties_dynamical}
\end{table}

\section{Lipschitz formulations} 
\label{section:lipschitz}
\subsection{Spectral distance-based formulation} \label{subsection:lipschitz_spectral}

One of the original works in non-commutative optimal transport was the spectral distance defined by Connes \cite[introductory chapter 5]{connes-1994}. This defines a quantum analogue of the classical Kantorovich dual formulation of classical optimal transport as follows.

\begin{defn}
    A \textbf{spectral triple} is a triple $(M, \mathcal{H},D)$ of a non-commutative algebra $M$ acting on a Hilbert space $\mathcal{H}$, and $D$ a self-adjoint operator on $\mathcal{H}$ such that for all $a \in M$ we have that the commutator $[D,a]$ is bounded as an operator on $\mathcal{H}$ and for all $\lambda \notin \text{Sp}(D)$ that $a[D-\lambda \mathbb{I}]^{-1}$ is compact as an operator on $\mathcal{H}$. For states $\rho,\sigma$ on $M$, the \textbf{spectral distance} is
    \begin{equation}
        W_1^D(\rho,\sigma) = \sup_{a \in M} \left\{ |\rho(a) - \sigma(a)| : \| [D,a] \|_{\mathcal{H}} \leq 1 \right\}.
    \end{equation}
    where $\norm{\cdot}_{\mathcal{H}}$ is the operator norm on $\mathcal{H}$.
\end{defn}

The standard use of this definition is the case where $D$ is some differential operator, such as the Dirac operator in quantum mechanics or the sum $\text{d} + \text{d}^\dagger$ of the differential and its Hodge adjoint of  differential forms on some space \cite[introductory chapter 5]{connes-1994}. In the latter case, taking $(\mathcal{X}, g)$ to be some complete connected boundaryless Riemannian manifold and $(C_0^\infty(\mathcal{X},\Omega^{\bullet} (\mathcal{X}), \text{d} + \text{d}^\dagger)$ to be our spectral triple, where $\Omega^{\bullet}(\mathcal{X})$ is the Hilbert space of square integrable differential forms on $\mathcal{X}$, the distance $W_1^{\text{d} + \text{d}^\dagger}(\delta_x, \delta_y)$ agrees with the geodesic distance on $(\mathcal{X},g)$ \cite[Proposition 1.119]{connes-1994} and so this can be considered as a non-commutative analogue of the classical Wasserstein distance arising from the geodesic distance. For a more detailed treatment, see \cite{pretoria-thesis}.

\subsection{Hamming distance-based formulations} \label{subsection:lipschitz_giacomo}

In another celebrated direction, the work~\cite{GdP-20-W1}
introduced a first-order quantum Wasserstein distance on $n$-qudit systems generalising the classical Wasserstein distance on an $n$-dit system with the Hamming distance $H$. This was extended in~\cite{GdP-22-W1SS}
to translation-invariant states on infinite lattices, and built upon further in~\cite{GdP-23-ST}
to better reflect locality and local distinguishability. The framework of these quantities is a quantisation of the Kantorovich dual formula of first-order classical Wasserstein distances:
\begin{equation}
    \mathcal{W}_1^d(\mu,\nu) = \sup_{f: \mathcal{X} \to \mathbb{R} \text{ 1-Lipschitz}}\int_{\mathcal{X}} f(x)\mathrm{d}(\mu-\nu)(x).
\end{equation}

\subsubsection{Definitions}
The original definition is as follows. We restrict to $\mathcal{H} = \left(\mathbb{C}^d \right)^{\otimes n}$, with the qudits indexed by $1 \leq i \leq n$.
\begin{defn} \label{defn:lipschitz_giacomo_finite}
    We say that two density matrices $\rho', \sigma'$ are \textbf{neighbouring} if there exists a qudit $i$ for which $\text{Tr}_i [\rho'-\sigma'] = 0$. We can then define a norm $\lVert \cdot \rVert_{W_1^H}$ on the set of traceless Hermitian operators on $\mathcal{H}$, called the \textbf{first-order quantum Wasserstein distance on qudits}, whose unit ball is chosen to be the convex hull of the set $\{ \rho' - \sigma' : \rho', \sigma' \text{ neighbouring}$.
    Between pairs of density matrices $\rho, \sigma$, this can also be defined by \cite[Definition 7]{GdP-20-W1}
\begin{equation} \label{eq:lipschitz_w1_ci_defn}
    \lVert \rho - \sigma \rVert_{W_1^H} = \min \left\{\sum_{i=1}^n c_i : \sum_{i=1}^n c_i (\rho_i - \sigma_i) = 0, \text{Tr}_i [\rho_i - \sigma_i] = 0 \right\}.
\end{equation}
\end{defn}
 This directly mimics the classical setting: any two distributions $\mu, \nu$ on an $n$-dit space whose marginals on some $n-1$ dits agree have classical Wasserstein distance $\mathcal{W}_1^H(\mu,\nu)$ with respect to the Hamming distance $H$, know as Ornstein's $\bar{d}$-distance \cite{Ornstein-1973}. And for states $\rho = \sum_{x \in \{0,1\}^n} \mu(x)\ketbra{x}{x}$, $\sigma = \sum_{y \in \{0,1\}^n} \nu(y)\ketbra{y}{y}$ diagonal in the computational basis, we get $\norm{\rho-\sigma}_{W_1^H} = \mathcal{W}_1^H (\mu,\nu)$.

This definition also integrates seamlessly the formulation of $\mathcal{W}_1$ distances via Kantorovich duality \eqref{eq:classical-lipschitz}. Indeed, for a qudit $i$ and Hermitian operator $H$ on $\mathcal{H}$, we define its {\em dependence on site $i$} as
\begin{equation} \label{eq:lipschitz-dependence_on_site}
    \partial_i H = 2 \inf_{H_i} \norm{H - H_i \otimes I_i}_{\infty}
\end{equation}
where the infimum is over all Hermitian operators $H_i$ acting on all qudits except qudit $i$.
We then have the definition of the quantum Lipschitz constant. This constant is defined analogously for all operators, but it is defined as a norm on the set of traceless operators.

\begin{defn}
     The \textbf{Lipschitz constant} of a traceless Hermitian operator $H$ on $\mathcal{H}$ is
\begin{equation} \label{eq:lipschitz-def_lipschitz_constant}
    \norm{H}_L = \max_{1 \leq i \leq n} \partial_i H.
\end{equation}
\end{defn}

The work~\cite{GdP-22-W1SS}
extends this definition from states on a finite number of qubits, or spins, to translation-invariant states on infinite lattices. In this setting, we consider a $D$-dimensional lattice $\mathbb{Z}^D$ in which each site is equipped with a copy of $\mathbb{C}^d$. By saying a state $\rho$ is translation-invariant, we mean that for all automorphisms $\tau_y$ of $\mathbb{Z}^D$ sending each site $\mathbb{C}_x^d$ to $\mathbb{C}_{x+y}^d$, we have $\tau_x \rho = \rho$.

\begin{defn} \label{defn:lipschitz_giacomo_spin_systems} \textbf{The specific quantum $W_1$ distance} between translation-invariant states $\rho$ and $\sigma$ on $\mathbb{Z}^D$ is
\begin{equation}
    w_1(\rho,\sigma) = \lim_{a \to \infty} \frac{\norm{\rho_{\Lambda_a} - \sigma_{\Lambda_a}}_{W_1}}{|\Lambda_a|}
\end{equation}
for subsystems $\Lambda_a = [-a,a]^D \cap \mathbb{Z}^D$. \end{defn}

This is once again a distance, and relates strongly to the quantity $\frac{1}{n}\norm{\rho-\sigma}_{W_1}$ in the finite setting, as we will see from its properties later.

We also have an equivalent formulation similar to equation \eqref{eq:lipschitz_w1_ci_defn} as follows:

\begin{equation}
    w_1(\rho,\sigma) = \inf \left\{ c \geq 0: \exists \rho', \sigma' \text{ s.t. } \forall \text{ finite } \Lambda \subseteq \mathbb{Z}^d, \rho_\Lambda - \sigma_\Lambda = c \sum_{y \in \Lambda} (\tau_y \rho - \tau_y \sigma)_\Lambda \right\}.
\end{equation}
Note that for this definition we still take $\rho,\sigma$ to be translation invariant, but $\rho', \sigma'$ need not be.

The definition of the Lipschitz constant also extends to this setting. For a self-adjoint operator $H$ on the lattice, we define its {\em dependence on site $x$} as 
\begin{equation} \label{eq:lipschitz_lattice_dependence_site}
    \partial_x H = 2 \inf_{H_x \text{ on } \mathbb{Z}^D \backslash x} \norm{H - H_x}_{\infty}.
\end{equation}

\begin{defn}
For a formal sum of local operators $H = \sum_{\text{finite } \Lambda \subseteq \mathbb{Z}^D} H_\Lambda$, its \textbf{Lipschitz constant} is
\begin{equation} \label{eq:lipschitz_lattice_lipschitz}
    \norm{H}_L = \partial_0 \sum_{\text{finite } \Lambda \ni 0} H_\Lambda.
\end{equation}
\end{defn}

Some works \cite{EB-24-WP, GdP-23-ST, GdP-22-VQA} have discussed the operational interpretations of these distances. Classically, both the Hamming distance and Ornstein's $\bar{d}$-distance encode local distinguishability, in the sense that $\mu$ and $\nu$ are on average locally distinguishable if and only if $\mathcal{W}_1^H(\mu,\nu)$ is large. This means \cite{GdP-22-VQA} that we might expect $W_1^H$ to encode local distinguishability of quantum states. However, it should also be noted that on average the Hamming distance between two uniformly random bit strings of length $n$ is small ($O(1)$), and so we may expect this property to carry over to the quantum setting, noting the difference that random bit strings are on average locally distinguishable but random pure quantum states are on average not.

This interpretation was resolved in \cite{EB-24-WP} and concurrently in \cite{GdP-23-ST}, noting that the hypothesis of \cite{GdP-22-VQA}, that the norm $\norm{\rho - \sigma}_{W_1^H}$ reflects the local distinguishability of two states $\rho$ and $\sigma$, is incorrect. Indeed, it was noted that the $W_1^H$ norm between the projectors of two independent Haar-random pure states $\ket{\psi}$ and $\ket{\varphi}$ grows linearly in expectation in the number $n$ of qudits, meaning that most locally indistinguishable states are still almost maximally far apart in $W_1^H$ norm.

As a version which does reflect local distinguishability, the following `local' version of the quantum Lipschitz constant was proposed in \cite{GdP-23-ST}.

\begin{defn} \label{defn:lipschitz_giacomo_local}
    For traceless operators $H$ on $\mathcal{H} = \left( \mathbb{C}^d\right)^{\otimes n}$, the \textbf{local Lipschitz norm} is
\begin{equation} \label{eq:defn_lipschitz_loc_norm}
    \norm{H}_{\text{loc}} = 2 \min \left\{ \max_{1 \leq i \leq n} \sum_{[n] \supseteq \Lambda \ni i} c_{|\Lambda|} \norm{H_{\Lambda}}_\infty : H = \sum_{\Lambda \subseteq [n]} H_\Lambda, H_{\Lambda} \text{ acts on } \Lambda \right\}
\end{equation}
where $c_n \geq \dots \geq c_1 = 1$ are chosen constants. The \textbf{local quantum $W_1$ norm on qudits} $\norm{\cdot}_{W_1^H,\text{loc}}$ is the dual of this local Lipschitz norm. 
\end{defn}
This norm still recovers the Hamming distance for states in the computational basis, however it is suppressed for locally indistinguishable states. Indeed, for states whose marginals agree on every region of size $\leq k$, the local quantum $W_1$ norm between them is at most $\frac{n}{c_k k}$.

The main motivation of this adaptation seems to be to show that the classical shadow protocol with $\mathcal{O}(\log n)$ copies of the state gives estimations of expectation values local observables converging to their true expectation values. However, this distance also has relevance to qWGANs, which will be discussed in Section \ref{section:globalapps}.

\subsubsection{Properties}

For properties of $W_1^H$, we note the following from \cite{GdP-20-W1} and \cite{GdP-22-W1SS}:
\begin{itemize}
    \item $\norm{\cdot}_{W_1^H}$ is a true norm, so the induced distance is a true metric.
    \item In the case $n=1$, the classical Hamming distance reduces to the discrete metric, and $\norm{\cdot}_{W_1^H}$ recovers the trace distance, so this construction recovers the generalisation of the total variation distance to the trace distance.
    \item For states $\rho = \sum_{x \in \{0,\dots,d-1\}^n} \mu(x) \ketbra{x}{x}$ and $\sigma = \sum_{x \in \{0,\dots,d-1\}^n} \nu(x) \ketbra{x}{x}$ in the computational basis, we have $\norm{\rho-\sigma}_{W_1^H} = \mathcal{W}_1^H(\mu,\nu)$ and so the classical distance is recovered.
    \item This quantity is subadditive: for states $\rho, \sigma$, and subsystems $\mathcal{H} = \mathcal{H}_1 \otimes \mathcal{H}_2$ covering the first $k$ and final $n-k$ qudits respectively, we have
    \begin{equation}
        \norm{\text{Tr}_{\mathcal{H}_1} [\rho-\sigma]}_{W_1^H} + \norm{\text{Tr}_{\mathcal{H}_1} [\rho-\sigma]}_{W_1^H} \leq \norm{\rho-\sigma}_{W_1^H}
    \end{equation}
    with equality when $\rho,\sigma$ are product states over $\mathcal{H}_1 \otimes \mathcal{H}_2$.
    \item The norm is always within a factor of $n$ of the trace norm:
    \begin{equation} \label{eq:lipschitz_bound_by_trace_norm_property}
        \frac{1}{2}\norm{\rho-\sigma}_1 \leq \norm{\rho-\sigma}_{W_1^H} \leq \frac{n}{2}\norm{\rho-\sigma}_1.
    \end{equation}
    \item For channels $\Phi$ acting on at most $k$ qudits, 
    \begin{equation} \label{eq:lipschitz_k_local_channel_property}
    \norm{\Phi(\rho) - \rho}_{W_1^H} \leq 2k\frac{d^2-1}{d^2}.
    \end{equation}
    \item For the von Neumann entropy $S(\rho) = -\text{Tr}[\rho \log \rho]$ and $h_2(x) = -x \log x -(1-x) \log (1-x)$, we have
    \begin{equation}
        \frac{1}{n}|S(\rho) - S(\sigma)| 
        \leq h_2\left( \frac{\norm{\rho-\sigma}_{W_1^H}}{n}\right) + \frac{\norm{\rho-\sigma}_{W_1^H}}{n} \log (d^2-1)  .
    \end{equation}
\end{itemize}
This final inequality is an exact analogue of a classical inequality for $\mathcal{W}_1^H$, with the exception that the $\log(d^2-1)$ term becomes $\log(d-1)$ in the classical setting.

In practice, the exact value of dependence on a site can be tricky to calculate. The optimisation itself is an SDP, though the dimension is $2^n$ and so impractical for large $n$. More practically, we show here that
\begin{equation}
    \frac{d^2}{d^2-1}\max_i \norm{H - \text{Tr}_i[H] \otimes I_i}_{\infty} \leq \norm{H}_L \leq 2\max_i \norm{H - \text{Tr}_i[H] \otimes I_i}_\infty
\end{equation}
and so $\norm{H}_L$ can be expressed up to a factor of $2(d^2-1)/d^2$ in terms of $\infty$-norms. While this is still computationally the same order of complexity to approximate (polynomial in dimension), it allows us to recover operators which achieve the supremum to within a factor of $2$ and gives a deeper understanding of the behaviour and nature of these Lipschitz norms.

The upper bound is clear taking $H_i = \text{Tr}_i[H]$. The lower bound comes from the following construction of $\rho,\sigma$ such that $\text{Tr}_i[\rho-\sigma] = 0$ and $\text{Tr}[H(\rho-\sigma)] \geq \frac{d^2}{d^2-1}\norm{H - \text{Tr}_i[H] \otimes I_i}_{\infty}$. For each $i$, let $\ket{\psi}$ be an eigenvector of $H - \text{Tr}_i[H] \otimes I_i$ with eigenvalue (choosing sign w.l.o.g.) $\norm{H- \text{Tr}_i[H] \otimes I_i}_\infty$. Let $\rho = \ketbra{\psi}{\psi}$. Consider for $X$ the ``phase shift'' and $Z$ the ``clock shift'' operator, the channel with Kraus operators $\left\{ \frac{1}{d}X^jZ^k \right\}_{j,k=0}^{d-1}$. This gives the completely depolarising channel, and so 
\begin{equation}
    \sum_{j,k = 0}^{d-1} (X^jZ^k)_i \otimes I_{i^c} \ketbra{\psi}{\psi} (X^jZ^k)_i^{\dagger} \otimes I_{i^c} = d \mathbb{I}_i \otimes \text{Tr}_i \rho.
\end{equation}
Noting that $\text{Tr}\left[ (H - \text{Tr}_i[H] \otimes I)d\mathbb{I}_i \otimes \text{Tr}_i \rho\right]  = 0$, and that the term above with $j = k = 0$ contributes an amount $\norm{H - \text{Tr}_i[H] \otimes I_i}_{\infty}$ to this trace, there must be some term above contributing at most $-\frac{1}{d^2-1}\norm{H - \text{Tr}_i[H] \otimes I_i}_{\infty}$. Let that term have $j = \hat{j}$, $k = \hat{k}$, and we define $\sigma = (X^{\hat{j}}Z^{\hat{k}})_i \otimes I_{i^c} \ketbra{\psi}{\psi} (X^{\hat{j}}Z^{\hat{k}})_i^{\dagger} \otimes I_{i^c}$. This gives $\text{Tr}_i \sigma = \text{Tr}_i \rho$, and so $\text{Tr}[H(\rho-\sigma)] \geq (1 + \frac{1}{d^2-1})\norm{H - \text{Tr}_i[H] \otimes I_i}_{\infty}$.

\subsection{Resource-based Lipschitz formulations} \label{subsection:lipschitz_resource}

As an extension to the spectral distance in Section \ref{subsection:lipschitz_spectral}, work \cite{araiza2025resourcedependentcomplexityquantumchannels} defined the notion of a commutator-based Lipschitz constant of an operator for a resource set $R$.
\begin{defn} \label{defn:lipschitz_resource}
    For a finite-dimensional von Neumann algebra $M \subseteq \mathcal{B}(\mathcal{H})$, and set $R \subseteq \mathcal{B}(\mathcal{H})$, the \textbf{resource-based Lipschitz semi-norm} is
    \begin{equation}
        \norm{f}_R = \sup_{r \in R} \norm{[f,r]}_{\infty}
    \end{equation}
    where $\norm{\cdot}_{\infty}$ is as an operator on $\mathcal{H}$. This gives a \textbf{resource-based first-order quantum Wasserstein distance} via a Lipschitz formulation as follows:
    \begin{equation}
        W_1^R(\rho,\sigma) = \sup_{\norm{f}_R \leq 1} |\rho(f) - \sigma(f)|.
    \end{equation}
\end{defn}

This, in turn, allows the definition of the \textit{Lipschitz complexity of a quantum channel} $\Phi$: that is, its distance from the identity on the set of Lipschitz elements of the algebra \cite[equation (4.2)]{Ding_2025}
\begin{equation}
    C_R(\Phi) = \sum_{\norm{f}_R \leq 1} \norm{\Phi^*(f) - f}_{\infty}
\end{equation}
and extended to the \textit{complete Lipschitz complexity} $C^{cb}_R(\Phi)$ via the addition of auxiliary matrix systems, see \cite[equation (4.6)]{Ding_2025} for details. 

These are called \textit{resource-based} quantities because of their connection to complexity: \cite{araiza2025resourcedependentcomplexityquantumchannels} shows that the Lipschitz complexity of a unitary conjugation quantum channel induced by $\norm{\cdot}_R$ is closely linked to the approximate complexity of the unitary in terms of compositions of elements in $R$. The canonical example of this distance is taking $R$ to be a Pauli gate set, in which case $W_1^R$ is equivalent to the norm $\norm{\cdot}_{W_1^H}$ discussed in Section \ref{subsection:lipschitz_giacomo}.

\subsection{Applications of the Hamming distance and resource-based Lipschitz formulations} \label{subsection:lipschitz_applications}

These definitions of $W_1$, in particular the original definition for finite domains based on the Hamming distance, have been used to great effect in a broad range of contexts. 

\subsubsection{Concentration inequalities} \label{subsubsection:lipschitz_concentration_inequalities}

In the first instance, $\norm{\cdot}_{W_1^H}$ was shown to verify some quantum analogues of classical transportation-cost inequalities. Analogously to the classical case for the Hamming distance on the hypercube $\{ 0, 1,\dots, d\}$, for product states $\sigma$ this norm verifies $TC_1 \left(\frac{n}{2}\right)$:
\begin{equation}
    \norm{\rho-\sigma}_{W_1^H} \leq \sqrt{\frac{n}{2}D(\rho||\sigma)}.
\end{equation}

This then leads to the following Gaussian concentration inequality for Lipschitz observables:
\begin{equation}
    \frac{1}{d^n} \text{Tr} \exp \left[t \left( O - \text{Tr}[O] \mathbb{I}/d^n \right) \right] \leq \exp \frac{n^2 t^2 \norm{O}_L^2}{8}.
\end{equation}

The consequence of this is that the proportion of eigenvalues of $O$ lying outside an interval of size $\sqrt{\epsilon n}\norm{O}_L$ decays exponentially as $\epsilon$ increases.

The paper \cite{GPCR-21-CI} explored further sufficient conditions on $\sigma$ for this transportation-cost inequality to hold. For a state $\sigma$, we say that $C(\sigma)$ is the best constant for which $TC_1(C(\sigma))$ holds. They gave bounds for $C(\sigma)$ in the case of
\begin{itemize}
    \item quantum Markov states whose recovery maps \cite[equation (1)]{Sutter_2016} are contractions in trace distance,
    \item states with exponentially decaying correlations \cite[equation (2)]{Brand_o_2014},
    \item quantum Gibbs states of commuting Hamiltonians on hypergraphs \cite[equation (25)]{gibbs_commuting},
    \item quantum Gibbs states with inverse temperature \cite[page 244]{Dalzell_McArdle} below a critical inverse temperature  for the Hamiltonian,
    \item states satisfying local indistinguishability \cite[page 2]{Ma_2014} conditions.
\end{itemize}
They then convert these transportation-cost inequalities into Gaussian concentration inequalities for observables $O$ and their concentration with respect to $\sigma$.

In another direction, the local distance from \cite{GdP-23-ST} has been used to develop concentration inequalitites. In \cite{GdP-24-CITE}, they discuss the case where the penalty constants are either all $1$, or $c_l = 1$ for $l \leq k$ and $c_l = \infty$ otherwise (noting that these are the same for $k$-local observables).

In the first instance, they show that for a $k$-local observable $H$ and product state $\rho$:
\begin{equation}
    \langle e^H \rangle_\rho \leq \exp \left(\langle H \rangle_\rho + \frac{n}{k}\left( e^{k||H||_{\text{loc}}}-1-\frac{k ||H||_{\text{loc}}}{2} \right)||H||_{\text{loc}} \right).
\end{equation}
Similarly, for $\rho$ no longer product but with exponentially decaying correlations of length $\eta$ with respect to some arrangement of $[n]$ into a spin lattice,
\begin{equation}
    \langle e^H \rangle_\rho \leq \exp \left(\langle H \rangle_\rho + \frac{n}{k}\left( e^{k(Al^D)^2||H||_{\text{loc}}}-1-\frac{k (Al^D)^2 ||H||_{\text{loc}}}{2} \right)||H||_{\text{loc}} \right) + C e^{\langle H \rangle + n||H||_{\text{loc}} - \frac{1}{\xi}}
\end{equation}
for any $l > 0$. Constants $D$ and $A$ bound the dimensionality of the spin lattice in the sense that for any spin $v$ and radius $r$, we have $|B_r(v)| \leq Ar^D$.

They use these results to show new convergence properties of thermodynamical ensembles, namely partial results toward the eigenstate thermalisation hypothesis (ETH). The ETH is a key open problem in quantum thermodynamics which describes how systems which are far from equilibrium can evolve in time to those which appear to be in equilibrium. The \textit{strong} ETH \cite{strong_ETH} asserts that \textit{any} two eigenstates with approximately the same energy value yield approximately the same expectation on local observables, and the \textit{weak} ETH \cite{weak_ETH_1, weak_ETH_2} asserts that \textit{most} pairs of eigenstates do. Work \cite{GdP-24-CITE} shows that in the case of exponentially decaying correlations, the microcanonical and canonical ensembles are equivalent in the limit $n \to \infty$, and that any convex combination of eigenstates which make an exponentially small fraction of the total number of eigenstates in the microcanonical shell is close in this local norm to the Gibbs state. This gives an exponential improvement over the weak ETH for Gibbs states with exponentially decaying correlations, but leaves unsolved the strong version.

\subsubsection{Limitations of VQAs and simulation of open quantum systems}

In recent years, advances in technology have led to the creation of physical quantum devices which have begun to outpace those being simulated on classical computers. However, due to a lack of implementation of error correction these quantum devices are subject to noise. It is nonetheless of interest to see whether or not these devices can be of use to us, and a class of algorithms for which it has been suggested that noisy quantum devices may outperform classical devices is the class of variational quantum algorithms (VQAs). 

However, \cite{GdP-22-VQA} showed a number of limitations to the effectiveness of VQAs on noisy devices using concentration inequalities derived from the $W_1^H$ distance. More precisely:
\begin{itemize}
    \item They show that solving the MaxCut problem for certain classes of bipartite $D$-regular graphs via a quantum approximate optimisation algorithm (QAOA) requires, in some instances, circuits of depth at least $\frac{1}{2 \log(D+1)} \log \frac{n}{576}$ to outperform classical algorithms.
    \item They show that when trying to solve combinatorial optimisation problems under depolarising noise $p$ on each qubit, it is exponentially unlikely that a circuit of depth $\mathcal{O}(p^{-1})$ will outperform a classical computer.
    \item They extend the methods used to prove Poincaré-style concentration inequalities for the outputs of noisy shallow circuits, where the variance of an observable $O$ on an output state $\sigma$ is bounded as 
    \begin{equation}
        \text{Var}_{\sigma}(O) \leq Cn\norm{O}_L^2
    \end{equation}
    for some constant $C$ and a number $n$ of qubits. 
\end{itemize}
The high-level consequence of these results is that, at least for noisy intermediate-scale quantum devices of the type likely to be developed over the next decade or so, where noise makes simulation of only shallow quantum circuits feasible, variational quantum algorithms cannot outperform classical algorithms for a broad class of problems. This is because for a small number of qubits, classical algorithms are not yet burdened by fast-growing complexity, and for a larger number of qubits, QAOA is limited by noise.

In a similar vein, work \cite{Ding_2025} uses the resource-based Lipschitz constant to show lower bounds for the circuit complexity of simulating open quantum systems. For an open quantum system whose dynamics in the Schrödinger picture are modelled by the Lindbladian
\begin{equation}
    \mathcal{L}^{\dagger}\rho = -i[H,\rho] + \sum_{j} \left[V_j \rho V_j^{\dagger} - \frac{1}{2}\{V^{\dagger}V, \rho\} \right],
\end{equation}
assuming ergodicity (meaning that there is some quantum channel $E_\infty$ for which $\lim_{t \to \infty} \int_0^t \left(e^{s\mathcal{L}}\right)^\dagger \text{d}s = E_\infty$), and assuming there is some resource set $R$ compatible with the accessible gate set $\mathcal{U}$, the complexity of simulating the system is lower bounded by $\frac{C_R^{cb}(E_\infty)}{F(\epsilon, t_{\text{mix}})}$ where $\epsilon$ is the simulation accuracy, $t_{\text{mix}}$ is the mixing time of the evolution $e^{t\mathcal{L}}$, and $F$ is some function. These computational lower bounds highlight difficulties in simulation of open quantum systems by quantum computers, which is a key potential application of quantum computing \cite{feynman}. 

\subsubsection{GKP codes}

Bosonic error-correction codes are a key class of codes in quantum error correction, in which logical qubits are encoded in physical systems known as bosonic modes (equivalent to quantum harmonic oscillators) which admit infinitely many energy levels. This is opposed to a traditional encoding system using the ground state for $\ket{0}$ and the first excited state for $\ket{1}$, which is prone to noise degradation. For an overview of bosonic codes, see \cite{bosonic}. GKP (Gottesman-Kitaev-Preskill) codes \cite{og-gkp} are a key class of bosonic codes in which a single qubit is encoded in a single harmonic oscillator. The formal definition in its simplest form is that of a code whose code space is the fixed points of the abelian group generated by displacement operators
\begin{equation}
D(\xi) = \exp \left( -i \sqrt{2 \pi} \xi^T J x \right), \qquad \xi = \binom{u}{v}, \quad x = \binom{p}{q}, \quad J = \begin{pmatrix}
    0 & 1 \\ -1 & 0
\end{pmatrix}
\end{equation}
These have recently been implemented on an ultra-low-loss
integrated photonic chip \cite{boson-implementation}, in a way that shows critical features for fault-tolerant scaling.

Inspired by the $\norm{\cdot}_{W_1^H}$ distance, \cite{RK-23-GKP} proposed a tweak to the definition of the Lipschitz constant, giving the {\em bosonic Lipschitz constant}, defining instead the dependence on site $i$ as
\begin{equation}
    \partial_i H = \sup \max \{ |\braket{\psi | [P_i, H] \otimes I_R | \varphi}|, |\braket{\psi | [Q_i, H] \otimes I_R | \varphi}|\}
\end{equation}
for quadratures $P_i, Q_i$ position and momentum, and where the supremum is over all reference systems $R$ and all pure states $\ket{\psi}, \ket{\varphi}$ with finite total photon number on $\mathcal{H}$. Dualising this bosonic Lipschitz constant gives their definition of a bosonic quantum Wasserstein norm.

This distance was then used to show depolarising decay in concatenated GKP-stabiliser error correction. They show that, for one cycle being the noise channel combined with the local GKP recovery map, the 1-norm between the output under $T$ cycles of two different states decays exponentially in $T$.

\subsubsection{Computational complexity} \label{subsubsection:lipschitz_applications_complexity}

A prominent open problem \cite{Haferkamp_2022} in quantum computing is the question of whether or not a random quantum circuit of exponential depth has, on average, exponential circuit complexity. It is known \cite{Jia_2023} that random $n$-qubit unitary matrices have exponential circuit complexity on average, though \cite{Jia_2023} also shows that explicit examples are difficult to find. The work \cite{Haferkamp2022randomquantum} shows that random quantum circuits of depth linear in $n$ and polynomial in $t$ approximate the Haar measure on $n$-qubit unitary matrices up to agreement in $t^{\text{th}}$ moments. These results together strongly suggest that the complexity of an exponential-depth random quantum circuit should, on average, be exponential, though this remains an open question. The qudit structure underpinning $W_1^H$, the property of data processing under single-qubit channels, and the relation \eqref{eq:lipschitz_k_local_channel_property}
linking the locality of a channel and the change in $W_1^H$, strongly suggest that $\norm{\rho-\sigma}_{W_1^H}$ has a connection to computational complexity. And indeed, \cite{LL-22-WC}
does give a linear lower bound for the complexity of a shallow (linear depth) circuit using the $W_1^H$ norm. However, the bound \eqref{eq:lipschitz_bound_by_trace_norm_property}
shows that any bound between computational complexity and $W_1^H$ norm cannot gain a factor of more than $n$ over any bound arising from the trace distance - in other words, analysis using $W_1^H$ will only be effective for circuits of at most linear complexity.

The work \cite{araiza2025resourcedependentcomplexityquantumchannels} gives similar applications of $W_1^R$ to toric codes and random circuits, though as with $\norm{\cdot}_{W_1^H}$ the linear upper bound of $W_1^R$ in the number of qubits restricts this discussion to the complexity of shallow circuits.

\subsection{Lipschitz formulations conclusion: desirable properties}

We give in Table \ref{tab:desirable_properties_lipschitz} the desirable properties detailed in Section \ref{section:quantumbg} and which are satisfied by each of the quantum Wasserstein distances arising from a Lipschitz formulation.

\begin{table}[htbp]
    \centering
    \begin{tabular}{|c|c|c|c|c|c|} \hline
        Property & \hyperref[subsection:lipschitz_spectral]{$W_1^D$} & \hyperref[defn:lipschitz_giacomo_finite]{$\norm{\cdot}_{W_1^H}$} & \hyperref[defn:lipschitz_giacomo_spin_systems]{$w_1^H$} & \hyperref[defn:lipschitz_giacomo_local]{$\norm{\cdot}_{W_1^H,loc}$}  & \hyperref[defn:lipschitz_resource]{$W_1^R$}\\ \hline
        Semi-metricity &  Yes& Yes & Yes & Yes & Yes \\
        Triangle inequality  & Yes & Yes & Yes & Yes & Depends on $S$   \\
        Link to couplings &  No & No & No & No&  No \\
        Link to duality &  Yes & Yes & Yes & Yes &  Yes \\
        Link to dynamics & Partial &  No & No & No  & No  \\
        Continuity & Yes  & Yes & Yes & Yes & Yes  \\
        Recovers $\norm{\rho-\sigma}_1$ for $p=1$ & No & Yes & Yes & Yes & Yes \\
        Data processing &  ? & Yes & Yes & Yes &  ?  \\
        Ease of calculation & No & SDP & No & Yes  &  Depends on $S$  \\
        Flexibility in definition & Yes  & No & No & No & Yes   \\
        Functional inequalities & ? & Yes & No & No  & No  \\
        Applicability & ? & Yes & Yes & Yes  &  Yes \\
        Variety of application & ? & Yes & Yes & Yes & ?   \\ \hline
        
    \end{tabular}
    \caption{Desirable properties satisfied by those quantum Wasserstein distances given by a Lipschitz formulation}
    \label{tab:desirable_properties_lipschitz}
\end{table}
\section{Global applications: learning and GANs} 
\label{section:globalapps}

As we have seen, the generalisation of the optimal transport problem to the quantum setting comes in many different flavours, due to the impracticality of direct translation of the classical definitions. This means that, in general, each application of quantum optimal transport is associated to one specific generalisation. As these generalisations come from completely different directions, and in some cases have been `cooked up' with one specific application in mind, it is very difficult to draw links between different definitions and it is unlikely that two different distances will give rise to the same application.

The notable exception to this is in the field of machine learning, specifically in generative adversarial networks (GANs). A GAN, introduced in the classical setting in \cite{og-gans}, is a machine learning architecture in which a network is trained to generate data which is indistinguishable from the training data via a zero-sum game consisting of two parts, the {\em generator} and the {\em discriminator}.

As this section brings together quantum Wasserstein distances from many different frameworks, we will keep this section light on notation in order to avoid confusion. We will give a higher-level overview of this topic than in much of the rest of the review to highlight the definitional links between these architectures: for precise mathematical details, see the works defining these objects themselves (that is, \cite{og-gans, og_qgans, pmlr-v70-arjovsky17a, SC-19-GAN, GdP-23-ST, EB-24-WP, SB-20-QSL}).

\begin{defn}
    Let $R$ be a source generating data according to a target distribution $p_{\text{data}}$ on state space $\mathcal{X}$. Let $G_{\theta_g}$ with law $p_{\theta_g}$, parameterised by $\theta_g$, be a random variable on $X$ called the generator, and $D_{\theta_d}:\mathcal{X}\to [0,1]$, parameterised by $\theta_d$, a discriminator. A \textbf{generative adversarial network (GAN)} is an iterative process in which each round updates $\theta_g$ and $\theta_d$ according to the minmax game
    
\begin{equation}
    \min_{\theta_g} \max_{\theta_d} \mathbb{E}_{X \sim p_{\text{data}}} [ \log D_{\theta_d}(X)] + \mathbb{E}_{G_{\theta_g} \sim p_{\theta_g}} [\log (1-D_{\theta_d}(G_{\theta_g}))].
\end{equation}

\end{defn}

The aim is for $G_{\theta_g}$ to replicate samples drawn from $p_{\text{data}}$. $D_{\theta_d}(z)$ is the probability that a sample $z$ came from $p_{\text{data}}$. Optimising over $\theta$, the generator aims to minimise the probability that the discriminator will correctly identify the origin of a sample, while the discriminator aims to maximise this probability.

In the quantum setting, the framework is very similar~\cite{og_qgans}, and we are able to define quantum GANs (quGANs).

\begin{defn}
    Let $R$ be a quantum source which, on label $\ket{\lambda}$ from $\ket{1}$ to $\ket{\Lambda}$, outputs density matrix $\rho_\lambda^R$. Let  $G$ be a quantum source, called the generator, parameterised by $\theta_g$ which, on label $\ket{\lambda , z}$, outputs state $\rho_{\lambda, z}^{G_{\theta_g}}$. Let $D_{\theta_d}$ be a quantum circuit parameterised by $\theta_d$, called the discriminator. Letting $U_{\theta_d}$ be the unitary determined by the discrimination circuit and $Z_1$ its measurement without loss of generality, a \textbf{quantum generative adversarial network (quGAN)} is an iterative process in which each round updates $\theta_g$ and $\theta_d$ according to the minmax game
    \begin{equation}
    \min_{\theta_g} \max_{\theta_d} \frac{1}{2} + \frac{1}{4\Lambda} \sum_{\lambda = 1}^\Lambda \text{Tr} \left[U_{\theta_d}\left( \rho_\lambda^R - \rho_{\lambda, z}^{G_{\theta_g}}\right) U_{\theta_d}^{\dagger}Z_1 \right].
\end{equation}
\end{defn}
As before, $G$ tries to mimic $R$ and $D$ tries to distinguish between $R$ and $G$. The extra register in $z$ is optional and acts as a source of unstructured noise which can later be fine-tuned. $G$ is generally also a parameterised quantum circuit.
This discrimination assumes even priors between $R$ and $G$, though the objective function can also be tuned to assume arbitrary priors.

Standard (both quantum and classical) GANs are known to suffer from a few convergence problems, notably mode collapse in the classical case, where entire chunks of the training data are missed out, and vanishing gradients and barren plateaus in both cases, where the size of the gradient decays meaning the network cannot converge.

In the classical case, Wasserstein GANs (WGANs) were proposed to solve these issues. The traditional GAN aims for convergence in Kullback-Leibler divergence, which is a particularly strong condition and is therefore difficult to guarantee. WGANs instead aim for convergence of $p_{\theta_g}$ to $p_{\text{data}}$ in first-order Wasserstein $\mathcal{W}_1$ distance, a weaker but still practically meaningful condition~\cite{pmlr-v70-arjovsky17a}. 
The choice of $\mathcal{W}_1$ over other orders $\mathcal{W}_p$ comes from the use of Kantorovich duality.

\begin{defn}
    A \textbf{Wasserstein GAN (WGAN)} is a GAN where the discriminator $D_{\theta_d} \in \{ f_{\theta_d}\}$ is restricted to a set of functions which are 1-Lipschitz, and the minmax game becomes
\begin{equation}
    \min_{\theta_g} \max_{\theta_d} \mathbb{E}_{X \sim p_{\text{data}}} [f_{\theta_d}(X)] - \mathbb{E}_{G_{\theta_g} \sim p_{\theta_g}} \left[ f_{\theta_d} (G_{\theta_g} )\right].
\end{equation}
\end{defn}
The  work~\cite{pmlr-v70-arjovsky17a}
gives examples of effective practical implementation of this algorithm.
It is natural, then, to consider how the proposed definitions of first-order quantum Wasserstein distances could be used to eliminate the issue of barren plateaus in quGANs, in the form of a \textit{quantum Wasserstein generative adversarial network (qWGAN)}. This appears to be the only application yet studied for which quantum $\mathcal{W}$ proposals from different origins and frameworks come together for the same purpose.

\begin{itemize}
    \item 
Work~\cite{SC-19-GAN}
was, as discussed, the origin of the definition 
\eqref{eq:defn_W2_asym}
in section
\ref{subsection:coupling_asymmetric_projectors}.
For a parameterised generator $G(\theta)$ generating state $\rho$ on $\mathcal{H}_A$, and target state $\sigma$ on $\mathcal{H}_B$, the associated optimisation is
\begin{equation} \label{eq:qwgan_sc19_primal}
    \min_{G} \min_{\tau \in \mathcal{C}(\rho,\sigma)} \text{Tr}[\tau P_{\text{asym}}]
\end{equation}
leading to the following minmax game by Kantorivich duality:
\begin{align} \label{eq:qwgan_sc19_dual}
    \min_G &\max_{\phi,\psi} \text{Tr}[\sigma \psi] - \text{Tr}[\rho \phi] \\
    &\text{s.t.} \mathbb{I}_{A} \otimes \psi - \phi \otimes \mathbb{I}_B \leq P_{\text{asym}}.
\end{align}

As the constraint $\mathbb{I}_{A} \otimes \psi - \phi \otimes \mathbb{I}_B \leq P_{\text{asym}}$ is difficult to enforce, they also propose a regularised version with tunable parameter $\lambda$ by adding $\lambda I_{\tau}(A;B)$ for $I_{\tau}$ the quantum mutual information with regard to $\tau$. As before, $G$ is generally parameterised as a quantum circuit. The simplicity of $P_{\text{asym}}$ means that calculating gradients and optimising is feasible in this setting, and they give a rough estimate that one optimisation step is efficient on quantum machines. As an experimental proof-of-concept, they show their techniques can generate a 3-qubit quantum circuit which approximates a circuit to simulate a 1-D Hamiltonian. The circuit generated by their methods uses around 50 gates, whereas other methods to approximate this circuit require over $10^4$ gates.
\item 
The $W_1^H$ norm of De Palma et. al. has also been used to construct qWGANs~\cite{BTK-20-LQ}.
The advantage of this is that it retains high gradients even as the number of qubits increases, effectively tackling the issue of barren plateaus. For $G$, $\rho$, and $\sigma$ as above, the theoretical associated minmax game is
\begin{equation}
    \min_G \max_{O : \norm{O}_L \leq 1} \text{Tr}[O(\rho-\sigma)].
\end{equation}
As optimisation over the set of Lipschitz operators is difficult, not least because of its dimension, the authors instead restrict to optimising over linear combinations of Pauli operators with support on at most 2 qubits, with a restriction to those whose vector of coefficients has 1-norm at most $1/2$. They use this to effectively learn the GHZ state, which previously was inaccessible due to barren plateaus in the standard qGAN architecture.
\cite{GdP-23-ST}
gave operational improvements to this method by showing that the classical shadows of the state could be used for learning instead of the state itself, allowing for a more efficient implementation of this algorithm.

However, the restricted set of Lipschitz operators chosen does not accurately reflect the whole set of Lipschitz operators in the qualitative sense. Indeed, as discussed \cite{EB-24-WP},
showed that the $W_1^H$ norm does not reflect local distinguishability of states, but rather a more exotic transport property. The choice linear combinations of Pauli operators with small support, however, does restrict to a set reflecting the property of local distinguishability. While able to achieve quantitative convergence through eliminating barren plateaus for some states, it is not clear that this quantitative convergence reflects a qualitative convergence in the desired properties of the states.
\item The paper~\cite{EB-24-WP}
discusses solutions to this issue, by constructing qWGANs using instead $W_1^d$ for an arbitrary underlying distance $d$. The flexibility of this definition in $d$ allows for a wide range of properties to be represented qualitatively depending on the aim of the learning algorithm. Though this discussion is entirely theoretical, its eventual implementation could allow for solutions to a much broader class of learning problems and fine-tuning of the convergence parameters.
\item In the dynamical framework, $W_2^{\mathcal{L}}$ has also been employed in learning problems via GANs. The differential structure, however, makes the flavour of this application significantly different. The strategy of~\cite{SB-20-QSL}
is instead to pull back the metric $W_2^{\mathcal{L}}$ from the state space to the parameter space, and to calculate gradients directly in the parameter space. The ergodicity assumption $\text{ker}(\mathcal{L}) = \text{span} \{ \text{id} \}$ allows this pullback, and the resulting Riemannian metric is known as the {\em Wasserstein information matrix} $g_W(\theta)$. This gives update $\dot(\theta)(t) = -\nabla_g R(\theta(t))$ for an objective function $R$, allowing learning directly in the parameter space.
\end{itemize}

It is important here to acknowledge the ethical issues present in this line of research. In recent years, a huge number of ethical concerns have been raised in the field of generative modelling, including but not limited to copyright claims and violation of intellectual property rights \cite{Bondari_2025, Darroch.2017}, deepfakes of a defamatory or exploitative nature \cite{deepfakes1,mitra2024worldgenerativeaideepfakes}, generation of inappropriate and illegal content \cite{inappropriate1, inappropriate2}, and misinformation and censorship across the internet \cite{misinfo1, misinfo2}. Any continued research in this field must consider and address how to mitigate the impact of their work on these issues.

\section{Open directions} 
\label{section:perspectives}
Though substantial progress has been made in this field to quantise optimal mass transport and to recover the results and applications therein, there is still a broad range of directions and topics which we believe will merit great attention over the next few years. Chief among them are the following.

\paragraph{Linking different approaches.} One of the main open questions in this field is what links or overarching frameworks can be drawn between existing proposals for quantum Wasserstein distances. It is an artefact of the definitional frameworks that they do not, at first glance, relate to one another well. As we discussed, the inherent difficulties in quantising the classical definition of Wasserstein distances led to three separate broad approaches: coupling, dynamical, and Lipschitz. These each quantise a different definition of the Wasserstein distances, which in the classical setting are equivalent but which in the quantum setting have diverged. An important ongoing area is the harmonisation of these three frameworks to allow for more applications, for example the extension of existing concentration inequalities (such as $TC_1 \implies TC_2$) from quantum Markov semigroups in detailed balance to other quantum systems. This could also allow the extension of hypercontractivity from the formulation in \cite{EB-24-WP} to other definitions of the quantum Wasserstein distance. Hypercontractivity requires us to compare quantum Wasserstein distances of two different orders $W_{p_1}$ and $W_{p_2}$, but the Lipschitz and dynamical frameworks generalise the classical distances for only $p=1$ and $p=2$ respectively. Linking these frameworks will allow us to compare orders $p=1$ and $p=2$ in a broader class of quantum systems.

\paragraph{Linking definitions within similar approaches.}
Additionally, even within frameworks we see discord that could be resolved. For example, in the coupling framework where the optimal cost is given by $\inf_{\pi \in \mathcal{C}(\rho,\sigma)}\text{Tr}[C\pi]$ for some $C$, a general method to translate an underlying geometric structure into a cost matrix is as yet unpresented. This means that existing choices of coupling matrix $C$ give distances that do not relate to one another well. A broader method for choosing a cost matrix or operator would allow us to relate distances to one another more easily.

\paragraph{Quantitative equivalences between definitions.} While definitional connections are important, another open area is quantitative comparison between definitions. So far, we have very little in the way of equivalence between two definitions via inequalities (in the sense of $cx \leq y \leq Cx$ for constants $c$, $C$). Such inequalities, particularly for constants which are dimensionally independent or have little dependence on dimension, would allow us to pull together multiple definitions in one application and use the properties of those different definitions, allowing us to circumvent the requirement to 'sacrifice' some desirable property in each definition.

\paragraph{Expansion of definitions to new quantum structures.} In a similar vein, the expansion of existing rigid definitions to more general settings would help expand existing applications to a wider area. For example, the family of Lipschitz $W_1^H$ norms satisfies almost all desirable properties except for flexibility: the only geometries as yet represented by these distances are the Hamming metric on either a finite or infinite qudit lattice. The same is true of the $P_{\text{asym}}$ matricial distance, whose geometry is restricted to the discrete metric on a single qudit. Extension of these to more exotic geometries, such as those considering also the distance between lattice points, could allow their applications to be replicated in other areas such as the study of physically local noise in quantum computers, of quantum Gibbs states with decaying correlations, and many more.

\paragraph{Numerical computation and approximation.} One of the more practical open areas is in the numerical computation of these distances. On one side, it will be important to be able to efficiently compute or approximate these distances. Many of them, such as $\norm{\cdot}_{W_1^H}$ or any coupling-based quantity using a linear cost operator, can be written as SDPs and so calculation in low dimension is feasible. However, in higher dimensions or for other quantities this remains a difficult task. 
On the other side, implementation of the qWGANs discussed is also an important open problem. While some models have had success with toy examples, wider implementation is needed for practical use.

\paragraph{Triangle inequality.} In terms of properties, we have discussed that in many definitions, the validation or violation of the triangle inequality is still unknown. The triangle inequality is a key part in the proof of many applications of classical optimal transport, and so establishing whether or not it holds is vitally important for each proposal.

\paragraph{Applications from classical optimal transport.} There are also applications from classical mass transport that, as far as we are aware, have not yet been solved in the quantum setting. For example, \cite{path_coupling} discusses the path coupling method from classical optimal transport used to prove rapid mixing of Markov chains, which doesn't yet have a quantum counterpart. It is unclear whether any of the existing distances could be employed to solve this problem, or whether this highlights a gap in the literature on quantum generalisations of Wasserstein distances. This is one of the applications for which the triangle inequality is integral to the classical proof. However, it seems that the proof in the quantum setting would also rely heavily on putting an underlying geometry on a qudit system which represents a graph distance. None of the distances proposed seem to have both the triangle inequality and the ability to adapt to this more exotic geometry. 

\paragraph{Other hypothesised applications.} A number of the papers themselves also mention potential unexplored applications linked to their specific implementations of quantum Wasserstein distances. Work \cite{GdP-20-W1} suggests that $\norm{\cdot}_{W_1^H}$ could be used in the analysis of the robustness of quantum machine learning algorithms, to extend ideas of quantum differential privacy beyond product states, in the study of quasi-local Hamiltonians, and in tomography. The method of quadratures with partial transposes \cite{GPDT-19-QC}, taking
could be applied in the theory of quantum rate-distortion coding, and the SWAP-fidelity introduced in \cite{SF-21-MK} is likely to extend the ways of discussing proximity between quantum states.

\section{Concluding remarks} 
\label{section:conclusion}
In this work, we have presented the state of the art in the theory of quantum Wasserstein distances. We have discussed existing proposals to generalise the classical Wasserstein distances to quantities relating quantum states, categorised into three major frameworks motivating their definitions: coupling methods, dynamical methods, and dual or Lipschitz methods. We have discussed major properties associated with these distances and presented their applications. We have also discussed the pitfalls arising from each framework, and how these have been circumvented to give greater utility to the quantities defined.

With each new definition proposed, and armed with certain impossibility results on the recovery of desirable properties by direct methods, it becomes clearer and clearer that there is unlikely to be a ``correct'' definition of the quantum Wasserstein distances. Each proposal has chosen something to sacrifice in order to achieve a desired property, outcome, or application, and the theory becomes richer with each new development.

This area of research is constantly undergoing new developements and ideas, with offshoots into entropic regularisation, quantum chemistry, quantum algorithms, and so many more. There are also swathes of unanswered questions in this area, each of which may have many possible ``correct'' answers. The field of classical optimal transport is also a hotbed of research, and with each new application in the classical setting comes the potential for new tools in quantum optimal transport to improve our understanding of the behaviour of quantum states. 

We liken the state of the art to a half-finished patchwork quilt: each new definition is like a patch added to a quilt to cover a hole, and it appears that almost all of the holes are covered. However, there is currently very little thread in between the patches holding it together. Without such thread, it is difficult to say the theory is anywhere near complete.

\section*{Acknowledgements}

This project received support from the PEPR integrated project EPiQ ANR-22-PETQ-0007 part of Plan France 2030. Part of this review was written while the author was visiting the Institute for Pure and Applied Mathematics (IPAM), which is supported by the National Science Foundation (Grant No. DMS-1925919). The author would like to thank Daniel Stilck França for valuable discussions and feedback on numerous early drafts of this work, as well as Mohammad Ahmadpoor, Augusto Gerolin, and Lorenzo Portinale for helpful comments.

\section*{Conflicts of interest}

The author declares no conflicts of interest.

\printbibliography

@article{AC-20-MLSI,
  title={The modified logarithmic Sobolev inequality for quantum spin systems: classical and commuting nearest neighbour interactions},
  author={'Angela Capel and Cambyse Rouz'e and Daniel Stilck Francca},
  journal={arXiv:2009.11817},
  year={2020},
  url={https://api.semanticscholar.org/CorpusID:221879260}
}

@article{BTK-20-LQ,
   title={Learning quantum data with the quantum earth mover’s distance},
   volume={7},
   ISSN={2058-9565},
   url={http://dx.doi.org/10.1088/2058-9565/ac79c9},
   DOI={10.1088/2058-9565/ac79c9},
   number={4},
   journal={Quantum Science and Technology},
   publisher={IOP Publishing},
   author={Kiani, Bobak Toussi and De Palma, Giacomo and Marvian, Milad and Liu, Zi-Wen and Lloyd, Seth},
   year={2022},
   month={7}, pages={045002} }

@article{CM-12-2W,
      title={An Analog of the 2-Wasserstein Metric in Non-commutative Probability under which the Fermionic Fokker-Planck Equation is Gradient Flow for the Entropy}, 
      author={Eric A. Carlen and Jan Maas},
      year={2014},
      journal={Communications in Mathematical Physics},
      volume={331},
pages={887-926}
}

@article{CM-16-GF,
title = {Gradient flow and entropy inequalities for quantum Markov semigroups with detailed balance},
journal = {Journal of Functional Analysis},
volume = {273},
number = {5},
pages = {1810-1869},
year = {2017},
issn = {0022-1236},
doi = {https://doi.org/10.1016/j.jfa.2017.05.003},
url = {https://www.sciencedirect.com/science/article/pii/S0022123617301878},
author = {Eric A. Carlen and Jan Maas}
}

@article{CM-18-OT,
   title={Non-commutative Calculus, Optimal Transport and Functional Inequalities in Dissipative Quantum Systems},
   volume={178},
   ISSN={1572-9613},
   url={http://dx.doi.org/10.1007/s10955-019-02434-w},
   DOI={10.1007/s10955-019-02434-w},
   number={2},
   journal={Journal of Statistical Physics},
   publisher={Springer Science and Business Media LLC},
   author={Carlen, Eric A. and Maas, Jan},
   year={2019},
   month={11}, pages={319–378} }

@article{CRND-17-CI,
      title={Concentration of quantum states from quantum functional and transportation cost inequalities}, 
      author={Cambyse Rouzé and Nilanjana Datta},
      year={2019},
month={1},
journal={Journal of Mathematical Physics},
volume={60},
number={1}
}

@article{CRND-17-HWI,
      title={Relating relative entropy, optimal transport and Fisher information: a quantum HWI inequality}, 
      author={Cambyse Rouzé and Nilanjana Datta},
      year={2020},
month={2},
journal={Annales Hénri Poicaré},
volume={21},
pages={2115-2150}
}

@article{DH-19-FDC,
  title={Quantum optimal transport for approximately finite-dimensional C* algebras},
  author={David F. Hornshaw},
  journal={arXiv:1910.03312},
  year={2019},
  url={https://api.semanticscholar.org/CorpusID:203902379}
}

@article{DP-22-GM,
      title={Gibbs {M}anifolds}, 
      author={Dmitrii Pavlov and Bernd Sturmfels and Simon Telen},
      year={2022},
month={11},
journal={arXiv:2211.15490}
}

@article{EB-24-WP,
   title={Order p Quantum Wasserstein Distances from Couplings},
   ISSN={1424-0661},
   url={http://dx.doi.org/10.1007/s00023-025-01557-z},
   DOI={10.1007/s00023-025-01557-z},
   journal={Annales Henri Poincaré},
   publisher={Springer Science and Business Media LLC},
   author={Beatty, Emily and Stilck França, Daniel},
   year={2025},
   month=mar }

@article{FG-15-QM,
   title={On the Mean Field and Classical Limits of Quantum Mechanics},
   volume={343},
   ISSN={1432-0916},
   url={http://dx.doi.org/10.1007/s00220-015-2485-7},
   DOI={10.1007/s00220-015-2485-7},
   number={1},
   journal={Communications in Mathematical Physics},
   publisher={Springer Science and Business Media LLC},
   author={Golse, François and Mouhot, Clément and Paul, Thierry},
   year={2016},
   month={1}, pages={165–205} }

@article{FG-15-SE,
   title={The Schrödinger Equation in the Mean-Field and Semiclassical Regime},
   volume={223},
   ISSN={1432-0673},
   url={http://dx.doi.org/10.1007/s00205-016-1031-x},
   DOI={10.1007/s00205-016-1031-x},
   number={1},
   journal={Archive for Rational Mechanics and Analysis},
   publisher={Springer Science and Business Media LLC},
   author={Golse, François and Paul, Thierry},
   year={2016},
   month={8}, pages={57–94} }

@article{FG-17-WP,
   title={Wave packets and the quadratic Monge–Kantorovich distance in quantum mechanics},
   volume={356},
   ISSN={1778-3569},
   url={http://dx.doi.org/10.1016/j.crma.2017.12.007},
   DOI={10.1016/j.crma.2017.12.007},
   number={2},
   journal={Comptes Rendus. Mathématique},
   publisher={Cellule MathDoc/Centre Mersenne},
   author={Golse, François and Paul, Thierry},
   year={2018},
   month={1}, pages={177–197} }

@article{FG-19-CH,
      title={Quantum optimal transport is cheaper}, 
      author={François Golse and Emanuele Caglioti and Thierry Paul},
      year={2020},
month={5},
journal={Journal of Statistical Physics},
volume={181},
pages={149-162}
}

@article{FG-21-MK,
author = {Caglioti, Emanuele and Golse, Francois and Paul, Thierry},
year = {2021},
month = {01},
pages = {},
title = {Towards Optimal Transport for Quantum Densities},
journal={arXiv:2101.03256}
}

@inproceedings{FG-21-QS,
  title={Quantum and Semiquantum Pseudometrics and Applications},
  author={Franccois Golse and Thierry Paul},
  year={2022},
  journal = {Journal of Functional Analysis},
}

@article{GB-24-MP,
  title = {Metric property of quantum Wasserstein divergences},
  author = {Bunth, Gergely and Pitrik, J\'ozsef and Titkos, Tam\'as and Virosztek, D\'aniel},
  journal = {Phys. Rev. A},
  volume = {110},
  issue = {2},
  pages = {022211},
  numpages = {14},
  year = {2024},
  month = {8},
  publisher = {American Physical Society},
  doi = {10.1103/PhysRevA.110.022211},
  url = {https://link.aps.org/doi/10.1103/PhysRevA.110.022211}
}

@article{GdP-20-W1,
  author={De Palma, Giacomo and Marvian, Milad and Trevisan, Dario and Lloyd, Seth},
  journal={IEEE Transactions on Information Theory}, 
  title={The Quantum Wasserstein Distance of Order 1}, 
  year={2021},
  volume={67},
  number={10},
  pages={6627-6643},
  doi={10.1109/TIT.2021.3076442}}

@article{GdP-22-VQA,
  title = {Limitations of Variational Quantum Algorithms: A Quantum Optimal Transport Approach},
  author = {De Palma, Giacomo and Marvian, Milad and Rouz\'e, Cambyse and Fran\ifmmode \mbox{\c{c}}\else \c{c}\fi{}a, Daniel Stilck},
  journal = {PRX Quantum},
  volume = {4},
  issue = {1},
  pages = {010309},
  numpages = {30},
  year = {2023},
  month = {1},
  publisher = {American Physical Society},
  doi = {10.1103/PRXQuantum.4.010309},
  url = {https://link.aps.org/doi/10.1103/PRXQuantum.4.010309}
}

@article{GdP-22-W1SS,
   title={The Wasserstein Distance of Order 1 for Quantum Spin Systems on Infinite Lattices},
   volume={24},
   ISSN={1424-0661},
   url={http://dx.doi.org/10.1007/s00023-023-01340-y},
   DOI={10.1007/s00023-023-01340-y},
   number={12},
   journal={Annales Henri Poincaré},
   publisher={Springer Science and Business Media LLC},
   author={De Palma, Giacomo and Trevisan, Dario},
   year={2023},
   month={6}, pages={4237–4282} }

@article{GdP-23-ST,
    author = {De Palma, Giacomo and Klein, Tristan and Pastorello, Davide},
    title = {Classical shadows meet quantum optimal mass transport},
    journal = {Journal of Mathematical Physics},
    volume = {65},
    number = {9},
    pages = {092201},
    year = {2024},
    month = {09},
    issn = {0022-2488},
    doi = {10.1063/5.0178897},
    url = {https://doi.org/10.1063/5.0178897}
}

@article{GdP-24-CITE,
      title={Quantum concentration inequalities and equivalence of the thermodynamical ensembles: an optimal mass transport approach}, 
      author={Giacomo De Palma and Davide Pastorello},
      year={2024},
      journal={arXiv:2403.18617},
month={3}
}

@article{GPCR-21-CI,
   title={Quantum Concentration Inequalities},
   volume={23},
   ISSN={1424-0661},
   url={http://dx.doi.org/10.1007/s00023-022-01181-1},
   DOI={10.1007/s00023-022-01181-1},
   number={9},
   journal={Annales Henri Poincaré},
   publisher={Springer Science and Business Media LLC},
   author={De Palma, Giacomo and Rouzé, Cambyse},
   year={2022},
   month={4}, pages={3391–3429} }

@article{GPDT-19-QC,
   title={Quantum Optimal Transport with Quantum Channels},
   volume={22},
   ISSN={1424-0661},
   url={http://dx.doi.org/10.1007/s00023-021-01042-3},
   DOI={10.1007/s00023-021-01042-3},
   number={10},
   journal={Annales Henri Poincaré},
   publisher={Springer Science and Business Media LLC},
   author={De Palma, Giacomo and Trevisan, Dario},
   year={2021},
   month={3}, pages={3199–3234} }

@article{GPG-22-QWI,
title = {Quantum Wasserstein isometries on the qubit state space},
journal = {Journal of Mathematical Analysis and Applications},
volume = {522},
number = {2},
pages = {126955},
year = {2023},
issn = {0022-247X},
doi = {https://doi.org/10.1016/j.jmaa.2022.126955},
url = {https://www.sciencedirect.com/science/article/pii/S0022247X22009696},
author = {György Pál Gehér and József Pitrik and Tamás Titkos and Dániel Virosztek}
}

@article{GT-22-SS,
   title={Quantum Wasserstein distance based on an optimization over separable states},
   volume={7},
   ISSN={2521-327X},
   url={http://dx.doi.org/10.22331/q-2023-10-16-1143},
   DOI={10.22331/q-2023-10-16-1143},
   journal={Quantum},
   publisher={Verein zur Forderung des Open Access Publizierens in den Quantenwissenschaften},
   author={Tóth, Géza and Pitrik, József},
   year={2023},
   month={10}, pages={1143} }

@article{KZ-22-MN,
title={Monotonicity of a quantum 2-Wasserstein distance},
   volume={56},
   ISSN={1751-8121},
   url={http://dx.doi.org/10.1088/1751-8121/acb9c8},
   DOI={10.1088/1751-8121/acb9c8},
   number={9},
   journal={Journal of Physics A: Mathematical and Theoretical},
   publisher={IOP Publishing},
   author={Bistroń, R and Eckstein, M and Życzkowski, K},
   year={2023},
   month={2}, pages={095301} }

@article{KZ-21-QOT,
   title={On Quantum Optimal Transport},
   volume={26},
   ISSN={1572-9656},
   url={http://dx.doi.org/10.1007/s11040-023-09456-7},
   DOI={10.1007/s11040-023-09456-7},
   number={2},
   journal={Mathematical Physics, Analysis and Geometry},
   publisher={Springer Science and Business Media LLC},
   author={Cole, Sam and Eckstein, Michał and Friedland, Shmuel and Życzkowski, Karol},
   year={2023},
   month={6} }

@article{LL-22-WC,
  title={Wasserstein Complexity of Quantum Circuits},
  author={Lu Li and Kaifeng Bu and Dax Enshan Koh and Arthur Jaffe and Seth Lloyd},
  journal={arXiv:2208.06306},
  year={2022},
month={6}
}

@article{MH-22-MN,
      title={On the monotonicity of a quantum optimal transport cost}, 
      author={Alexander Müller-Hermes},
      year={2022},
month={11},
journal={arXiv:2211.11713}
}

@article{MI-24-ES,
   title={Enhanced Stability in Quantum Optimal Transport Pseudometrics: From Hartree to Vlasov–Poisson},
   volume={191},
   ISSN={1572-9613},
   url={http://dx.doi.org/10.1007/s10955-024-03367-9},
   DOI={10.1007/s10955-024-03367-9},
   number={12},
   journal={Journal of Statistical Physics},
   publisher={Springer Science and Business Media LLC},
   author={Iacobelli, Mikaela and Lafleche, Laurent},
   year={2024},
   month={11} }

@article{NY-18-QE,
   title={Quantum earth mover’s distance, a no-go quantum Kantorovich–Rubinstein theorem, and quantum marginal problem},
   volume={63},
   ISSN={1089-7658},
   url={http://dx.doi.org/10.1063/5.0068344},
   DOI={10.1063/5.0068344},
   number={10},
   journal={Journal of Mathematical Physics},
   publisher={AIP Publishing},
   author={Zhou, Li and Yu, Nengkun and Ying, Shenggang and Ying, Mingsheng},
   year={2022},
   month={10} }

@article{RD-18-TP,
      title={Quadratic Wasserstein metrics for von Neumann algebras via transport plans}, 
      author={Rocco Duvenhage},
      year={2020},
month={12},
      journal={arXiv:2012.03564}
}

@article{RD-21-WD,
title = {Wasserstein distance between noncommutative dynamical systems},
journal = {Journal of Mathematical Analysis and Applications},
volume = {527},
number = {1, Part 2},
pages = {127353},
year = {2023},
issn = {0022-247X},
doi = {https://doi.org/10.1016/j.jmaa.2023.127353},
url = {https://www.sciencedirect.com/science/article/pii/S0022247X23003566},
author = {Rocco Duvenhage}
}

@article{RD-22-DB,
      title={Extending quantum detailed balance through optimal transport}, 
      author={Rocco Duvenhage and Samuel Skosana and Machiel Snyman},
      year={2022},
month={6},
      journal={arXiv:2206.15287}
}

@article{RD-23-QC,
author = {Duvenhage, Rocco and Mapaya, Mathumo},
year = {2023},
month = {04},
pages = {},
title = {Quantum Wasserstein distance of order 1 between channels},
volume = {26},
journal = {Infinite Dimensional Analysis, Quantum Probability and Related Topics},
doi = {10.1142/S0219025723500066}
}

@article{RK-23-GKP,
      title={Limitations of local update recovery in stabilizer-GKP codes: a quantum optimal transport approach}, 
      author={Robert König and Cambyse Rouzé},
      year={2023},
month={9},
      journal={arXiv:2309.16241}
}

@article{SB-20-QSL,
   title={Quantum Statistical Learning via Quantum Wasserstein Natural Gradient},
   volume={182},
   ISSN={1572-9613},
   url={http://dx.doi.org/10.1007/s10955-020-02682-1},
   DOI={10.1007/s10955-020-02682-1},
   number={1},
   journal={Journal of Statistical Physics},
   publisher={Springer Science and Business Media LLC},
   author={Becker, Simon and Li, Wuchen},
   year={2021},
   month={1} }

@inbook{SC-19-GAN,
author = {Chakrabarti, Shouvanik and Huang, Yiming and Li, Tongyang and Feizi, Soheil and Wu, Xiaodi},
title = {Quantum wasserstein GANs},
year = {2019},
publisher = {Curran Associates Inc.},
address = {Red Hook, NY, USA},
booktitle = {Proceedings of the 33rd International Conference on Neural Information Processing Systems},
articleno = {609},
numpages = {12}
}

@article{SF-21-MK,
   title={Quantum Monge-Kantorovich Problem and Transport Distance between Density Matrices},
   volume={129},
   ISSN={1079-7114},
   url={http://dx.doi.org/10.1103/PhysRevLett.129.110402},
   DOI={10.1103/physrevlett.129.110402},
   number={11},
   journal={Physical Review Letters},
   publisher={American Physical Society (APS)},
   author={Friedland, Shmuel and Eckstein, Michał and Cole, Sam and Życzkowski, Karol},
   year={2022},
   month={9} }

@article{XQ-23-QW,
  title = {Quantum Wasserstein distance between unitary operations},
  author = {Qiu, Xinyu and Chen, Lin and Zhao, Li-Jun},
  journal = {Phys. Rev. A},
  volume = {110},
  issue = {1},
  pages = {012412},
  numpages = {12},
  year = {2024},
  month = {7},
  publisher = {American Physical Society},
  doi = {10.1103/PhysRevA.110.012412},
  url = {https://link.aps.org/doi/10.1103/PhysRevA.110.012412}
}

@article{XS-23-CQ,
  title = {Coherence quantifier based on the quantum optimal transport cost},
  author = {Shi, Xian},
  journal = {Phys. Rev. A},
  volume = {109},
  issue = {5},
  pages = {052443},
  numpages = {7},
  year = {2024},
  month = {5},
  publisher = {American Physical Society},
  doi = {10.1103/PhysRevA.109.052443},
  url = {https://link.aps.org/doi/10.1103/PhysRevA.109.052443}
}

@article{LL-23-WT,
      title={Quantum Optimal Transport and Weak Topologies}, 
      author={Laurent Lafleche},
      year={2023},
      journal={arXiv:2306.12944}
}

@article{MW-18-QMS,
      title={A Noncommutative Transport Metric and Symmetric Quantum Markov Semigroups as Gradient Flows of the Entropy}, 
      author={Melchior Wirth},
      year={2018},
    month = {8},
      journal={arXiv:1808.05419}
}

@article{MW-20-GE,
   title={Complete Gradient Estimates of Quantum Markov Semigroups},
   volume={387},
   ISSN={1432-0916},
   url={http://dx.doi.org/10.1007/s00220-021-04199-4},
   DOI={10.1007/s00220-021-04199-4},
   number={2},
   journal={Communications in Mathematical Physics},
   publisher={Springer Science and Business Media LLC},
   author={Wirth, Melchior and Zhang, Haonan},
   year={2021},
   month=aug, pages={761–791} }

@article{MW-21-DF,
   title={A Dual Formula for the Noncommutative Transport Distance},
   volume={187},
   ISSN={1572-9613},
   url={http://dx.doi.org/10.1007/s10955-022-02911-9},
   DOI={10.1007/s10955-022-02911-9},
   number={2},
   journal={Journal of Statistical Physics},
   publisher={Springer Science and Business Media LLC},
   author={Wirth, Melchior},
   year={2022},
   month=apr }

@article{BV-01-FP,
author = {Biane, Philippe and Voiculescu, Dan},
year = {2001},
month = {12},
pages = {1125-1138},
title = {A Free Probability Analogue of the Wasserstein Metric on the Trace-State Space},
volume = {11},
journal = {Geometric And Functional Analysis},
doi = {10.1007/s00039-001-8226-4}
}

@book{serafini,
title={ Quantum Continuous Variables: A Primer of Theoretical Methods},
author={Alessio Serafini},
year={2023},
month={8},
isbn={978-1032157238},
publisher={CRC Press}
}

@book{funcanaquantinfo,
title = {The Functional Analysis of Quantum Information Theory},
author={Gilles Pisier and K. R. Parthasarathy and Vern Paulsen and Andreas Winter and Ved Prakash Gupta and Prabha Mandayam and V. S. Sunder},
publisher = {Springer Cham},
isbn={978-3-319-16717-6},
year={2015},
doi={https://doi.org/10.1007/978-3-319-16718-3}
}

@article{weak_ETH_2,
   title={Eigenstate Thermalization from the Clustering Property of Correlation},
   volume={124},
   ISSN={1079-7114},
   url={http://dx.doi.org/10.1103/PhysRevLett.124.200604},
   DOI={10.1103/physrevlett.124.200604},
   number={20},
   journal={Physical Review Letters},
   publisher={American Physical Society (APS)},
   author={Kuwahara, Tomotaka and Saito, Keiji},
   year={2020},
   month=may }

@article{weak_ETH_1,
   title={Effect of Rare Fluctuations on the Thermalization of Isolated Quantum Systems},
   volume={105},
   ISSN={1079-7114},
   url={http://dx.doi.org/10.1103/PhysRevLett.105.250401},
   DOI={10.1103/physrevlett.105.250401},
   number={25},
   journal={Physical Review Letters},
   publisher={American Physical Society (APS)},
   author={Biroli, Giulio and Kollath, Corinna and Läuchli, Andreas M.},
   year={2010},
   month=dec }

@article{strong_ETH,
  title = {Quantum statistical mechanics in a closed system},
  author = {Deutsch, J. M.},
  journal = {Phys. Rev. A},
  volume = {43},
  issue = {4},
  pages = {2046--2049},
  numpages = {0},
  year = {1991},
  month = {2},
  publisher = {American Physical Society},
  doi = {10.1103/PhysRevA.43.2046},
  url = {https://link.aps.org/doi/10.1103/PhysRevA.43.2046}
}

@article{nonot_schrodinger_convergence_4,
author = {Bardos, Claude and Golse, Francois and Mauser, Norbert},
year = {2000},
month = {01},
pages = {},
title = {Weak coupling limit of the N-particle Schrödinger equation},
volume = {7},
journal = {Methods and Applications of Analysis},
doi = {10.4310/MAA.2000.v7.n2.a2}
}

@article{nonot_schrodinger_convergence_3,
   title={On the rate of convergence for the mean field approximation of Bosonic many-body quantum dynamics},
   volume={14},
   ISSN={1945-0796},
   url={http://dx.doi.org/10.4310/CMS.2016.v14.n5.a9},
   DOI={10.4310/cms.2016.v14.n5.a9},
   number={5},
   journal={Communications in Mathematical Sciences},
   publisher={International Press of Boston},
   author={Ammari, Zied and Falconi, Marco and Pawilowski, Boris},
   year={2016},
   pages={1417–1442} }

@article{nonot_schrodinger_convergence_2,
  title = {Kinetic equations from Hamiltonian dynamics: Markovian limits},
  author = {Spohn, Herbert},
  journal = {Rev. Mod. Phys.},
  volume = {52},
  issue = {3},
  pages = {569--615},
  numpages = {0},
  year = {1980},
  month = {7},
  publisher = {American Physical Society},
  doi = {10.1103/RevModPhys.52.569},
  url = {https://link.aps.org/doi/10.1103/RevModPhys.52.569}
}

@misc{nonot_schrodinger_convergence_1,
      title={Quantum Fluctuations and Rate of Convergence towards Mean Field Dynamics}, 
      author={Igor Rodnianski and Benjamin Schlein},
      year={2007},
      eprint={0711.3087},
      archivePrefix={arXiv},
      primaryClass={math-ph},
      url={https://arxiv.org/abs/0711.3087}, 
}

@article{og-gkp,
   title={Encoding a qubit in an oscillator},
   volume={64},
   ISSN={1094-1622},
   url={http://dx.doi.org/10.1103/PhysRevA.64.012310},
   DOI={10.1103/physreva.64.012310},
   number={1},
   journal={Physical Review A},
   publisher={American Physical Society (APS)},
   author={Gottesman, Daniel and Kitaev, Alexei and Preskill, John},
   year={2001},
   month=jun }

@article{boson-implementation,
title={Integrated photonic source of Gottesman–Kitaev–Preskill qubits},
author={M.V. Larsen and J.E. Bourassa and S. Kocsis},
year={2025},
journal={Nature},
}

@article{bosonic,
title={Bosonic coding: introduction and use cases},
author = {Victor V. Albert},
journal ={arXiv:2211.05714},
year={2022},
month={11}
}

@article{Ornstein-1973,
title={An Application of Ergodic Theory to Probability Theory},
author = {David S. Ornstein},
year = {1973},
month = {2},
journal = {The Annals of Probability},
volume = {1},
number = {1},
pages = {43--58}
}

@article{Hu_2018,
   title={Quantum coherence and geometric quantum discord},
   volume={762–764},
   ISSN={0370-1573},
   url={http://dx.doi.org/10.1016/j.physrep.2018.07.004},
   DOI={10.1016/j.physrep.2018.07.004},
   journal={Physics Reports},
   publisher={Elsevier BV},
   author={Hu, Ming-Liang and Hu, Xueyuan and Wang, Jieci and Peng, Yi and Zhang, Yu-Ran and Fan, Heng},
   year={2018},
   month={11}, pages={1–100} }

@article{coherence_use,
  title = {Resource Theory of Quantum States Out of Thermal Equilibrium},
  author = {Brand\~ao, Fernando G. S. L. and Horodecki, Micha\l{} and Oppenheim, Jonathan and Renes, Joseph M. and Spekkens, Robert W.},
  journal = {Phys. Rev. Lett.},
  volume = {111},
  issue = {25},
  pages = {250404},
  numpages = {5},
  year = {2013},
  month = {12},
  publisher = {American Physical Society},
  doi = {10.1103/PhysRevLett.111.250404},
  url = {https://link.aps.org/doi/10.1103/PhysRevLett.111.250404}
}

@article{deutsch-josza,
title = {Rapid solution of problems by quantum computation},
author = {David Deutsch and Richard Josza},
journal = {Proceedings of the Royal Society A},
year = {1992},
month = {12}
}

@article{Haapasalo_2021,
   title={Quantum marginal problem and incompatibility},
   volume={5},
   ISSN={2521-327X},
   url={http://dx.doi.org/10.22331/q-2021-06-15-476},
   DOI={10.22331/q-2021-06-15-476},
   journal={Quantum},
   publisher={Verein zur Forderung des Open Access Publizierens in den Quantenwissenschaften},
   author={Haapasalo, Erkka and Kraft, Tristan and Miklin, Nikolai and Uola, Roope},
   year={2021},
   month=jun, pages={476} }

@article{feynman,
title={Simulating physics with computers},
author={Feynman, R.P},
journal={Int J Theor Phys},
volume={21},
pages={467–-488},
year={1982},
url={https://doi.org/10.1007/BF02650179}}

@article{Ding_2025,
   title={Lower Bound for Simulation Cost of Open Quantum Systems: Lipschitz Continuity Approach},
   volume={406},
   ISSN={1432-0916},
   url={http://dx.doi.org/10.1007/s00220-025-05240-6},
   DOI={10.1007/s00220-025-05240-6},
   number={3},
   journal={Communications in Mathematical Physics},
   publisher={Springer Science and Business Media LLC},
   author={Ding, Zhiyan and Junge, Marius and Schleich, Philipp and Wu, Peixue},
   year={2025},
   month=feb }

@article{Haferkamp2022randomquantum,
  doi = {10.22331/q-2022-09-08-795},
  url = {https://doi.org/10.22331/q-2022-09-08-795},
  title = {Random quantum circuits are approximate unitary {$t$}-designs in depth {$O\left(nt^{5+o(1)}\right)$}},
  author = {Haferkamp, Jonas},
  journal = {{Quantum}},
  issn = {2521-327X},
  publisher = {{Verein zur F{\"{o}}rderung des Open Access Publizierens in den Quantenwissenschaften}},
  volume = {6},
  pages = {795},
  month = sep,
  year = {2022}
}

@article{Jia_2023,
   title={Hay from the Haystack: Explicit Examples of Exponential Quantum Circuit Complexity},
   volume={402},
   ISSN={1432-0916},
   url={http://dx.doi.org/10.1007/s00220-023-04720-x},
   DOI={10.1007/s00220-023-04720-x},
   number={1},
   journal={Communications in Mathematical Physics},
   publisher={Springer Science and Business Media LLC},
   author={Jia, Yifan and Wolf, Michael M.},
   year={2023},
   month=may, pages={141–156} }

@article{Haferkamp_2022,
   title={Linear growth of quantum circuit complexity},
   volume={18},
   ISSN={1745-2481},
   url={http://dx.doi.org/10.1038/s41567-022-01539-6},
   DOI={10.1038/s41567-022-01539-6},
   number={5},
   journal={Nature Physics},
   publisher={Springer Science and Business Media LLC},
   author={Haferkamp, Jonas and Faist, Philippe and Kothakonda, Naga B. T. and Eisert, Jens and Yunger Halpern, Nicole},
   year={2022},
   month=mar, pages={528–532} }

@article{Ma_2014,
   title={Non-commutativity and Local Indistinguishability of Quantum States},
   volume={4},
   ISSN={2045-2322},
   url={http://dx.doi.org/10.1038/srep06336},
   DOI={10.1038/srep06336},
   number={1},
   journal={Scientific Reports},
   publisher={Springer Science and Business Media LLC},
   author={Ma, Teng and Zhao, Ming-Jing and Wang, Yao-Kun and Fei, Shao-Ming},
   year={2014},
   month=sep }

@inbook{Dalzell_McArdle, place={Cambridge}, title={Gibbs sampling}, booktitle={Quantum Algorithms: A Survey of Applications and End-to-end Complexities}, publisher={Cambridge University Press}, author={Dalzell, Alexander M. and McArdle, Sam and Berta, Mario and Bienias, Przemyslaw and Chen, Chi-Fang and Gilyén, András and Hann, Connor T. and Kastoryano, Michael J. and Khabiboulline, Emil T. and Kubica, Aleksander and et al.}, year={2025}, pages={243–249}}

@article{gibbs_commuting,
    author = {Kastoryano, M.J. and Brandão, F.G.S.L},
    title = {Quantum Gibbs Samplers: The Commuting Case },
    journal = {Commun. Math. Phys.},
volume = {344},
pages = {915--957},
    year = {2016}
}

@article{Sutter_2016,
   title={Universal recovery map for approximate Markov chains},
   volume={472},
   ISSN={1471-2946},
   url={http://dx.doi.org/10.1098/rspa.2015.0623},
   DOI={10.1098/rspa.2015.0623},
   number={2186},
   journal={Proceedings of the Royal Society A: Mathematical, Physical and Engineering Sciences},
   publisher={The Royal Society},
   author={Sutter, David and Fawzi, Omar and Renner, Renato},
   year={2016},
   month=feb, pages={20150623} }

@article{Brand_o_2014,
   title={Exponential Decay of Correlations Implies Area Law},
   volume={333},
   ISSN={1432-0916},
   url={http://dx.doi.org/10.1007/s00220-014-2213-8},
   DOI={10.1007/s00220-014-2213-8},
   number={2},
   journal={Communications in Mathematical Physics},
   publisher={Springer Science and Business Media LLC},
   author={Brandão, Fernando G. S. L. and Horodecki, Michał},
   year={2014},
   month=nov, pages={761–798} }

@Inbook{Gartner2012,
author={Gärtner, Bernd
and Matoušek, Jiří},
title={Semidefinite Programming},
bookTitle={Approximation Algorithms and Semidefinite Programming},
year={2012},
publisher={Springer Berlin Heidelberg},
pages={15--25},
isbn={978-3-642-22015-9},
doi={10.1007/978-3-642-22015-9_2},
url={https://doi.org/10.1007/978-3-642-22015-9_2}
}

@InProceedings{caro-learning,
  title = 	 {Information-theoretic generalization bounds for learning from quantum data},
  author =       {Caro, Matthias C. and Gur, Tom and Rouz{\'e}, Cambyse and Stilck Fran\c{c}a, Daniel and Subramanian, Sathyawageeswar},
  booktitle = 	 {Proceedings of Thirty Seventh Conference on Learning Theory},
  pages = 	 {775--839},
  year = 	 {2024},
  editor = 	 {Agrawal, Shipra and Roth, Aaron},
  volume = 	 {247},
  series = 	 {Proceedings of Machine Learning Research},
  month = 	 {7},
  publisher =    {PMLR},
  url = 	 {https://proceedings.mlr.press/v247/caro24a.html}
}

@article{Rouz__2024,
   title={Learning quantum many-body systems from a few copies},
   volume={8},
   ISSN={2521-327X},
   url={http://dx.doi.org/10.22331/q-2024-04-30-1319},
   DOI={10.22331/q-2024-04-30-1319},
   journal={Quantum},
   publisher={Verein zur Forderung des Open Access Publizierens in den Quantenwissenschaften},
   author={Rouzé, Cambyse and Stilck França, Daniel},
   year={2024},
   month=apr, pages={1319} }

@article{ALICKI1976249,
title = {On the detailed balance condition for non-hamiltonian systems},
journal = {Reports on Mathematical Physics},
volume = {10},
number = {2},
pages = {249-258},
year = {1976},
issn = {0034-4877},
doi = {https://doi.org/10.1016/0034-4877(76)90046-X},
url = {https://www.sciencedirect.com/science/article/pii/003448777690046X},
author = {Robert Alicki}
}

@misc{lucys_blog, title={Sinkhorn Algorithm}, journal={Lucy's Blog}, publisher={https://lucyliu-ucsb.github.io/posts/Sinkhorn-algorithm/}, author={Lucy Liu}, year={2020}, month={4}}

@article{COTFNT,
year = {2019},
volume = {11},
journal = {Foundations and Trends in Machine Learning},
title = {Computational Optimal Transport},
number = {5-6},
pages = {355--607},
author = {Gabriel Peyré and Marco Cuturi}
}

@misc{landsman1998lecturenotescalgebrashilbert,
      title={Lecture notes on C*-algebras, Hilbert C*-modules, and quantum mechanics}, 
      author={N. P. Landsman},
      year={1998},
    journal={arXiv:math-ph/9807030}
}

@phdthesis{pretoria-thesis,
    author = {Werndly Jakobus van Staden},
    title = {Metric aspects of noncommutative geometry},
    school = {University of Pretoria},
    year = {2019}
}

@misc{araiza2025resourcedependentcomplexityquantumchannels,
      title={Resource-Dependent Complexity of Quantum Channels}, 
      author={Roy Araiza and Yidong Chen and Marius Junge and Peixue Wu},
      year={2025},
      eprint={2303.11304},
      archivePrefix={arXiv},
      primaryClass={quant-ph},
      url={https://arxiv.org/abs/2303.11304}, 
}

@book{connes-1994,
title={Noncommutative geometry},
author={Alain Connes},
year= {1994},
publisher={Academic Press},
ISBN={9780121858605}
}

@book{vnas,
title={Operator Algebras: Theory of C*-Algebras and von Neumann Algebras},
author={Bruce Blackadar},
publisher={Springer},
ISBN={9783540284864},
year={2006}
}

@article{inappropriate2,
    author = {Natasha Singer},
    title = {Teen Girls Confront an Epidemic of Deepfake Nudes in Schools},
    journal = {New York Times},
    year = {2024},
    month = {4}
}

@article{inappropriate1,
author={Ben Finley},
title={Disgruntled high school athletic director uses AI to clone principal’s voice in racist, antisemitic deep fake},
journal = {Fortune},
year = {2024},
month = {4}
}

@article{mitra2024worldgenerativeaideepfakes,
      title={The World of Generative AI: Deepfakes and Large Language Models}, 
      author={Alakananda Mitra and Saraju P. Mohanty and Elias Kougianos},
      year={2024},
month={2},
    journal={arXiv:2402.04373}
}

@article{deepfakes1,
    author = {Derek Leben},
    title = {Deepfakes and the Ethics of Generative AI},
    journal = {Tepper School of Business, Carnegie Mellon University},
    year = {2025},
month = {8}
}

@report{misinfo2,
author = {Allie Funk and Adrian Shahbaz and Kian Vesteinsson},
title = {The Repressive Power of Artificial Intelligence},
institution = {Freedom House},
year = {2023},
}

@article{misinfo1,
author = {Jaidka, Kokil and Chen, Tsuhan and Chesterman, Simon and Hsu, Wynne and Kan, Min-Yen and Kankanhalli, Mohan and Lee, Mong Li and Seres, Gyula and Sim, Terence and Taeihagh, Araz and Tung, Anthony and Xiao, Xiaokui and Yue, Audrey},
title = {Misinformation, Disinformation, and Generative AI: Implications for Perception and Policy},
year = {2025},
issue_date = {March 2025},
publisher = {Association for Computing Machinery},
address = {New York, NY, USA},
volume = {6},
number = {1},
url = {https://doi.org/10.1145/3689372},
doi = {10.1145/3689372},
journal = {Digit. Gov.: Res. Pract.},
month = feb,
articleno = {11},
numpages = {15}
}

@Online{Darroch.2017,
 author = {Gil Appel and Juliana Neelbauer and David A. Schweidel},
 year = {2025},
 title = {Generative AI Has an Intellectual Property Problem},
 journal = {Harvard Business Review},
 url = {https://hbr.org/2023/04/generative-ai-has-an-intellectual-property-problem},
 urldate = {2025-05-22}
}

@misc{Bondari_2025, title={AI, Copyright, and the Law: The Ongoing Battle Over Intellectual Property Rights}, journal={IP & Technology Law Society Blog}, publisher={Gould School of Law, University of Southern California}, author={Bondari, Negar}, year={2025}, month={2}}

@book{budapest-school,
title={Optimal Transport on Quantum Structures},
authors = {Jan Maas and Simone Rademacher and Tamás Titkos and Dániel Virosztek},
publisher = {Springer Cham},
year = {2024}
}

@article{dario-invitation,
    author = {Dario Trevisan},
    title = {Quantum optimal transport: an invitation},
    journal = {Boll Unione Mat Ital},
volume = {18},
pages={34-360},
    year = {2025}
}

@book{Ollivier_Pajot_Villani_2014, place={Cambridge}, series={London Mathematical Society Lecture Note Series}, title={Optimal Transport: Theory and Applications}, publisher={Cambridge University Press}, year={2014}, collection={London Mathematical Society Lecture Note Series}}

@article{Benoist_2022,
   title={Deviation bounds and concentration inequalities for quantum noises},
   volume={6},
   ISSN={2521-327X},
   url={http://dx.doi.org/10.22331/q-2022-08-04-772},
   DOI={10.22331/q-2022-08-04-772},
   journal={Quantum},
   publisher={Verein zur Forderung des Open Access Publizierens in den Quantenwissenschaften},
   author={Benoist, Tristan and Hänggli, Lisa and Rouzé, Cambyse},
   year={2022},
   month=aug, pages={772} }

@book{ci_concmeasure_book,
title={The Concentration of Measure Phenomenon},
author={Michael Ledoux},
year={2001},
publisher={American Mathematical Society}

}

@book{raginsky2015concentrationmeasureinequalitiesinformation,
      title={Concentration of Measure Inequalities in Information Theory, Communications and Coding (Second Edition)}, 
      author={Maxim Raginsky and Igal Sason},
      year={2015}, 
}

@book{ci_geo,
author = {Guédon, Olivier and Nayar, Piotr and Tkocz, Tomasz},
year = {2014},
month = {01},
pages = {},
title = {Concentration inequalities and geometry of convex bodies}
}

@article{ci_states,
   title={Concentration Inequalities for Statistical Inference},
   volume={37},
   ISSN={1674-5647},
   url={http://dx.doi.org/10.4208/cmr.2020-0041},
   DOI={10.4208/cmr.2020-0041},
   number={1},
   journal={Communications in Mathematical Research},
   publisher={Global Science Press},
   author={Huiming Zhang, Huiming Zhang and Songxi Chen, Songxi Chen},
   year={2021},
   month=jan, pages={1–85} }

@InCollection{wasserstein-encyc-maths,
author       =  {European Mathematical Society},
title        = {Wasserstein metric},
booktitle    =  {Encyclopedia of Mathematics},
howpublished =  {\url{http://encyclopediaofmath.org/index.php?title=Cauchy_inequality\&oldid=28864}}
}

@article{paulin2015mixingconcentrationriccicurvature,
      title={Mixing and Concentration by Ricci Curvature}, 
      author={Daniel Paulin},
      year={2014},
    month={4},
      journal={arXiv:1404.2802},
}

@article{mcmc_numerint,
    author = {Press, William H. and Farrar, Glennys R.},
    title = {Recursive Stratified Sampling for Multidimensional Monte Carlo Integration},
    journal = {Computer in Physics},
    volume = {4},
    number = {2},
    pages = {190-195},
    year = {1990},
    month = {03},
    issn = {0894-1866},
    doi = {10.1063/1.4822899},
    url = {https://doi.org/10.1063/1.4822899},
    eprint = {https://pubs.aip.org/aip/cip/article-pdf/4/2/190/12118327/190\_1\_online.pdf},
}

@ARTICLE{mcmc_sysbio,
       author = {{Gupta}, Ankur and {Rawlings}, James B.},
        title = "{Comparison of parameter estimation methods in stochastic chemical kinetic models: Examples in systems biology}",
      journal = {AIChE Journal},
         year = 2014,
        month = apr,
       volume = {60},
       number = {4},
        pages = {1253-1268},
          doi = {10.1002/aic.14409},
       adsurl = {https://ui.adsabs.harvard.edu/abs/2014AIChE..60.1253G}
}

@article{mcmc_mathphys,
   title={Retrieving fields from proton radiography without source profiles},
   volume={100},
   ISSN={2470-0053},
   url={http://dx.doi.org/10.1103/PhysRevE.100.033208},
   DOI={10.1103/physreve.100.033208},
   number={3},
   journal={Physical Review E},
   publisher={American Physical Society (APS)},
   author={Kasim, M. F. and Bott, A. F. A. and Tzeferacos, P. and Lamb, D. Q. and Gregori, G. and Vinko, S. M.},
   year={2019},
   month=sep }

@article{tc_gotze,
title = {Exponential Integrability and Transportation Cost Related to Logarithmic Sobolev Inequalities},
journal = {Journal of Functional Analysis},
volume = {163},
number = {1},
pages = {1-28},
year = {1999},
author = {S.G Bobkov and F Götze}
}

@inproceedings{path_coupling,
author = {Bubley, R. and Dyer, M.},
title = {Path coupling: A technique for proving rapid mixing in Markov chains},
year = {1997},
isbn = {0818681977},
publisher = {IEEE Computer Society},
address = {USA},
booktitle = {Proceedings of the 38th Annual Symposium on Foundations of Computer Science},
pages = {223},
series = {FOCS '97}
}

@book{mcmt,
title = {Markov Chains and Mixing Times, Second Edition}, 
author = {David A. Levin and Yuval Peres},
year = 2017,
publisher = {American Mathematical Society},
url = {https://pages.uoregon.edu/dlevin/MARKOV/mcmt2e.pdf}
}

@article{BRANNAN2022108129,
title = {Complete logarithmic Sobolev inequalities via Ricci curvature bounded below},
journal = {Advances in Mathematics},
volume = {394},
pages = {108129},
year = {2022},
issn = {0001-8708},
doi = {https://doi.org/10.1016/j.aim.2021.108129},
url = {https://www.sciencedirect.com/science/article/pii/S0001870821005685},
author = {Michael Brannan and Li Gao and Marius Junge}
}

@article{Zyczkowski_1998,
   title={The Monge distance between quantum states},
   volume={31},
   ISSN={1361-6447},
   url={http://dx.doi.org/10.1088/0305-4470/31/45/009},
   DOI={10.1088/0305-4470/31/45/009},
   number={45},
   journal={Journal of Physics A: Mathematical and General},
   publisher={IOP Publishing},
   author={Zyczkowski, Karol and Slomczynski, Wojeciech},
   year={1998},
   month=nov, pages={9095–9104} }

@article{caputo2024quantumoptimaltransportconvex,
      title={Quantum optimal transport with convex regularization}, 
      author={Emanuele Caputo and Augusto Gerolin and Nataliia Monina and Lorenzo Portinale},
      year={2024},
    month={9},
      journal={arXiv:2409.03698}
}

@article{gerolin2024noncommutativeoptimaltransportsemidefinite,
      title={Non-commutative Optimal Transport for semi-definite positive matrices}, 
      author={Augusto Gerolin and Nataliia Monina},
      year={2023},
    month = {9},
      journal={arXiv:2309.04846}
}

@article{Nielsen_2006,
   title={Quantum Computation as Geometry},
   volume={311},
   ISSN={1095-9203},
   url={http://dx.doi.org/10.1126/science.1121541},
   DOI={10.1126/science.1121541},
   number={5764},
   journal={Science},
   publisher={American Association for the Advancement of Science (AAAS)},
   author={Nielsen, Michael A. and Dowling, Mark R. and Gu, Mile and Doherty, Andrew C.},
   year={2006},
   month=feb, pages={1133–1135} }

@article{monge-1781,
title={Mémoire sur la théorie des déblais et des remblais},
author = {Monge, G.},
year = {1781},
journal={Histoire de l’Académie Royale des Sciences de Paris, avec les Mémoires de Mathématique et de Physique pour la même année},
pages={666–704}}

@article{kantorovich-1942,
title={On the translocation of masses},
author={Kantorovich, L.},
year={1942},
journal={Dokl. Akad. Nauk SSSR},
volume={133},
number={7-8},
pages={227–229}}

@book{figalli-book-2023,
  title={An invitation to optimal transport, Wasserstein distances, and gradient flows},
  author={Figalli, Alessio and Glaudo, Federico},
  year={2023}
}

@article{benamou-brenier-2000,
  title={A computational fluid mechanics solution to the Monge-Kantorovich mass transfer problem},
  author={Benamou, Jean-David and Brenier, Yann},
  journal={Numerische Mathematik},
  volume={84},
  number={3},
  pages={375--393},
  year={2000}
}

@article{buttazzo-2009,
title={A Mass Transportation Model for the Optimal Planning
of an Urban Region},
author={Giuseppe Buttazzo and Filippo Santambrogio},
year={2009},
journal={SIAM Review},
volume={51},
number={3},
pages={593-610}}

@misc{santambrogio-2009,
title={Models and applications of Optimal Transport Theory [lecture notes]},
author={Filippo Santambrogio},
year={2009},
url={https://math.univ-lyon1.fr/~santambrogio/Grenoble%20Models.pdf}}

@article{15_min_city,
title={15-Minute City: Decomposing the New Urban Planning Eutopia},
author={Georgia Pozoukidou and Zoi Chatziyiannaki},
journal={Sustainability},
year={2021},
volume={13},
number={2},
url={https://doi.org/10.3390/su13020928}
}

@book{villani2008optimal,
  title={Optimal Transport: Old and New},
  author={Villani, C.},
  isbn={9783540710493},
  lccn={2008932183},
  series={Grundlehren der mathematischen Wissenschaften},
  url={https://books.google.co.uk/books?id=NZXiNAEACAAJ},
  year={2008},
  publisher={Springer Berlin Heidelberg}
}

@article{sinkhorn-1964,
author = {Richard Sinkhorn},
title = {{A Relationship Between Arbitrary Positive Matrices and Doubly Stochastic Matrices}},
volume = {35},
journal = {The Annals of Mathematical Statistics},
number = {2},
publisher = {Institute of Mathematical Statistics},
pages = {876 -- 879},
year = {1964},
doi = {10.1214/aoms/1177703591},
URL = {https://doi.org/10.1214/aoms/1177703591}
}

@article{otto-jordan-1988,
author = {Jordan, Richard and Kinderlehrer, David and Otto, Felix},
title = {The Variational Formulation of the Fokker--Planck Equation},
journal = {SIAM Journal on Mathematical Analysis},
volume = {29},
number = {1},
pages = {1-17},
year = {1998}
}

@article{schrodinger-connections-2013,
author = {Léonard, Christian},
year = {2013},
month = {08},
pages = {},
title = {A survey of the Schrödinger problem and some of its connections with optimal transport},
volume = {34},
journal = {Discrete and Continuous Dynamical Systems},
doi = {10.3934/dcds.2014.34.1533}
}

@article{schrodinger-connections-meanfield-2020,
author = {Backhoff, Julio and Conforti, Giovanni and Gentil, Ivan and Léonard, Christian},
year = {2020},
month = {10},
pages = {},
title = {The mean field Schrödinger problem: ergodic behavior, entropy estimates and functional inequalities},
volume = {178},
journal = {Probability Theory and Related Fields},
doi = {10.1007/s00440-020-00977-8}
}

@article{marton-1996,
 ISSN = {00911798, 2168894X},
 URL = {http://www.jstor.org/stable/2244952},
 author = {K. Marton},
 journal = {The Annals of Probability},
 number = {2},
 pages = {857--866},
 publisher = {Institute of Mathematical Statistics},
 title = {Bounding d-Distance by Informational Divergence: A Method to Prove Measure Concentration},
 urldate = {2025-03-05},
 volume = {24},
 year = {1996}
}

@article{otto-villani-2000,
title = {Generalization of an Inequality by Talagrand and Links with the Logarithmic Sobolev Inequality},
journal = {Journal of Functional Analysis},
volume = {173},
number = {2},
pages = {361-400},
year = {2000},
issn = {0022-1236},
doi = {https://doi.org/10.1006/jfan.1999.3557},
url = {https://www.sciencedirect.com/science/article/pii/S0022123699935577},
author = {F. Otto and C. Villani}
}

@misc{jog-2024,
title={Part III Concentration Inequalities [lecture notes]},
author={Jog, V.},
year={2024},
publisher={University of Cambridge}}

@article{milman-1983,
  title={A topological application of the isoperimetric inequality},
  author={V. D. Milman and Mikhael Gromov},
  journal={American Journal of Mathematics},
  year={1983},
  volume={105},
  pages={843},
  url={https://api.semanticscholar.org/CorpusID:14234600}
}

@article{erbar-maas-2012,
   title={Ricci Curvature of Finite Markov Chains via Convexity of the Entropy},
   volume={206},
   ISSN={1432-0673},
   url={http://dx.doi.org/10.1007/s00205-012-0554-z},
   DOI={10.1007/s00205-012-0554-z},
   number={3},
   journal={Archive for Rational Mechanics and Analysis},
   publisher={Springer Science and Business Media LLC},
   author={Erbar, Matthias and Maas, Jan},
   year={2012},
   month={8}, pages={997–1038} }

@article{gentil-hwi-2020,
title = {An entropic interpolation proof of the HWI inequality},
journal = {Stochastic Processes and their Applications},
volume = {130},
number = {2},
pages = {907-923},
year = {2020},
issn = {0304-4149},
doi = {https://doi.org/10.1016/j.spa.2019.04.002},
url = {https://www.sciencedirect.com/science/article/pii/S0304414918303454},
author = {Ivan Gentil and Christian Léonard and Luigia Ripani and Luca Tamanini}
}

@article{lott-villani-2005,
author = {Lott, John and Villani, Cedric},
year = {2005},
month = {01},
pages = {},
title = {Ricci curvature for metric-measure spaces via optimal transport},
volume = {169},
journal = {Annals of Mathematics},
doi = {10.4007/annals.2009.169.903}
}

@article{opalgs,
author={Landsman, N. P.},
title={Lecture Notes on Operator Algebras},
publisher={Radboud Universiteit},
year={2011},
url={https://www.math.ru.nl/~landsman/oa2011.pdf}
}

@inbook{otqs-palma-trevisan,
   title={Quantum Optimal Transport: Quantum Channels and Qubits},
   ISSN={2947-9460},
   booktitle={Optimal Transport on Quantum Structures},
   publisher={Springer Nature Switzerland},
   author={De Palma, Giacomo and Trevisan, Dario},
   year={2024},
   pages={203–239} }

@article{baumgratz-2014,
  title = {Quantifying Coherence},
  author = {Baumgratz, T. and Cramer, M. and Plenio, M. B.},
  journal = {Phys. Rev. Lett.},
  volume = {113},
  issue = {14},
  pages = {140401},
  numpages = {5},
  year = {2014},
  month = {9},
  publisher = {American Physical Society},
  doi = {10.1103/PhysRevLett.113.140401},
  url = {https://link.aps.org/doi/10.1103/PhysRevLett.113.140401}
}

@article{sturm-2005,
author = {Sturm, Karl-Theodor and Renesse, Max},
year = {2005},
month = {07},
pages = {},
title = {Transport inequalities, gradient estimates, entropy and Ricci curvature},
volume = {58},
journal = {Communications on Pure and Applied Mathematics - COMMUN PURE APPL MATH},
doi = {10.1002/cpa.20060}
}

@article{kastoryano-2012,
author = {Kastoryano, Michael and Temme, Kristan},
year = {2012},
month = {07},
pages = {},
title = {Quantum logarithmic Sobolev inequalities and rapid mixing},
volume = {54},
journal = {Journal of Mathematical Physics},
doi = {10.1063/1.4804995}
}

@article{erbar-2016,
title = {Poincaré, modified logarithmic Sobolev and isoperimetric inequalities for Markov chains with non-negative Ricci curvature},
journal = {Journal of Functional Analysis},
volume = {274},
number = {11},
pages = {3056-3089},
year = {2018},
issn = {0022-1236},
doi = {https://doi.org/10.1016/j.jfa.2018.03.011},
url = {https://www.sciencedirect.com/science/article/pii/S0022123618301101},
author = {Matthias Erbar and Max Fathi}
}

@article{og-gans,
author = {Goodfellow, Ian and Pouget-Abadie, Jean and Mirza, Mehdi and Xu, Bing and Warde-Farley, David and Ozair, Sherjil and Courville, Aaron and Bengio, Y.},
year = {2014},
month = {06},
pages = {},
title = {Generative Adversarial Networks},
volume = {3},
journal = {Advances in Neural Information Processing Systems},
doi = {10.1145/3422622}
}

@article{og_qgans,
   title={Quantum generative adversarial networks},
   volume={98},
   ISSN={2469-9934},
   url={http://dx.doi.org/10.1103/PhysRevA.98.012324},
   DOI={10.1103/physreva.98.012324},
   number={1},
   journal={Physical Review A},
   publisher={American Physical Society (APS)},
   author={Dallaire-Demers, Pierre-Luc and Killoran, Nathan},
   year={2018},
   month={7} }

@book{Nielsen_Chuang, place={Cambridge}, title={Quantum Computation and Quantum Information: 10th Anniversary Edition}, publisher={Cambridge University Press}, author={Nielsen, Michael A. and Chuang, Isaac L.}, year={2010}}

@article{classical_applications_signals_machinelearning,
author = {Kolouri, Soheil and Park, Serim and Thorpe, Matthew and Slepcev, Dejan and Rohde, Gustavo},
year = {2017},
month = {07},
pages = {43-59},
title = {Optimal Mass Transport: Signal processing and machine-learning applications},
volume = {34},
journal = {IEEE Signal Processing Magazine},
doi = {10.1109/MSP.2017.2695801}
}

@InProceedings{pmlr-v70-arjovsky17a,
  title = 	 {{W}asserstein Generative Adversarial Networks},
  author =       {Martin Arjovsky and Soumith Chintala and L{\'e}on Bottou},
  booktitle = 	 {Proceedings of the 34th International Conference on Machine Learning},
  pages = 	 {214--223},
  year = 	 {2017},
  editor = 	 {Precup, Doina and Teh, Yee Whye},
  volume = 	 {70},
  series = 	 {Proceedings of Machine Learning Research},
  month = 	 {8},
  publisher =    {PMLR},
  url = 	 {https://proceedings.mlr.press/v70/arjovsky17a.html}
}
\end{document}